\documentclass[aps,prd,superscriptaddress,preprint]{revtex4-1}

\usepackage{amsmath,amssymb,bm}
\usepackage{graphicx,graphics,color}
\usepackage[colorlinks=true,linkcolor=blue,bookmarks=false]{hyperref}
\usepackage{epsfig}
\usepackage{latexsym}
\usepackage{rotating}
\usepackage{subfigure}
\usepackage{centernot}
\usepackage{longtable}
\usepackage{hhline,multirow,tabularx}  
\usepackage{float}
\usepackage{mathrsfs}
\usepackage{xr}
\usepackage{adjustbox}

\newcommand{\RNum}[1]{\uppercase\expandafter{\romannumeral #1\relax}}
\newcommand{\an}{\alpha^{(n)}}
\newcommand{\bn}{\beta^{(n)}}
\newcommand{\sn}{\sigma^{(n)}}
\newcommand{\tn}{\theta^{(n)}}
\newcommand{\rn}{\rho^{(n)}}

\newcommand{\ban}{\bar{\alpha}^{(n)}}
\newcommand{\bbn}{\bar{\beta}^{(n)}}
\newcommand{\bsn}{\bar{\sigma}^{(n)}}
\newcommand{\btn}{\bar{\theta}^{(n)}}
\newcommand{\brn}{\bar{\rho}^{(n)}}
\newcommand{\bon}{\bar{\omega}^{(n)}}

\begin{document}

\preprint{JLAB-THY-19-2996, ADP-19-15/T1095}

\title{Parton distributions from nonlocal chiral SU(3) effective theory:
	Flavor asymmetries}

\author{Y. Salamu}
\affiliation{Institute of High Energy Physics, CAS,
	Beijing 100049, China}
\affiliation{School of Physical Sciences, University of Chinese Academy
	of Sciences, Beijing 100049, China}
\author{Chueng-Ryong Ji}
\affiliation{Department of Physics, North Carolina State University,
	Raleigh, NC 27695, USA}
\author{W. Melnitchouk}
\affiliation{Jefferson Lab, Newport News,
	Virginia 23606, USA}
\author{A. W. Thomas}
\affiliation{CoEPP and CSSM, Department of Physics, University of Adelaide,
	Adelaide SA 5005, Australia}
\author{P. Wang}
\affiliation{Institute of High Energy Physics, CAS,
	Beijing 100049, China}
\affiliation{Theoretical Physics Center for Science Facilities, CAS,
	Beijing 100049, China}
\author{X. G. Wang}
\affiliation{CoEPP and CSSM, Department of Physics, University of Adelaide,
	Adelaide SA 5005, Australia}

\begin{abstract}
Using recently derived results for one-loop hadronic splitting functions
from a nonlocal implementation of chiral effective theory, we study
the contributions from pseudoscalar meson loops to flavor asymmetries
in the proton.
Constraining the parameters of the regulating functions by inclusive
production of $n$, $\Delta^{++}$, $\Lambda$ and $\Sigma^{*+}$ baryons in
$pp$ collisions, we compute the shape of the light antiquark asymmetry
$\bar{d}-\bar{u}$ in the proton and the strange asymmetry $s-\bar{s}$
in the nucleon sea.
With these constraints, the magnitude of the $\bar{d}-\bar{u}$ asymmetry
is found to be compatible with that extracted from the Fermilab E866
Drell-Yan measurement, with no indication of a sign change at large
values of $x$, and an integrated value in the range
    $\langle \bar d-\bar u \rangle \approx 0.09-0.17$.
The $s-\bar s$ asymmetry is predicted to be positive at $x > 0$,
with compensating negative contributions at $x=0$, and an integrated
$x$-weighted moment in the range
    $\langle x (s-\bar s) \rangle \approx (0.9-2.5) \times 10^{-3}$.
\end{abstract}

\date{\today}
\maketitle

\section{Introduction}
\label{sec:intro}

It is well known that a complete characterization of nucleon
substructure must go beyond three valence quarks.  One of the great
challenges of modern hadron physics is to unravel the precise role
of hidden flavors in the structure of the nucleon.  The observation
of the $\bar{d}-\bar{u}$ flavor asymmetry in the light quark sea
of the proton~\cite{Arneodo:1994sh, Ackerstaff:1998sr, Baldit,
Towell:2001nh}, following its prediction by Thomas a decade
earlier~\cite{Thomas:1983fh} on the basis of chiral symmetry
breaking~\cite{Thomas:1982kv, Thomas:1981vc}, has been one of
the seminal results in hadronic physics over the past two decades.
It has led to a major reevaluation of our understanding of the
role of the non-valence components of the nucleon and their
origin in QCD~\cite{Geesaman:2018ixo}.

The role that strange quarks, in particular, play in the nucleon has
also been the focus of attention in hadronic physics for many years.
Early polarized deep-inelastic scattering (DIS) experiments suggested
that a surprisingly large fraction of the proton's spin might be
carried by strange quarks~\cite{Ashman}, in contrast to the naive quark
model expectations~\cite{Ellis}. One of the guiding principles for
understanding the nonperturbative features of strange quarks and
antiquarks in the nucleon sea has been chiral symmetry breaking in QCD.
While the generation of $s\bar{s}$ pairs through perturbative gluon
radiation typically produces symmetric $s$ and $\bar{s}$ distributions
(at least up to two loop corrections~\cite{Catani04}), any
significant difference between the momentum dependence of the
$s$ and $\bar{s}$ distributions would be a clear signal of
nonperturbative effects~\cite{Signal87, Malheiro97, Sufian18}.

In the previous paper~\cite{nonlocal-I}, we presented the proton $\to$
pseudoscalar meson ($\phi$) $+$ baryon splitting functions for the
intermediate octet ($B$) and decuplet ($T$) baryon configurations
in nonlocal chiral effective theory~\cite{He1, He2}.
From the calculated splitting functions, the parton distribution
functions (PDFs) of the nucleon are obtained as convolutions of
these with PDFs of the intermediate state mesons and
baryons~\cite{Chen02, XGWangPLB, XGWangPRD}.
Here we apply the results from~\cite{nonlocal-I} to compute,
for the first time within the nonlocal theory, sea quark PDF
asymmetries in the proton, including the light antiquark
flavor asymmetry $\bar{d}-\bar{u}$ and the strange quark
asymmetry $s-\bar{s}$.
Using SU(3) relations for the intermediate state hadron PDFs, the
only parameters in the calculation of the asymmetries are the mass
parameters appearing in the ultraviolet regulator functions.
These will be determined by fitting cross section data from
inclusive baryon production in high energy $pp$ scattering, using
the same splitting functions that appear in the PDF asymmetries.

We begin in Sec.~\ref{sec:convolution} by summarizing the convolution
formulas for the quark and antiquark PDFs in terms of the fluctuations
of the nucleon into its meson-baryon light-cone components.
The calculation of the PDFs of the intermediate state baryons
and mesons in the chiral theory is discussed in detail in
Sec.~\ref{sec:barePDF}.
Numerical results for the sea quark asymmetries are presented in
Sec.~\ref{sec.results}, where we compare the results for
$\bar d - \bar u$ with those extracted from Drell-Yan and
semi-inclusive DIS measurements, and compare predictions for
$s - \bar s$ asymmetries with some recent PDF parametrizations.
Finally, in Sec.~\ref{sec.conclusion} we summarize our results
and discuss future measurements which could further constrain
the PDF asymmetries experimentally.

\section{Convolution formulas}
\label{sec:convolution}

Using the crossing symmetry properties of the spin-averaged PDFs,
$q(-x) = -\bar q(x)$, the $n$-th Mellin moment ($n \geq 1$) of
the distribution for a given flavor $q$ ($q = u, d, s, \ldots$)
is defined by
\begin{eqnarray}
Q^{(n-1)}
&=& \int_0^1 dx\, x^{n-1}
    \big[ q(x) + (-1)^n\, \bar{q}(x) \big].
\label{eq:mom_def}
\end{eqnarray}
In the operator product expansion, the moments $Q^{(n-1)}$ are
related to matrix elements of local twist-two, spin-$n$ operators
$\mathcal{O}_q^{\mu_1 \cdots \mu_n}$ between nucleon states with
momentum~$p$,
\begin{equation}
\langle N(p) |\, \mathcal{O}_q^{\mu_1 \cdots \mu_n}\, | N(p) \rangle
= 2\, Q^{(n-1)}\, p^{\mu_1} \cdots p^{\mu_n},
\end{equation}
where the operators are given by
\begin{equation}
\label{eq:Oq}
\mathcal{O}^{\mu_1 \cdots \mu_n}_q
= i^{n-1}\, \bar{q} \gamma^{ \{ \mu_1 }\overleftrightarrow{D}^{\mu_2}
  \cdots \overleftrightarrow{D}^{ \mu_n \} } q\, ,
\end{equation}
with
$\overleftrightarrow{D}
 = \frac{1}{2} \big( \overrightarrow{D} - \overleftarrow{D} \big)$,
and the braces $\{\, \cdots \}$ denote symmetrization of
Lorentz indices.
The effective theory allows the quark operators ${\cal O}_q$
to be matched to hadronic operators ${\cal O}_j$ with the same
quantum numbers~\cite{Chen02},
\begin{equation}
\mathcal{O}^{\mu_1 \cdots \mu_n}_q
= \sum_{j} c^{(n)}_{q/j}\ \mathcal{O}^{\mu_1 \cdots \mu_n}_j,
\label{eq:match}
\end{equation}
where the coefficients $c^{(n)}_{q/j}$ are the $n$-th moments of
the PDF $q_j(x)$ in the hadronic configuration $j$,
\begin{eqnarray}
\label{eq:cnqj}
c^{(n)}_{q/j}
&=& \int_{-1}^1 dx\, x^{n-1}\, q_j(x)\,
\equiv\, Q^{(n-1)}_j\ .
\end{eqnarray}
The nucleon matrix elements of the hadronic operators
${\cal O}_j^{\mu_1 \cdots \mu_n}$ are given in terms of moments
of the splitting functions $f_j(y)$,
\begin{equation}
\langle N(p) | \mathcal{O}_j^{\mu_1 \cdots \mu_n} | N(p) \rangle
= 2\, f_j^{(n)}\,
  p^{ \{ \mu_1 } \cdots p^{ \mu_n\} },
\label{eq:fjn_def}
\end{equation}
where
\begin{eqnarray}
f_j^{(n)} &=& \int_{-1}^{1} dy\, y^{n-1} f_j(y),
\label{eq:fjn}
\end{eqnarray}
with $y$ the light-cone momentum fraction of the nucleon
carried by the hadronic state $j$.
The operator relation in Eq.~(\ref{eq:match}) then gives rise to
the convolution formula for the PDFs~\cite{Chen02, XGWangPRD},
\begin{eqnarray}
q(x)
&=& \sum_j \big[ f_j \otimes q_j^v \big](x)\
\equiv\ \sum_j
        \int_0^1 dy \int_0^1 dz\, \delta(x-yz)\, f_j(y)\, q_j^v(z),
\label{eq:conv}
\end{eqnarray}
where $q_j^v \equiv q_j - \bar{q}_j$ is the valence distribution
for the quark flavor $q$ in the hadron $j$.
The complete set of splitting functions $f_j(y)$ for octet
and decuplet baryons is given in Ref.~\cite{nonlocal-I}.

\begin{figure}[t]
\centering
\includegraphics[width=0.6\textwidth]{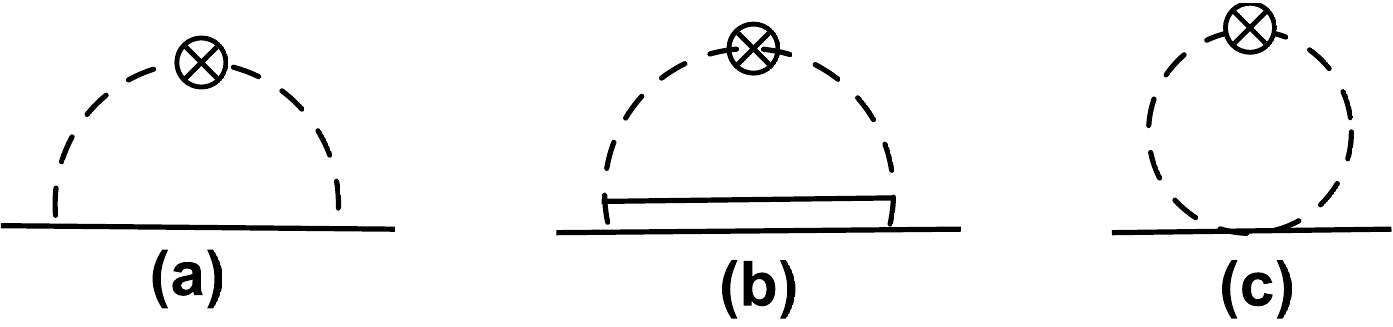}
\caption{One-meson loop diagrams contributing to quark and
	antiquark PDFs in the nucleon, representing
   {\bf (a)}~rainbow diagram with octet baryon (solid lines)
	    intermediate state;
   {\bf (b)}~rainbow diagram with decuplet baryon (double solid lines)
	    intermediate state;
   {\bf	(c)}~meson (dashed lines) bubble diagram.
	The symbol ``$\otimes$'' represents an operator insertion.}
\label{fig:loop_meson}
\end{figure}

In the present analysis we work under the basic assumption that
the bare baryon states are composed of three valence quarks plus
quark-antiquark pairs that are generated perturbatively through
gluon radiation.  Such contributions will effectively cancel in
any differences of PDFs, such as $\bar d-\bar u$ or $s-\bar s$.
We therefore focus only on the nonperturbative contributions to
sea quark PDFs which arise from pseudoscalar meson loops.
In this approximation antiquark distributions arise only from
diagrams involving direct coupling to mesons, as in the meson
rainbow and bubble diagrams in Fig.~\ref{fig:loop_meson}.
The meson loop contribution to the antiquark PDFs in the nucleon
can then be written as
\begin{equation}
\begin{split}
\bar{q}(x)
&= \sum\limits_{B,T,\phi}
\big[
  \big( f^{\rm (rbw)}_{\phi B}
      + f^{\rm (rbw)}_{\phi T}
      + f^{\rm (bub)}_{\phi}
  \big) \otimes \bar{q}_\phi
\big](x),
\end{split}
\label{eq:antiquark}
\end{equation}
where $f^{\rm (rbw)}_{\phi B}$ and $f^{\rm (rbw)}_{\phi T}$
represent splitting functions from the rainbow diagrams with
octet and decuplet baryons in Fig.~\ref{fig:loop_meson}(a) and (b),
respectively, $f^{\rm (bub)}_{\phi}$ is the splitting function for
the meson bubble diagram in Fig.~\ref{fig:loop_meson}(c), and
${\bar q}_\phi(x)$ is the antiquark PDF in the meson.

\begin{figure}[t]
\centering
\includegraphics[width=\textwidth]{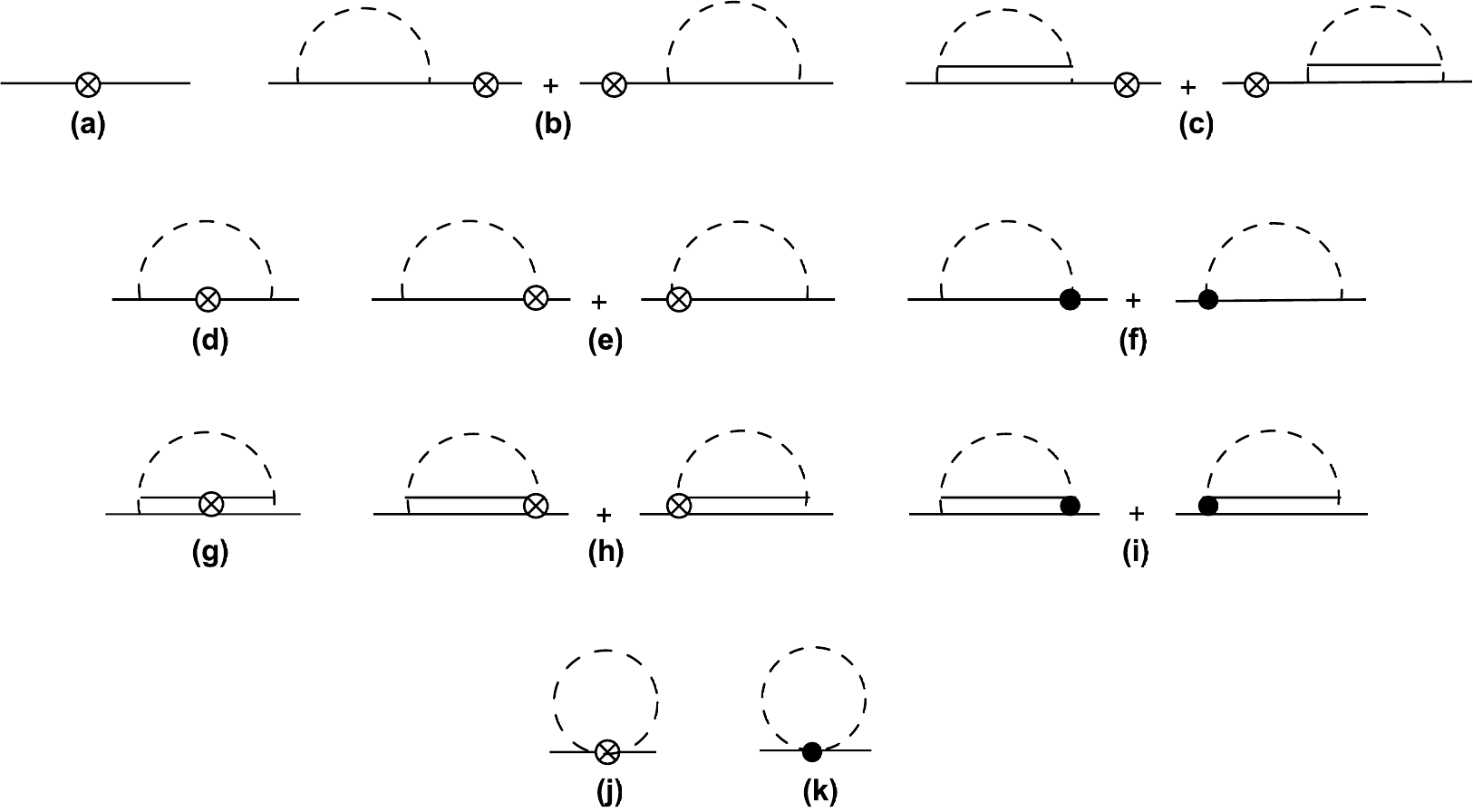}
\caption{Contributions to quark PDFs in the nucleon from baryon
	coupling diagrams, representing
   {\bf (a)}~coupling to the bare nucleon;
   {\bf	(b)}~and~{\bf (c)}~contributions from wave function
	    renormalization with octet and decuplet baryon
	    intermediate states;
   {\bf (d)}~rainbow diagram with octet baryon;
   {\bf (e)}~and~{\bf (f)}~Kroll-Ruderman, and gauge link
	    (filled circle) Kroll-Ruderman diagrams with octet baryon;
   {\bf (g)}~rainbow diagram with decuplet baryon;
   {\bf (h)}~and~{\bf (i)}~Kroll-Ruderman diagrams with decuplet baryon;
   {\bf (j)}~and~{\bf (k)}~meson tadpole and gauge link tadpole diagrams.}
\label{fig:loop_baryon}
\end{figure}

Contributions to quark PDFs can in principle come from both meson
coupling and baryon coupling diagrams.  The latter are illustrated
in Fig.~\ref{fig:loop_baryon}, and involve the
bare nucleon coupling [Fig.~\ref{fig:loop_baryon}(a)],
wave function renormalization [Fig.~\ref{fig:loop_baryon}(b) and (c),
with octet and decuplet baryon intermediate states, respectively],
baryon rainbow [Fig.~\ref{fig:loop_baryon}(d) and (g)],
Kroll-Ruderman [Fig.~\ref{fig:loop_baryon}(e) and (h)], and
meson tadpole [Fig.~\ref{fig:loop_baryon}(j)] diagrams,
along with gauge link dependent Kroll-Ruderman
[Fig.~\ref{fig:loop_baryon}(f) and (i)]
and tadpole [Fig.~\ref{fig:loop_baryon}(k)] diagrams.
Within the valence approximation, all of these diagrams will contribute
to the $u$ and $d$ quarks in the nucleon.  However, for the strange
quark the bare coupling and wave function renormalization diagrams
do not contribute.
The total nonperturbative contribution from meson loops to the
quark PDF in the nucleon can then be written
\begin{equation}
\begin{split}
q(x)
=&\ Z_2\, q^{(0)}(x)\
+\ \sum\limits_{B,T,\phi}
\bigg\{
   \big[
   \big( f^{\rm (rbw)}_{\phi B}
       + f^{\rm (rbw)}_{\phi T}
       + f^{\rm (bub)}_{\phi}
   \big) \otimes q_\phi
   \big](x)				\\
&
+\ \big[ {\bar f}^{\rm (rbw)}_{B\phi} \otimes q_B \big](x)
+\ \big[ {\bar f}^{\rm (KR)}_B \otimes q^{\rm (KR)}_B \big](x)
+\ \big[ \delta {\bar f}^{\rm (KR)}_B \otimes q^{(\delta)}_B \big](x) \\
&
+\ \big[ {\bar f}^{\rm (rbw)}_{T\phi} \otimes q_T \big](x)	
+\ \big[ {\bar f}^{\rm (KR)}_T \otimes q^{\rm (KR)}_T \big](x)
+\ \big[ \delta {\bar f}^{\rm (KR)}_T \otimes q^{(\delta)}_T \big](x) \\
&
+\ \big[ {\bar f}^{\rm (tad)}_\phi \otimes q^{\rm (tad)}_\phi \big](x)
+\ \big[ \delta {\bar f}^{\rm (tad)}_\phi \otimes q^{(\delta)}_\phi \big](x)
\bigg\},
\end{split}
\label{eq:quark}
\end{equation}
where $q^{(0)}$ is the quark PDF in the bare nucleon, and the
wave function renormalization $Z_2$ arises from the summation over
the diagrams in Figs.~\ref{fig:loop_baryon}(a)--(c)~\cite{Ji:2013bca}.
Following Ref.~\cite{XGWangPRD}, we will work in terms of the same
momentum fraction $y$ for all meson and baryon coupling diagrams in
Figs.~\ref{fig:loop_meson} and \ref{fig:loop_baryon}.  Using the same
definition of the convolution integral as in Eq.~(\ref{eq:conv}),
it will be convenient therefore to define for each of the splitting
functions in Eq.~(\ref{eq:quark}) involving the coupling to baryons
the shorthand notation ${\bar f}_j(y) \equiv f_j(1-y)$
[see Sec.~\ref{ssec:ssbar} below].
Explicit expressions for the splitting functions
  $f^{\rm (rbw)}_{B\phi}$, $f^{\rm (KR)}_B$, $\delta f^{\rm (KR)}_B$,
  $f^{\rm (rbw)}_{T\phi}$, $f^{\rm (KR)}_T$, $\delta f^{\rm (KR)}_T$,
  $f^{\rm (tad)}_\phi$ and $\delta f^{\rm (tad)}_\phi$,
which represent the diagrams in Figs.~\ref{fig:loop_baryon}(d)--(k),
respectively, are given in Ref.~\cite{nonlocal-I}.
The corresponding quark PDFs for the intermediate state octet and
decuplet baryons are discussed in the next section.

\section{Bare baryon and meson PDFs}
\label{sec:barePDF}

To calculate the contributions to the quark and antiquark distributions
in the proton in the convolution formulas (\ref{eq:antiquark}) and
(\ref{eq:quark}) requires the proton $\to$ baryon $+$ meson splitting
functions and the PDFs of the baryons and mesons to which the current
couples.  The full set of splitting functions was presented in our
previous paper, Ref.~\cite{nonlocal-I}.
In this section we derive the (valence) PDFs of the bare baryon
and meson intermediate states using the same chiral SU(3) EFT
framework that was used to compute the splitting functions.

\subsection{Operators and moments}

In the effective theory the quark level operators
are matched to a sum of hadronic level operators whose matrix
elements [see Eq.~(\ref{eq:match})] are given by the moments
of the splitting functions, as in Eq.~(\ref{eq:fjn_def}).
Identifying all possible contributions from octet and decuplet
baryon intermediate states that transform as vectors, the most
general expression for the quark vector operator
	$\mathcal{O}^{\mu_1 \cdots \mu_n}_q$
is given by~\cite{XGWangPRD, Shanahan:2013xw}
\begin{equation}
\begin{split}
\label{eq:vector-op}
\mathcal{O}^{\mu_1 \cdots \mu_n}_q
&= a^{(n)} i^n \frac{f_\phi^2}{4}
\Big\{
  \mathrm{Tr}
  \left[ U^\dag \lambda^q \partial_{\mu_1} \cdots \partial_{\mu_n} U
  \right]
+ \mathrm{Tr}
  \left[ U \lambda^q \partial_{\mu_1} \cdots \partial_{\mu_n} U^\dag
  \right]
\Big\}							\\
&+
\Big[
  \alpha^{(n)}
  (\overline{\cal B} \gamma^{\mu_1} {\cal B} \lambda^q_+)
+ \beta^{(n)}
  (\overline{\cal B} \gamma^{\mu_1} \lambda^q_+ {\cal B})
+ \sigma^{(n)}
  (\overline{\cal B} \gamma^{\mu_1} {\cal B})\,
  \mathrm{Tr}[\lambda^q_+]
\Big]\, p^{\mu_2} \cdots p^{\mu_n}			\\
&+
\Big[
  \bar\alpha^{(n)}
  (\overline{\cal B} \gamma^{\mu_1} \gamma_5 {\cal B} \lambda^q_-)
+ \bar\beta^{(n)}
  (\overline{\cal B} \gamma^{\mu_1} \gamma_5 \lambda^q_- {\cal B})
+ \bar\sigma^{(n)}
  (\overline{\cal B} \gamma^{\mu_1} \gamma_5 {\cal B})\,
  \mathrm{Tr}[\lambda^q_-]
\Big]\, p^{\mu_2} \cdots p^{\mu_n}			\\
&+
\Big[
  \theta^{(n)}
  (\overline{\cal T}_\alpha \gamma^{\alpha\beta\mu_1}
  \lambda^q_+ {\cal T}_\beta)
+ \rho^{(n)}
  (\overline{\cal T}_\alpha \gamma^{\alpha\beta\mu_1}
  {\cal T}_\beta)\, {\rm Tr}[\lambda^q_+]
\Big]\, p^{\mu_2} \cdots p^{\mu_n}			\\
&+
\Big[
  \bar\theta^{(n)}
  (\overline{\cal T}_\alpha \gamma^{\mu_1} \gamma_5
  \lambda^q_- {\cal T}^\alpha)
+ \bar\rho^{(n)}
  (\overline{\cal T}_\alpha \gamma^{\mu_1} \gamma_5
  {\cal T}^\alpha\, {\rm Tr}[\lambda^q_-]
\Big]\, p^{\mu_2} \cdots p^{\mu_n}			\\
&-
\sqrt{\frac{3}{2}}\, \bar\omega^{(n)}
\Big[
  (\overline{\cal B} \Theta^{\mu_1\mu} \lambda^q_- {\cal T}_\mu)
+ (\overline{\cal T}_\mu \Theta^{\mu\mu_1} \lambda^q_- {\cal B})
\Big]\, p^{\mu_2} \cdots p^{\mu_n}			\\
&+
\mathrm{permutations} - \mathrm{Tr},
\end{split}
\end{equation}
where ``Tr'' denotes traces over Lorentz indices.
In the first term of Eq.~(\ref{eq:vector-op}), the operator $U$
represents pseudoscalar meson fields $\phi$,
\begin{equation}
U = \exp \left( i \frac{\sqrt2\, \phi}{f_\phi} \right),
\end{equation}
where $f_\phi$ is the pseudoscalar decay constant, and the coefficients
$a^{(n)}$ are related to moments of quark and antiquark PDFs in the
pseudoscalar mesons.
The flavor operators $\lambda_\pm^q$ are defined by
\begin{equation}
\lambda_\pm^q
= \frac{1}{2}
  \big( u \lambda^q u^\dag \pm u^\dag \lambda^q u \big) \, ,
\end{equation}
where $\lambda^q = {\rm diag}(\delta_{qu}, \delta_{qd}, \delta_{qs})$
are diagonal $3 \times 3$ quark flavor matrices.
%

In the remaining terms of Eq.~(\ref{eq:vector-op}), the operators
$\cal B$ and ${\cal T}_\alpha$ represent octet and decuplet baryon
fields, respectively, and we define the Dirac tensors
  $\gamma^{\alpha\beta\rho}
  = \frac{1}{2} \{\gamma^{\mu\nu}, \gamma^\alpha\}$
and
  $\gamma^{\mu\nu}
  = \frac{1}{2} [\gamma^\mu,\gamma^\nu]$.
The coefficients
  $\{ \alpha^{(n)}, \beta^{(n)}, \sigma^{(n)} \}$
and
  $\{ \bar\alpha^{(n)}, \bar\beta^{(n)}, \bar\sigma^{(n)} \}$
are related to moments of the spin-averaged and spin-dependent PDFs
in octet baryons, while
  $\{ \theta^{(n)}, \rho^{(n)} \}$
and
  $\{ \bar\theta^{(n)}, \bar\rho^{(n)} \}$
are related to moments of spin-averaged and spin-dependent PDFs
in decuplet baryons, respectively.
The coefficients $\bar\omega^{(n)}$ are given in terms of moments
of spin-dependent octet--decuplet transition PDFs, where the
octet--decuplet transition tensor operator $\Theta^{\mu\nu}$
is defined~as
\begin{equation}
\Theta^{\mu\nu}
= g^{\mu\nu} - \big( Z+\tfrac12 \big) \gamma^\mu \gamma^\nu.
\label{eq:Theta}
\end{equation}
Here $Z$ is the decuplet off-shell parameter, and since physical
quantities do not depend on $Z$, it is convenient to choose
$Z = 1/2$ to simplify the form of the spin-3/2 propagator
\cite{Hacker:2005fh, Nath:1971wp}.

For the Kroll-Ruderman diagrams in Fig.~\ref{fig:loop_baryon}(e), (f),
(h) and (i), the presence of the pseudoscalar field at the vertex
introduces hadronic axial vector operators, whose contribution to
the quark axial vector operator can in general be written as
\begin{equation}
\label{eq:axial-vector-op}
\begin{split}
\mathcal{O}^{\mu_1 \cdots \mu_n}_{\Delta q}
&=
\Big[
  \bar{\alpha}^{(n)}
  (\overline{\cal B} \gamma^{\mu_1}\gamma_5 {\cal B} \lambda^q_+)
+ \bar{\beta}^{(n)}
  (\overline{\cal B} \gamma^{\mu_1}\gamma_5 \lambda^q_+ {\cal B})
+ \bar{\sigma}^{(n)}
  (\overline{\cal B} \gamma^{\mu_1}\gamma_5 {\cal B})\,
  \mathrm{Tr}[\lambda^q_+]
\Big]\, p^{\mu_2} \cdots p^{\mu_n}			\\
&+
\Big[
  \alpha^{(n)}
  (\overline{\cal B} \gamma^{\mu_1} {\cal B} \lambda^q_-)
+ \beta^{(n)}
  (\overline{\cal B} \gamma^{\mu_1} \lambda^q_- {\cal B})
+ \sigma^{(n)}
  (\overline{\cal B} \gamma^{\mu_1} {\cal B})\,
  \mathrm{Tr}[\lambda^q_-]
\Big]\, p^{\mu_2} \cdots p^{\mu_n}			\\
&+
\Big[
  \bar\theta^{(n)}
  (\overline{\cal T}_\alpha \gamma^{\mu_1} \gamma_5
  \lambda^q_+ {\cal T}^\alpha)
+ \bar\rho^{(n)}
  (\overline{\cal T}_\alpha \gamma^{\mu_1} \gamma_5
  {\cal T}^\alpha) {\rm Tr}[\lambda^q_+]
\Big]\, p^{\mu_2} \cdots p^{\mu_n}			\\
&+
\Big[
  \theta^{(n)}
  (\overline{\cal T}_\alpha \gamma^{\alpha\beta\mu_1}
  \lambda^q_- {\cal T}_\beta)
+ \rho^{(n)}
  (\overline{\cal T}_\alpha \gamma^{\alpha\beta\mu_1}
  {\cal T}_\beta) {\rm Tr}[\lambda^q_-]
\Big]\, p^{\mu_2} \cdots p^{\mu_n}			\\
&-
\sqrt{\frac{3}{2}}\, \bar\omega^{(n)}
\Big[
  (\overline{\cal B} \Theta^{\mu_1\mu} \lambda^q_+ {\cal T}_\mu)
+ (\overline{\cal T}_\mu \Theta^{\mu\mu_1} \lambda^q_+ {\cal B})
\Big]\, p^{\mu_2} \cdots p^{\mu_n}			\\
&+
\mathrm{permutations} - \mathrm{Tr}.
\end{split}
\end{equation}
From the transformation properties of the operators
$\mathcal{O}^{\mu_1 \cdots \mu_n}_q$ and
$\mathcal{O}^{\mu_1 \cdots \mu_n}_{\Delta q}$ under
parity~\cite{Moiseeva13}, the sets of coefficients
  $\{ \alpha^{(n)}, \beta^{(n)}, \sigma^{(n)},
      \theta^{(n)}, \rho^{(n)} \}$
and
  $\{ \bar\alpha^{(n)}, \bar\beta^{(n)}, \bar\sigma^{(n)},
      \bar\theta^{(n)}, \bar\rho^{(n)} \}$
in~(\ref{eq:axial-vector-op}) are the same as those in the
spin-averaged operators in~(\ref{eq:vector-op}).

The operators $\cal B$ and $\cal T_\alpha$ appearing in
Eqs.~(\ref{eq:vector-op}) and (\ref{eq:axial-vector-op}) can be
written in terms of the SU(3) baryon octet fields
  $B^{ij}$
  (which include $N$, $\Lambda$, $\Sigma$ and $\Xi$ fields)
and decuplet baryon fields
  $T_\alpha^{ijk}$
  (which include $\Delta$, $\Sigma^*$, $\Xi^*$ and $\Omega$ fields)
using the relations~\cite{Labrenz:1996jy, XGWangPRD}
\begin{subequations}
\label{eq:Bidentities}
\begin{eqnarray}
( \overline{\cal B}\, {\cal B} )
&=& {\rm Tr}[ \overline{B} B],					\\
( \overline{\cal B}\, {\cal B}\, A)
&=& - \frac{1}{6}\, {\rm Tr}[\overline{B} B A]
    + \frac{2}{3}\, {\rm Tr}[\overline{B} A B]
    + \frac{1}{6}\, {\rm Tr}[\overline{B} B]\, {\rm Tr}[A],	\\
( \overline{\cal B}\, A\, {\cal B} )
&=& - \frac{2}{3}\, {\rm Tr}[\overline{B} B A]
    - \frac{1}{3}\, {\rm Tr}[\overline{B} A B]
    + \frac{2}{3}\, {\rm Tr}[\overline{B} B]\, {\rm Tr}[A],
\end{eqnarray}
\end{subequations}
and
\begin{subequations}
\label{eq:Tidentities}
\begin{eqnarray}
( \overline{\cal T}_{\!\alpha}\, {\cal T}_\beta )
&=& \overline{T}_\alpha^{kji}\, T_\beta^{ijk},			\\
( \overline{\cal T}_{\!\alpha}\, A\, {\cal T}_\beta )
&=& \overline{T}_\alpha^{kji}\, A^{il}\, T_\beta^{ljk},		\\
( \overline{\cal T}_{\!\alpha}\, A\, {\cal B} )
&=& -\sqrt{\frac23}\,
    \overline{T}_\alpha^{ijk} A^{il} B^{jm} \varepsilon^{klm},
\end{eqnarray}
\end{subequations}
where $\varepsilon^{klm}$ is the antisymmetric tensor.
Applying the relations (\ref{eq:Bidentities})
and (\ref{eq:Tidentities}), the vector operator
$\mathcal{O}^{\mu_1 \cdots \mu_n}_q$ in Eq.~(\ref{eq:vector-op})
can then be more intuitively rearranged in the form
\begin{equation}
\begin{split}
\label{eq:vector-operator}
\mathcal{O}^{\mu_1 \cdots \mu_n}_q
&=\ Q^{(n-1)}_\phi\,
    \mathcal{O}^{\mu_1 \cdots \mu_n}_\phi\
 +\ Q^{(n-1)}_B\,
    \mathcal{O}^{\mu_1 \cdots \mu_n}_{BB'}\
 +\ Q^{(n-1)}_T\,
    \mathcal{O}^{\mu_1 \cdots \mu_n}_{TT'}		\\
&+\ Q^{(n-1)}_{{\rm (tad)} B\phi\phi}\,
    \mathcal{O}^{\mu_1 \cdots \mu_n}_{B\phi\phi}\
 +\ Q_{{\rm (KR)} B}^{(n-1)}\,
    \mathcal{O}^{\mu_1 \cdots \mu_n}_{BB'\phi}\
 +\ Q_{{\rm (KR)} T}^{(n-1)}\,
    \mathcal{O}^{\mu_1 \cdots \mu_n}_{BT\phi}\, .
\end{split}
\end{equation}
The individual vector hadronic operators in (\ref{eq:vector-operator})
are given by
\begin{subequations}
\label{eq:had_ops}
\begin{eqnarray}
\mathcal{O}^{\mu_1 \cdots \mu_n}_\phi
&=& i^n
    \big( \bar\phi\, \partial^{\mu_1} \cdots \partial^{\mu_n} \phi
        - \phi\, \partial^{\mu_1} \cdots \partial^{\mu_n} \bar\phi
    \big),					\\
\mathcal{O}^{\mu_1 \cdots \mu_n}_{B B'}
&=& \big( \overline{B}' \gamma^{\mu_1} B \big)\,
    p^{\mu_2} \cdots p^{\mu_n},			\\
\mathcal{O}^{\mu_1 \cdots \mu_n}_{T T'}
&=& \big( \overline{T}'_\alpha \gamma^{\alpha\beta\mu_1} T_\beta \big)\,
    p^{\mu_2} \cdots p^{\mu_n},			\\
\mathcal{O}^{\mu_1 \cdots \mu_n}_{B \phi \phi}
&=& \frac{1}{f^2_\phi}
    \big( \overline{B} \gamma^{\mu_1} B \bar{\phi}\, \phi \big)\,
    p^{\mu_2} \cdots p^{\mu_n},			\\
\mathcal{O}^{\mu_1 \cdots \mu_n}_{B B' \phi}
&=& \frac{i}{f_\phi}
    \big( \overline{B}' \gamma^{\mu_1}\gamma_5 B \phi
        - \overline{B}  \gamma^{\mu_1}\gamma_5 B' \bar{\phi}
    \big)\,
    p^{\mu_2} \cdots p^{\mu_n},                 \\
\mathcal{O}^{\mu_1 \cdots \mu_n}_{B T \phi}
&=& \frac{i}{f_\phi}
    \big( \overline{B} \Theta^{\mu_1\nu} T_{\nu} \bar{\phi}
        - \overline{T}_\nu \Theta^{\nu \mu_1} B \phi \big)\,
    p^{\mu_2} \cdots p^{\mu_n},
\end{eqnarray}
\end{subequations}
and correspond to the insertions in the diagrams of
Figs.~\ref{fig:loop_meson},
\ref{fig:loop_baryon}(d),
\ref{fig:loop_baryon}(g),
\ref{fig:loop_baryon}(j),
\ref{fig:loop_baryon}(e),
and
\ref{fig:loop_baryon}(h),
respectively.
The coefficients $Q_j^{(n-1)}$ of each of the operators are defined
in terms of Mellin moments of the corresponding parton distributions
in the intermediate mesons and baryons, as in Eq.~(\ref{eq:cnqj}),
\begin{subequations}
\label{eq:Qn-1}
\begin{eqnarray}
Q^{(n-1)}_\phi
&=& \int_{-1}^1 dx\, x^{n-1}\, q_\phi(x),		\\
Q^{(n-1)}_B
&=& \int_{-1}^1 dx\, x^{n-1}\, q_B(x),			\\
Q^{(n-1)}_T
&=& \int_{-1}^1 dx\, x^{n-1}\, q_T(x),			\\
Q^{(n-1)}_{{\rm (tad)} B\phi\phi}
&=& \int_{-1}^1 dx\, x^{n-1}\, q_\phi^{(\rm tad)},	\\
Q_{{\rm (KR)} B}^{(n-1)}
&=& \int_{-1}^1 dx\, x^{n-1}\, q_B^{(\rm KR)},		\\
Q_{{\rm (KR)} T}^{(n-1)}
&=& \int_{-1}^1 dx\, x^{n-1}\, q_T^{(\rm KR)},
\end{eqnarray}
\end{subequations}
where the PDFs correspond to those appearing in the convolution
expressions in Eqs.~(\ref{eq:antiquark}) and (\ref{eq:quark}).
%
%
Each of the moments $Q^{(n-1)}_j$ can be expressed in terms of the
coefficients
   $\{ a^{(n)}; \alpha^{(n)}, \beta^{(n)}, \sigma^{(n)};
		\theta^{(n)}, \rho^{(n)} \}$
appearing in Eq.~(\ref{eq:vector-op}), as discussed below.

In particular, for the meson PDFs, the contributions to the
$U_\phi^{(n-1)}$, $D_\phi^{(n-1)}$ and $S_\phi^{(n-1)}$ moments
are listed in Table~\ref{tab:meson} for the $\phi = \pi^+, K^+$
and $K^0$ mesons.
Conservation of valence quark number fixes the normalization
of the $n=1$ moment of the meson distribution, such that
\begin{equation}
a^{(1)} = 2.
\end{equation}
Note that in the SU(3) symmetric limit, the $u$-quark moments in
$\pi^+$ and $K^+$ are equivalent, as are the $s$-quark moments
in $K^+$ and $K^0$, while the $d$-quark moments in $\pi^+$ and
$K^0$ have equal magnitude but opposite sign,
\begin{equation}
\label{eq:pi_mom}
U_{\pi^+}^{(n-1)} = -D_{\pi^+}^{(n-1)}
= U_{K^+}^{(n-1)}   = -S_{K^+}^{(n-1)}
= D_{K^0}^{(n-1)}   = -S_{K^0}^{(n-1)}
= \frac12 a^{(n)}\, .
\end{equation}
The results for other charge states ($\pi^-$, $\pi^0$, $K^-$ and
$\overline{K}^0$) are obtained from those in Table~\ref{tab:meson}
using charge symmetry.
Unlike in baryons, the sea quark distributions in mesons are
flavor symmetric.
In the simplest valence quark models the sea quark distributions
in pions and kaons are zero.

\begin{table}[t]
\begin{center}
\caption{Moments $Q_\phi^{(n-1)}$ of the quark distributions
	$q (= u, d, s)$ in the pseudoscalar mesons $\pi^+$, $K^+$
	and $K^0$.  The moments are normalized such that
	$a^{(1)} = 2$. \\}
\begin{tabular}{l|ccc} \hline
\ \ $\phi$	&\ \ $U^{(n-1)}_\phi$\ \
		&\ \ $D^{(n-1)}_\phi$\ \
		&\ \ $S^{(n-1)}_\phi$\ \	\\ \hline
\ $\pi^+$\ \	& $\frac12 a^{(n)}$
		& $-\frac12 a^{(n)}$
		& 0			\\
\ $K^+$\ \	& $\frac12 a^{(n)}$
		& 0
		& $-\frac12 a^{(n)}$	\\
\ $K^0$\ \	& 0
		& $\frac12 a^{(n)}$
		& $-\frac12 a^{(n)}$	\\ \hline
\end{tabular}
\label{tab:meson}
\end{center}
\end{table}

\begin{table}[h]
\begin{center}
\caption{Moments $Q_B^{(n-1)}$ of the unpolarized quark distributions
	for $q=u, d$ or $s$ for octet baryons~$B$.
	The spin-dependent moments $\Delta Q_B^{(n-1)}$	can be
	obtained from the entries here by the replacements
	$\{ \alpha^{(n)} \to \bar\alpha^{(n)},
            \beta^{(n)}  \to \bar\beta^{(n)},
            \sigma^{(n)} \to \bar\sigma^{(n)} \}$.\\}
\begin{tabular}{l|c|c|c} \hline
\ $B$		& $U^{(n-1)}_B$
		& $D^{(n-1)}_B$
		& $S^{(n-1)}_B$			\\ \hline
\ $p$\ \	& $\frac56\an + \frac13\bn + \sn$
	        & $\frac16\an + \frac23\bn + \sn$
	        & $\sn$					\\
\ $n$\ \	& $\frac16\an + \frac23\bn + \sn$
                & $\frac56\an + \frac13\bn + \sn$
		& $\sn$					\\
\ $\Sigma^+$\ \	& $\frac56\an + \frac13\bn + \sn$
	        & $\sn$
		& $\frac16\an + \frac23\bn + \sn$	\\
\ $\Sigma^0$\ \	& $\frac{5}{12}\an + \frac16\bn + \sn$
		& $\frac{5}{12}\an + \frac16\bn + \sn$
		& $\frac16\an + \frac23\bn + \sn$	\\
\ $\Sigma^-$\ \	& $\sn$
		& $\frac56\an + \frac13\bn + \sn$
		& $\frac16\an + \frac23\bn + \sn$	\\
\ $\Lambda$\ \	& $\frac14\an + \frac12\bn + \sn$
		& $\frac14\an + \frac12\bn + \sn$
		& $\frac12\an + \sn$			\\
 $\Lambda\Sigma^0$\ \ & $\frac{\sqrt{3}}{12}[\an-2\bn]$
		& $-\frac{\sqrt 3}{12}[\an-2\bn]$
		& 0					\\
\ $\Xi^0$\ \	& $\frac16\an + \frac23\bn + \sn$
		& $\sn$
		& $\frac56\an + \frac13\bn + \sn$	\\
\ $\Xi^-$\ \	& $\sn$
		& $\frac16\an + \frac23\bn + \sn$
		& $\frac56\an + \frac13\bn + \sn$	\\ \hline
\end{tabular}
\label{tab:oct}
\end{center}
\end{table}

For the moments of the quark PDFs in the intermediate state baryons,
the contributions from the $u$, $d$ and $s$ flavors to the octet
baryon moments $Q_B^{(n-1)}$ are given in terms of combinations of
$\{ \alpha^{(n)}, \beta^{(n)}, \sigma^{(n)} \}$ and listed in
Table~\ref{tab:oct} for baryons
	$B=p, n, \Sigma^{\pm,0}, \Lambda, \Xi^{-,0}$
as well as for the $\Lambda$-$\Sigma^0$ interference.
Solving for the coefficients, one can write these as linear combinations
of the individual $u$, $d$ and $s$ quark moments in the proton,
\begin{subequations}
\label{eq:Brelations}
\begin{eqnarray}
\an &=& \frac43 U_p^{(n-1)} - \frac23 D_p^{(n-1)} - \frac23 S_p^{(n-1)},\\
\bn &=&-\frac13 U_p^{(n-1)} + \frac53 D_p^{(n-1)} - \frac43 S_p^{(n-1)},\\
\sn &=& S_p^{(n-1)}.
\end{eqnarray}
\end{subequations}
Assuming the strangeness in the intermediate state nucleon to be zero
(or equivalently, that the $u$ content of $\Sigma^-$, for example,
vanishes), one finds for the lowest ($n=1$) moments,
\begin{equation}
\label{eq:Brelations-n1}
\alpha^{(1)} = 2,\ \ \
\beta^{(1)}  = 1,\ \ \
\sigma^{(1)} = 0.
\end{equation}

\begin{table}[tb]
\begin{center}
\caption{Moments $Q_T^{(n-1)}$ of the unpolarized quark distributions
	for $q=u, d$ or $s$ for decuplet baryons $T$.
	The results for the spin-dependent moments $\Delta Q_T^{(n-1)}$,
	can be obtained by the replacements
	$\{ \theta^{(n)} \to \bar\theta^{(n)},
            \rho^{(n)}   \to \bar\rho^{(n)} \}$.\\}
\begin{tabular}{l|c|c|c} \hline
\ \ $T$		& $U^{(n-1)}_T$
		& $D^{(n-1)}_T$
		& $S^{(n-1)}_T$			\\ \hline
\ $\Delta^{++}$	& $\tn + \rn$
		& $\rn$
	        & $\rn$					\\
\ $\Delta^+$	& $\frac23\tn + \rn$
		& $\frac13\tn + \rn$
	        & $\rn$					\\
\ $\Delta^0$	& $\frac13\tn + \rn$
		& $\frac23\tn + \rn$
	        & $\rn$					\\
\ $\Delta^-$	& $\rn$
		& $\tn + \rn$
	        & $\rn$					\\
\ $\Sigma^{*+}$	& $\frac23\tn + \rn$
		& $\rn$
	        & $\frac13\tn + \rn$			\\
\ $\Sigma^{*0}$	& $\frac13\tn + \rn$
		& $\frac13\tn + \rn$
	        & $\frac13\tn + \rn$			\\
\ $\Sigma^{*-}$	& $\rn$
		& $\frac23\tn + \rn$
	        & $\frac13\tn + \rn$			\\
\ $\Xi^{*0}$	& $\frac13\tn + \rn$
		& $\rn$
	        & $\frac23\tn + \rn$			\\
\ $\Xi^{*-}$	& $\rn$
		& $\frac13\tn + \rn$
	        & $\frac23\tn + \rn$			\\
\ $\Omega^-$	& $\rn$
		& $\rn$
	        & $\tn + \rn$				\\ \hline
\end{tabular}
\label{tab:dec}
\end{center}
\end{table}

For the quark PDFs in the decuplet baryon intermediate states $T$,
the moments $Q_T^{(n-1)}$ for the individual $u$, $d$ and $s$ flavors
are given in terms of combinations of $\{ \theta^{(n)}, \rho^{(n)} \}$,
and are listed in Table~\ref{tab:dec} for
	$T = \Delta, \Sigma^*, \Xi^*$ and $\Omega^-$.
Solving for the coefficients $\theta^{(n)}$ and $\rho^{(n)}$ in
terms of the moments in the $\Delta^+$ baryon, one has
\begin{subequations}
\label{eq:Trelations}
\begin{eqnarray}
\tn &=& 3 \big( D_{\Delta^+}^{(n-1)} - S_{\Delta^+}^{(n-1)} \big)\
     =  \frac32 \big( U_{\Delta^+}^{(n-1)} - S_{\Delta^+}^{(n-1)} \big), \\
\rn &=& S_{\Delta^+}^{(n-1)}.
\end{eqnarray}
\end{subequations}
Again, assuming zero strangeness in the $\Delta^+$, the $n=1$ moments
are given by
\begin{equation}
\label{eq:Trelations-n1}
\theta^{(1)} = 3,\ \ \
\rho^{(1)}   = 0.
\end{equation}

\begin{table}[tb]
\begin{center}
\caption{Moments of the unpolarized $u$, $d$ and $s$ quark distributions
	in octet baryons $B$ arising from the $BB\phi\phi$ tadpole
	vertex,	as in Fig.~\ref{fig:loop_baryon}(j).  The moments
	$U_{{\rm (tad)} B K^0 \overline{K}^0}^{(n-1)}$,
	$D_{{\rm (tad)} B K^+ K^-}^{(n-1)}$ and
	$S_{{\rm (tad)} B \pi^+ \pi^-}^{(n-1)}$ are zero for
	all baryons $B$, and are not listed in the table.}
\scalebox{0.95}{\begin{tabular}{l|c|c|c|c|c|c}	\hline
$B$\ \		& \multicolumn{2}{|c}{$U_{{\rm (tad)}B\phi\phi}^{(n-1)}$}
		& \multicolumn{2}{|c}{$D_{{\rm (tad)}B\phi\phi}^{(n-1)}$}
		& \multicolumn{2}{|c}{$S_{{\rm (tad)}B\phi\phi}^{(n-1)}$}
						\\ \hline
		& $\pi^+ \pi^-$
		& $K^+ K^-$
		& $\pi^+ \pi^-$
		& $K^0 \overline{K^0}$
		& $K^0 \overline{K^0}$
		& $K^+ K^-$			\\ \hline
$p$\ \		& $-\frac13\an + \frac16\bn$
		& $-\frac{5}{12}\an - \frac16\bn$
		& $ \frac13\an - \frac16\bn$
		& $-\frac{1}{12}\an - \frac13\bn$
		& $ \frac{1}{12}\an + \frac13\bn$
		& $ \frac{5}{12}\an + \frac16\bn$	\\
$n$\ \		& $ \frac13\an - \frac16\bn$
		& $-\frac{1}{12}\an - \frac13\bn$
		& $-\frac13\an + \frac16\bn$
		& $-\frac{5}{12}\an - \frac16\bn$
		& $ \frac{5}{12}\an + \frac16\bn$
		& $ \frac{1}{12}\an + \frac13\bn$	\\
$\Sigma^+$\	& $-\frac{5}{12}\an - \frac16\bn$
		& $-\frac13\an + \frac16\bn$
		& $ \frac{5}{12}\an + \frac16\bn$
		& $ \frac{1}{12}\an + \frac13\bn$
		& $-\frac{1}{12}\an - \frac13\bn$
		& $ \frac13\an - \frac16\bn$		\\
$\Sigma^0$\	& 0
		& $-\frac18\an + \frac14\bn$
		& 0
		& $-\frac18\an + \frac14\bn$
		& $ \frac18\an - \frac14\bn$
		& $ \frac18\an - \frac14\bn$		\\
$\Sigma^-$\ 	& $ \frac{5}{12}\an + \frac16\bn$
		& $ \frac{1}{12}\an + \frac13\bn$
		& $-\frac{5}{12}\an - \frac16\bn$
		& $-\frac13\an + \frac16\bn$
		& $ \frac13\an - \frac16\bn$
		& $-\frac{1}{12}\an - \frac13\bn$	\\
$\Lambda$\	& 0
		& $ \frac18\an - \frac14\bn$
		& 0
		& $ \frac18\an - \frac14\bn$
		& $-\frac18\an + \frac14\bn$
		& $-\frac18\an + \frac14\bn$		\\
$\Xi^0$\	& $-\frac{1}{12}\an - \frac13\bn$
		& $ \frac13\an - \frac16\bn$
		& $ \frac{1}{12}\an + \frac13\bn$
		& $ \frac{5}{12}\an + \frac16\bn$
		& $-\frac{5}{12}\an - \frac16\bn$
		& $-\frac13\an + \frac16\bn$		\\
$\Xi^-$\	& $ \frac{1}{12}\an + \frac13\bn$
		& $ \frac{5}{12}\an + \frac16\bn$
		& $-\frac{1}{12}\an - \frac13\bn$
		& $ \frac13\an - \frac16\bn$
		& $-\frac13\an + \frac16\bn$
		& $-\frac{5}{12}\an - \frac16\bn$	\\ \hline
\end{tabular}}
\label{tab:tad}
\end{center}
\end{table}

For the moments of the distributions generated by the tadpole diagrams
in Fig.~\ref{fig:loop_baryon}(j), in Table~\ref{tab:tad} we list the
contributions $Q^{(n-1)}_{{\rm (tad)} B\phi\phi}$ for the $u$, $d$
and $s$ flavors in each octet baryon $B$.  Note that the combinations
involving $K^0 \overline{K}^0$ do not contribute to the
$u$-quark moments, those involving $K^+ K^-$ do not contribute to the
$d$-quark moments, and the contributions from $\pi^+ \pi^-$ to the
$s$-quark moments are also zero.

\begin{table}[tb]
\begin{center}
\caption{Moments of the unpolarized $u$, $d$ and $s$ quark
	distributions from the Kroll-Ruderman vertex for
	transitions from a proton initial state to octet
	and decuplet baryon intermediate states, as in
	Figs.~\ref{fig:loop_baryon}(e) and (h), respectively. \\}
\begin{tabular}{l|c|c|c}	\hline
$B\phi$		& $U_{{\rm (KR)}B}^{(n-1)}$
		& $D_{{\rm (KR)}B}^{(n-1)}$
		& $S_{{\rm (KR)}B}^{(n-1)}$ 			\\ \hline
$n \pi^+$\	& $-\frac{\sqrt2}{3}\ban + \frac{\sqrt{2}}{6}\bbn$
		& $ \frac{\sqrt2}{3}\ban - \frac{\sqrt{2}}{6}\bbn$
		& 0						\\
$\Sigma^0 K^+$\	& $ \frac{1}{12}\ban + \frac13\bbn$
		& 0
		& $-\frac{1}{12}\ban - \frac13\bbn$		\\
$\Sigma^+ K^0$\	& 0
		& $ \frac{\sqrt2}{12}\ban + \frac{\sqrt{2}}{3}\bbn$
		& $-\frac{1}{12}\ban - \frac13\bbn$		\\
$\Lambda^0 K^+$	& $ \frac{\sqrt3}{4}\ban$
		& 0
		& $-\frac{\sqrt3}{4}\ban$			\\ \hline
$T\phi$		& $U_{{\rm (KR)}T}^{(n-1)}$
		& $D_{{\rm (KR)}T}^{(n-1)}$
		& $S_{{\rm (KR)}T}^{(n-1)}$ 			\\ \hline
$\Delta^0 \pi^+$& $ \frac{1}{\sqrt6}\bon$
		& $-\frac{1}{\sqrt6}\bon$
		& 0						\\
$\Delta^{++} \pi^-$& $\frac{1}{\sqrt2}\bon$
		& $-\frac{1}{\sqrt2}\bon$
		& 0						\\
$\Sigma^{*0} K^+$& $ \frac{1}{2\sqrt3}\bon$
		& 0
		& $-\frac{1}{2\sqrt3}\bon$			\\
$\Sigma^{*+} K^0$& 0
		& $-\frac{1}{\sqrt6}\bon$
		& $ \frac{1}{\sqrt6}\bon$			\\ \hline
\end{tabular}
\label{tab:kr}
\end{center}
\end{table}

Finally, to complete the set of the contributions to the
unpolarized PDFs, in Table~\ref{tab:kr} we list the moments
$Q_{{\rm (KR)}B}^{(n-1)}$ and $Q_{{\rm (KR)}T}^{(n-1)}$
of the Kroll-Ruderman induced quark distributions from
Fig.~\ref{fig:loop_baryon}(e) and (h), for the transitions
from a proton initial state to intermediate states including
octet $B$ and decuplet $T$ baryons, respectively.
(Similar results can be derived for other octet or decuplet baryon
initial states, but are not listed here to avoid unnecessary detail.)
Note that, unlike for all other contributions from the diagrams in
Fig.~\ref{fig:loop_baryon}, the moments $Q_{{\rm (KR)}B}^{(n-1)}$ and
$Q_{{\rm (KR)}T}^{(n-1)}$ are given in terms of the coefficients
$\ban$, $\bbn$ and $\bon$, which are related to moments of the
spin-dependent parton distributions.

\begin{table}[tb]
\begin{center}
\caption{Moments $\Delta Q^{(n-1)}_{BT}$ of the polarized $u$, $d$
	and $s$ quark distributions from the axial octet-decuplet
	transition.}
\begin{tabular}{l|c|c|c}	\hline
\ $BT$\				&
	$\Delta U^{(n-1)}_{BT}$ &
	$\Delta D^{(n-1)}_{BT}$ &
	$\Delta S^{(n-1)}_{BT}$ 		\\ \hline
\ $n\Delta^0$\			&
	$-\frac{1}{\sqrt3}\bon$ &
        $\frac{1}{\sqrt3}\bon$  &
	           0                    	\\
\ $p\Delta^+$\			&
        $-\frac{1}{\sqrt3}\bon$ &
        $\frac{1}{\sqrt3}\bon$  &
                   0                         	\\
\ $\Sigma^0\Sigma^{*0}$\ 	&
	$-\frac{1}{2\sqrt3}\bon$&
        $-\frac{1}{2\sqrt3}\bon$&
        $\frac{1}{\sqrt3}\bon$			\\
\ $\Sigma^+\Sigma^{*+}$\	&
	$\frac{1}{\sqrt3}\bon$	&
                   0		&
        $-\frac{1}{\sqrt3}\bon$			\\
\ $\Sigma^-\Sigma^{*-}$\	&
	           0            &
        $-\frac{1}{\sqrt3}\bon$ &
        $\frac{1}{\sqrt3}\bon$			\\
\ $\Lambda\Sigma^{*0}$\		&
        $\frac{1}{2}\bon$	&
        $-\frac{1}{2}\bon$	&
		   0		                \\ \hline
\end{tabular}
\label{tab:axial}
\end{center}
\end{table}

For the latter, recall that spin-dependent PDFs are related to
matrix elements of the axial vector operators
	$\mathcal{O}^{\mu_1 \cdots \mu_n}_{\Delta q}$
in Eq.~(\ref{eq:axial-vector-op}), which, using the relations
(\ref{eq:Bidentities}) and (\ref{eq:Tidentities}), can be expanded
in terms of hadronic axial vector operators with coefficients given
by moments $\Delta Q^{(n-1)}_j$ of the spin-dependent distributions.
In analogy to the expansion in Eq.~(\ref{eq:vector-operator}),
we therefore expand the axial vector operators as
\begin{equation}
\mathcal{O}^{\mu_1 \cdots \mu_n}_{\Delta q}\
 =\ \Delta Q^{(n-1)}_B\,
    \widetilde{\mathcal{O}}^{\mu_1 \cdots \mu_n}_{BB'}\
 +\ \Delta Q^{(n-1)}_T\,
    \widetilde{\mathcal{O}}^{\mu_1 \cdots \mu_n}_{TT'}\
 +\ \Delta Q^{(n-1)}_{BT}\,
    \widetilde{\mathcal{O}}^{\mu_1 \cdots \mu_n}_{BT}\
 +\ \cdots,
\label{eq:axialvector-operator}
\end{equation}
where only the operators relevant for the calculation of
unpolarized PDFs are listed.  (The remaining terms not listed
in Eq.~(\ref{eq:axialvector-operator}) will be relevant for the
calculation of spin-dependent PDFs in the proton~\cite{XDWang_spin}.)
More explicitly, the axial vector hadronic operators
in~(\ref{eq:axialvector-operator}) are given by
\begin{subequations}
\label{eq:had_opsax}
\begin{eqnarray}
\widetilde{\mathcal{O}}^{\mu_1 \cdots \mu_n}_{BB'}
&=& \big( \overline{B}' \gamma^{\mu_1} \gamma_5  B
    \big)\, p^{\mu_2} \cdots p^{\mu_n},			\\
\widetilde{\mathcal{O}}^{\mu_1 \cdots \mu_n}_{TT'}
&=& \big( \overline{T'}_\alpha \gamma^{\mu_1} \gamma_5
	  T^\alpha
    \big)\, p^{\mu_2} \cdots p^{\mu_n},			\\
\widetilde{\mathcal{O}}^{\mu_1 \cdots \mu_n}_{BT}
&=& \big( \overline{B} \Theta^{\mu_1\nu} T_\nu
	+ \overline{T}_\nu \Theta^{\nu \mu_1} B
    \big)\, p^{\mu_2} \cdots p^{\mu_n},
\end{eqnarray}
\end{subequations}
with the corresponding moments $\Delta Q_j^{(n-1)}$ of the
spin-dependent PDFs defined by
\begin{subequations}
\label{eq:DeltaQn-1}
\begin{eqnarray}
\Delta Q^{(n-1)}_B
&=& \int_{-1}^1 dx\, x^{n-1}\, \Delta q_B(x),		\\
\Delta Q^{(n-1)}_T
&=& \int_{-1}^1 dx\, x^{n-1}\, \Delta q_T(x),		\\
\Delta Q^{(n-1)}_{BT}
&=& \int_{-1}^1 dx\, x^{n-1}\, \Delta q_{BT}(x).
\end{eqnarray}
\end{subequations}

For simplicity, in Eq.~(\ref{eq:DeltaQn-1}) we restrict ourselves
to the diagonal octet ($B=B'$) and diagonal decuplet ($T=T'$)
cases, with respective spin-dependent PDFs $\Delta q_B(x)$ and
$\Delta q_T(x)$, and the octet--decuplet transition distribution,
$\Delta q_{BT}(x)$.
In particular, the moments $\Delta Q_B^{(n-1)}$ of the spin-dependent
PDFs in octet baryons can be obtained from the entries in
Table~\ref{tab:oct} by substituting
	$\{ \alpha^{(n)} \to \bar\alpha^{(n)},
	    \beta^{(n)}  \to \bar\beta^{(n)},
	    \sigma^{(n)} \to \bar\sigma^{(n)} \}$,
while the moments $\Delta Q_T^{(n-1)}$ of the spin-dependent PDFs
in decuplet baryons are obtained from Table~\ref{tab:dec} with the
replacements
	$\{ \theta^{(n)} \to \bar\theta^{(n)},
	    \rho^{(n)}   \to \bar\rho^{(n)} \}$.
For the octet-decuplet axial transition distribution, the moments,
$\Delta Q^{(n-1)}_{BT}$, are given in terms of the coefficient
$\bar\omega$ in Eq.~(\ref{eq:axial-vector-op}) and are listed
in Table~\ref{tab:axial}.
Solving for the octet coefficients in terms of the moments of the
spin-dependent proton PDFs in the {\it proton}, one has, in analogy
with Eq.~(\ref{eq:Brelations}), the relations
\begin{subequations}
\label{eq:BrelationsDel}
\begin{eqnarray}
\ban &=& \frac43 \Delta U_p^{(n-1)}
       - \frac23 \Delta D_p^{(n-1)}
       - \frac23 \Delta S_p^{(n-1)},	\\
\bbn &=&-\frac13 \Delta U_p^{(n-1)}
       + \frac53 \Delta D_p^{(n-1)}
       - \frac43 \Delta S_p^{(n-1)},	\\
\bsn &=& \Delta S_p^{(n-1)}.
\end{eqnarray}
\end{subequations}
Similarly, for the decuplet case, the coefficients $\btn$ and $\brn$
can be written in terms of the moments of the spin-dependent PDFs
of quarks in the $\Delta^+$ bayron,
\begin{subequations}
\label{eq:TrelationsDel}
\begin{eqnarray}
\btn
&=& 3
    \big(
    \Delta D_{\Delta^+}^{(n-1)} - \Delta S_{\Delta^+}^{(n-1)}
    \big)\
 =\ \frac32
    \big(
    \Delta U_{\Delta^+}^{(n-1)} - \Delta S_{\Delta^+}^{(n-1)}
    \big),					\\
\brn
&=& \Delta S_{\Delta^+}^{(n-1)}.
\end{eqnarray}
\end{subequations}
The moments of the octet--decuplet transition operators can be
related to the moments of the octet baryon operators via the
SU(3) relation
\begin{equation}
\label{eq:omegabarSU3}
\bar\omega^{(n)} = -\frac12 \bar\alpha^{(n)} + \bar\beta^{(n)},
\end{equation}
for all $n$.
For the $n=1$ octet baryon moments, in particular, the coefficients
are given in terms of axial vector charges $F$ and $D$,
\begin{equation}
\label{eq:BrelationsDel-n1}
\bar\alpha^{(1)} = \frac23 \big( 3 F + D \big),\ \ \
\bar\beta^{(1)}  = \frac13 \big( 3 F - 5 D \big),\ \ \
\bar\sigma^{(1)} = 0.
\end{equation}
In terms of moments of the spin-dependent proton PDFs, for the
octet--decuplet transition vertex, $\bar\omega^{(1)}$ is given
by the SU(3) symmetry relation~\cite{Shanahan:2013xw},
\begin{equation}
\label{eq:TrelationsDel-n1}
\bar\omega^{(1)}
= -\Delta U_p^{(0)}
+ 2\Delta D_p^{(0)}
-  \Delta S_p^{(0)},
\end{equation}
which also reproduces the relation
	${\cal C} = - 2D$
between the meson–octet–decuplet baryon coupling ${\cal C}$
and the meson-octet coupling $D$~\cite{Jenkins:1991ts}.
Note that through Eq.~(\ref{eq:omegabarSU3}) the quark distributions
in the Kroll-Ruderman diagrams with decuplet baryon intermediate states
in Fig.~\ref{fig:loop_baryon}(h) are related to the spin-dependent
distributions of quarks in proton.

This completes the discussion of the moments of the PDFs of the
various mesons and baryons appearing in the intermediate states
in the diagrams of Fig.~\ref{fig:loop_baryon}.
From these, in the next section we derive relations for the
$x$ dependence of the PDFs themselves.

\subsection{SU(3) relations for baryon and meson PDFs}

In the previous section we derived relations between the coefficients
of the various operators in $\mathcal{O}^{\mu_1 \cdots \mu_n}_q$ and
$\mathcal{O}^{\mu_1 \cdots \mu_n}_{\Delta q}$ and the $n$-th Mellin
moments of the quark distributions in
Eqs.~(\ref{eq:pi_mom})--(\ref{eq:Trelations-n1}) and
Eqs.~(\ref{eq:BrelationsDel})--(\ref{eq:TrelationsDel-n1}).
Since these relations are valid for all moments $n$, one can derive
from them explicit expressions for the $x$ dependence of the PDFs.

For the valence distributions in the pion and kaon, from
Eq.~(\ref{eq:pi_mom}) and Table~\ref{tab:meson} one has
\begin{equation}
\begin{split}
\overline{q}_\pi(x)
&\equiv u_{\pi^+}(x) = \overline{d}_{\pi^+}(x)
      = d_{\pi^-}(x) = \overline{u}_{\pi^-}(x)	\\
&     = u_{K^+}(x)   = \overline{s}_{K^+}(x)
      = d_{K^0}(x)   = \overline{s}_{K^0}(x),
\end{split}
\label{eq:quark35}
\end{equation}
for all values of $x$.
For the PDFs in the baryons, to simplify notations we shall label
the bare distributions in the proton without an explicit baryon
subscript, $q(x) \equiv q_p(x)$, and those in the $\Delta^+$ baryon
by $q_\Delta(x) \equiv q_{\Delta^+}(x)$.
Starting with the quark distributions in the SU(3) octet baryons,
from Table~\ref{tab:oct} the individual $u$-, $d$- and $s$-quark
flavor PDFs can be written in terms of the proton PDFs as
\begin{subequations}
\label{eq:PDFoct}
\begin{align}
u_n(x)		&= d(x),&
d_n(x)		&= u(x),&
s_n(x) 		&= s(x),	\\
u_{\Sigma^+}(x) &= u(x),&
d_{\Sigma^+}(x) &= s(x),&
s_{\Sigma^+}(x) &= d(x),	\\
u_{\Sigma^0}(x) &= \frac12 \big[ u(x) + s(x) \big],&
d_{\Sigma^0}(x) &= u_{\Sigma^0}(x),&
s_{\Sigma^0}(x) &= d(x),	\\
u_{\Sigma^-}(x) &= s(x),&
d_{\Sigma^-}(x) &= u(x),&
s_{\Sigma^-}(x) &= d(x),	\\
u_{\Lambda}(x)  &= \frac16 \big[ 4 d(x) + u(x) + s(x) \big],&
d_{\Lambda}(x)  &= u_{\Lambda}(x),&
s_{\Lambda}(x)  &= \frac13 \big[ 2 u(x) - d(x) + 2 s(x) \big].
\end{align}
\end{subequations}
For the quark distributions in the SU(3) decuplet baryons,
from Table~\ref{tab:dec} the $u$-, $d$- and $s$-quark PDFs
can be written in terms of the PDFs in the $\Delta^+$ as
\begin{subequations}
\label{eq:PDFdec}
\begin{align}
u_{\Delta^{++}}(x) &= u_\Delta(x) + d_\Delta(x) - s_\Delta(x),&
d_{\Delta^{++}}(x) &= s_\Delta(x),&
s_{\Delta^{++}}(x) &= s_\Delta(x),	\\
u_{\Delta^0}(x)    &= d_\Delta(x),&
d_{\Delta^0}(x)	   &= u_\Delta(x),&
s_{\Delta^0}(x)    &= s_\Delta(x),	\\
u_{\Delta^-}(x)	   &= s_\Delta(x),&
d_{\Delta^-}(x)    &= u_{\Delta^{++}}(x),&
s_{\Delta^-}(x)    &= s_\Delta(x),	\\
u_{\Sigma^{*+}}(x) &= u_\Delta(x),&
d_{\Sigma^{*+}}(x) &= s_\Delta(x),&
s_{\Sigma^{*+}}(x) &= d_\Delta(x),	\\
u_{\Sigma^{*0}}(x) &= d_\Delta(x),&
d_{\Sigma^{*0}}(x) &= d_\Delta(x),&
s_{\Sigma^{*0}}(x) &= d_\Delta(x),	\\
u_{\Sigma^{*-}}(x) &= s_\Delta(x),&
d_{\Sigma^{*-}}(x) &= u_\Delta(x),&
s_{\Sigma^{*-}}(x) &= d_\Delta(x).
\end{align}
\end{subequations}
In our actual numerical calculations, for simplicity we
approximate
	$q_\Delta(x) \approx q(x)$,
and assume valence quark dominance for the bare states, so that
	$s(x) \approx s_\Delta(x) \approx 0$.


For the PDFs arising from the tadpole diagrams in
Fig.~\ref{fig:loop_baryon}(j), from Table~\ref{tab:tad} the
$u$-, $d$- and $s$-quark distributions can be written as
\begin{subequations}
\label{eq:PDFtad}
\begin{align}
u^{\rm (tad)}_{\pi^+}(x) &= d^{\rm (tad)}_{\pi^+}(x)
  = u(x) - d(x),&
s^{\rm (tad)}_{\pi^+}(x) &= 0, 			\\
u^{\rm (tad)}_{K^+}(x)   &= s^{\rm (tad)}_{K^+}(x)
  = \frac12 \big[ u(x) - s(x) \big],&
d^{\rm (tad)}_{K^+}(x) &= 0,			\\
d^{\rm (tad)}_{K^0}(x)   &= s^{\rm (tad)}_{K^0}(x)
  = d(x) - s(x),&
u^{\rm (tad)}_{K^0}(x) &= 0.
\end{align}
\end{subequations}
The distributions associated with the tadpole gauge link diagrams
in Fig.~\ref{fig:loop_baryon}(g) turn out to be the same as those
for the regular tadpole diagrams,
\begin{equation}
q^{(\delta)}_\phi(x) = q^{\rm (tad)}_\phi(x).
\label{eq:PDFdelphi}
\end{equation}

Turning now to the Kroll-Ruderman diagrams in
Fig.~\ref{fig:loop_baryon}(e) and \ref{fig:loop_baryon}(h),
for a proton initial state the corresponding PDFs are expressed
in terms of spin-dependent PDFs in the proton,
	$\Delta q(x) \equiv \Delta q_p(x)$.
From Table~\ref{tab:kr}, for the octet baryon intermediate states
the $u$-, $d$- and $s$-quark distributions are given by
\begin{subequations}
\begin{align}
u^{\rm (KR)}_n(x) = d^{\rm (KR)}_n(x)
  &= \frac{\Delta u(x) - \Delta d(x)}{F+D},&
s^{\rm (KR)}_n(x)
  &= 0,					\\
d^{\rm (KR)}_{\Sigma^+}(x) = s^{\rm (KR)}_{\Sigma^+}(x)
  &= \frac{\Delta d(x) - \Delta s(x)}{F-D},&
u^{\rm (KR)}_{\Sigma^+}(x)
  &= 0,					\\
u^{\rm (KR)}_{\Sigma^0}(x) = s^{\rm (KR)}_{\Sigma^0}(x)
  &= \frac{\Delta d(x) - \Delta s(x)}{F-D},&
d^{\rm (KR)}_{\Sigma^0}(x)
  &= 0,					\\
u^{\rm (KR)}_{\Lambda}(x) = s^{\rm (KR)}_{\Lambda}(x)
  &= \frac{2\Delta u(x) - \Delta d(x) - \Delta s(x)}{3F+D},&
d^{\rm (KR)}_{\Lambda}(x)
  &= 0.
\end{align}
\end{subequations}
Similarly, for the decuplet baryon intermediate states the
individual quark flavor Kroll-Ruderman distributions are given by
\begin{subequations}
\begin{align}
u^{\rm (KR)}_{\Delta^{++}}(x) = d^{\rm (KR)}_{\Delta^{++}}(x)
  &= \frac{ \Delta u(x) - 2 \Delta d(x) + \Delta s(x)}{2D},&
s^{\rm (KR)}_{\Delta^{++}}(x)
  &= 0,							\\
u^{\rm (KR)}_{\Delta^0}(x) = d^{\rm (KR)}_{\Delta^0}(x)
  &= \frac{ \Delta u(x) - 2 \Delta d(x) + \Delta s(x)}{2D},&
s^{\rm (KR)}_{\Delta^0}(x)
  &= 0,							\\
d^{\rm (KR)}_{\Sigma^{*+}}(x) = s^{\rm (KR)}_{\Sigma^{*+}}(x)
  &= \frac{ \Delta u(x) - 2 \Delta d(x) + \Delta s(x)}{2D},&
u^{\rm (KR)}_{\Sigma^{*+}}(x)
  &= 0,							\\
u^{\rm (KR)}_{\Sigma^{*0}}(x) = s^{\rm (KR)}_{\Sigma^{*0}}(x)
  &= \frac{ \Delta u(x) - 2\Delta d(x) + \Delta s(x)}{2D},&
d^{\rm (KR)}_{\Sigma^{*0}}(x)
  &= 0.
\end{align}
\end{subequations}
The PDFs associated with the KR gauge link diagrams in
Figs.~\ref{fig:loop_baryon}(f) and \ref{fig:loop_baryon}(i)
are the same as those for the regular KR diagrams,
\begin{subequations}
\begin{align}
q^{(\delta)}_B(x) &= q^{\rm (KR)}_B(x),		\\
q^{(\delta)}_T(x) &= q^{\rm (KR)}_T(x).
\label{eq:PDFdelBT}
\end{align}
\label{eq:quark42}
\end{subequations}

With this set of distributions in the SU(3) octet and decuplet
baryons and mesons, and the proton $\to$ meson $+$ baryon splitting
functions from Ref.~\cite{nonlocal-I}, we can finally proceed with
the computation of the meson loop contributions to the quark and
antiquark PDFs in the proton, as in Eqs.~(\ref{eq:antiquark})
and (\ref{eq:quark}).
In the following section we focus on the calculation of specific
PDF asymmetries in the proton numerically.

\section{Sea quark asymmetries in the proton}
\label{sec.results}

To illustrate the calculation of the contributions to PDFs from
pseudoscalar meson loops within the nonlocal chiral effective theory
framework, we consider the examples of the flavor asymmetry in the
light antiquark sea in the proton, $\bar d-\bar u$, and the
strange--antistrange asymmetry in the nucleon, $s-\bar s$.
In both quantities perturbatively generated contributions from gluon
radiation effectively cancel, at least up to next-to-next-to-leading
order corrections in $\alpha_s$~\cite{Catani04}, so that
observation of large asymmetries may be indicative of
nonperturbative effects~\cite{Signal87, Malheiro97, Sufian18}.

For the numerical calculation of the meson--bayron splitting functions,
earlier work used various regularization prescriptions, including sharp
transverse momentum cutoffs, Pauli-Villars regularization, as well as
phenomenological vertex form factors~\cite{Holtmann96, MST98,
Burkardt13, Salamu15, XGWangPLB, XGWangPRD}.
At times the prescriptions have been imposed in rather {\it ad hoc}
ways, without necessarily ensuring that the relevant symmetries,
such as Lorentz, chiral, and local gauge symmetries, are necessarily
respected.
In the present work we for the first time perform the calculation
within nonlocal regularization, which is consistent with all of the
above symmetry requirements.
An advantage of the nonlocal method is that only a single parameter,
$\Lambda$, is needed to regulate {\it all} of the on-shell, off-shell
and $\delta$ functions associated with each of the diagrams in
Figs.~\ref{fig:loop_meson} and \ref{fig:loop_baryon}.

Following Ref.~\cite{nonlocal-I}, in the present analysis we adopt a
dipole shape in the meson virtuality $k^2$ for the regulator functions
for the one-loop contributions, parametrized by a cutoff parameter
$\Lambda$,
\begin{equation}
\widetilde{F}(k)
= \left( \frac{\Lambda^2-m_\phi^2}{D_\Lambda} \right)^2,
\label{eq:regulator}
\end{equation}
where $D_\Lambda = k^2 - \Lambda^2 + i\varepsilon$.
The cutoff $\Lambda$ can be determined by fitting the calculated
meson-exchange cross section to differential cross sections data
for inclusive baryon production in high-energy $pp$ scattering,
$pp \to B X$, for different species of baryon $B$.
Summing over the particles~$X$ in the final state, the differential
inclusive baryon production cross sections can be written as
\begin{equation}
\sigma(y, k_\bot^2)
\equiv E\, \frac{d^3\sigma}{d^3k}
= \frac{\bar{y}}{\pi}
  \frac{d^2\sigma}{dy\, dk_\bot^2},
\label{eq:sigdef}
\end{equation}
where $E$ is the incident proton energy and $\bar{y} \equiv 1-y$
is the longitudinal momentum fraction of the incident proton
carried by the produced baryon $B$.
In Eq.~(\ref{eq:sigdef}) we have used the fact that for spin-averaged
scattering the differential cross section is independent of the
azimuthal angle.
Available data exist on inclusive neutron and $\Delta^{++}$
production~\cite{Flauger:1976ju, Blobel:1978yj, Barish:1975uv},
as well as on $\Lambda$ and $\Sigma^{*+}$ production
\cite{Blobel:1978yj, Jaeger:1974pk, Bockmann:1977sc} in the
hyperon sector.
In principle, the cutoffs may depend on the baryon $B$,
although within the SU(3) symmetry framework we do not expect
large variations among the different $\Lambda$ values.

Once the cutoffs are determined and the one-loop splitting functions
are fixed, these can then be convoluted with the various meson and
baryon PDFs in Eqs.~(\ref{eq:antiquark}) and (\ref{eq:quark}) to
compute the contributions to the PDFs in the proton.
In the numerical calculations the input PDFs of the pion and kaon
are taken from Aicher {\it et al.}~\cite{Aicher:2010cb}.
The spin-averaged PDFs of the proton are from Ref.~\cite{Martin:1998sq},
while the spin-dependent PDFs are taken from Ref.~\cite{Leader:2010rb}.
Since the valence pion and proton PDFs are reasonably well determined,
at least compared with the sea quark distributions, using other
pion~\cite{Barry:2018ort, Owens84, SMRS92, GRV92, Wijesooriya05} or
proton~\cite{JMO13, ForteWatt13} parametrizations will not lead to
significant differences.

\subsection{$\bar{d}-\bar{u}$ asymmetry}

Turning to the light antiquark asymmetry in the proton sea, within the
chiral effective theory framework the primary source of the asymmetry
is the meson rainbow and bubble diagrams in Fig.~\ref{fig:loop_meson}.
In this approximation the $\bar d-\bar u$ difference does not depend
directly on the structure of the baryon coupling diagrams in
Fig.~\ref{fig:loop_baryon}, but only on the splitting functions and
the substructure of the pion.
More specifically, from Eq.~(\ref{eq:antiquark}) one can write the
contribution to the $\bar d-\bar u$ difference in the proton as the
convolution
\begin{equation}
\bar d(x) - \bar u(x)
= \Big[
  \big( f_{\pi^+ n}^{\rm (rbw)}
	+ f_{\pi^+ \Delta^0}^{\rm (rbw)}
	- f_{\pi^- \Delta^{++}}^{\rm (rbw)}
	+ f_\pi^{\rm (bub)}
  \big) \otimes \bar{q}_\pi
  \Big](x),
\label{eq:dubar}
\end{equation}
where the first (octet rainbow) term in the brackets is from
  Fig.~\ref{fig:loop_meson}(a),
the second and third (decuplet rainbow) terms correspond to
  Fig.~\ref{fig:loop_meson}(b),
and the fourth (bubble) term is from
  Fig.~\ref{fig:loop_meson}(c).
Using the notations of Ref.~\cite{nonlocal-I}, the splitting functions
in Eq.~(\ref{eq:dubar}) for the rainbow and bubble diagrams can be
expressed in terms of octet and decuplet basis functions.
In particular, for the $\pi N$ configuration the function
$f^{\rm (rbw)}_{\pi^+ n}$ is given by a sum of nucleon on-shell
and $\delta$-function contributions,
\begin{eqnarray}
f^{\rm (rbw)}_{\pi^+ n}(y)
&=& \frac{2 (D+F)^2 M^2}{(4\pi f)^2}
    \Big[ f^{\rm (on)}_N(y)
        + f^{(\delta)}_\pi(y)
        - \delta f^{(\delta)}_\pi(y)
    \Big],
\label{eq:piN-rbw}
\end{eqnarray}
where $D$ and $F$ are the SU(3) flavor coefficients,
and $f = 93$~MeV is the pseudoscalar meson decay constant.
Explicit forms for the basis functions are given in
Ref.~\cite{nonlocal-I} for the dipole regulator
$\widetilde{F}(k)$ in Eq.~(\ref{eq:regulator}).
The on-shell function $f^{\rm (on)}_N$ is nonzero for $y>0$,
while the local $f^{(\delta)}_\pi$ and nonlocal
$\delta f^{(\delta)}_\pi$ functions are proportional to
$\delta(y)$~\cite{nonlocal-I}, and therefore contribute to
the $\bar d-\bar u$ asymmetry only at $x=0$~\cite{Burkardt13,
Salamu15}.  In the pointlike limit, in which the form factor
cutoff $\Lambda \to \infty$, the nonlocal function
$\delta f^{(\delta)}_\pi$ vanishes; however, at finite
$\Lambda$ values it remains nonzero.

For the $\pi \Delta$ contributions to the asymmetry in
Eq.~(\ref{eq:dubar}), the splitting function for the rainbow
diagram in Fig.~\ref{fig:loop_meson}(b) includes several regular
and $\delta$-function terms,
\begin{eqnarray}
f^{\rm (rbw)}_{\pi^- \Delta^{++}}(y)\
=\ 3 f^{\rm (rbw)}_{\pi^+ \Delta^0}(y)\
&=& \frac{{\cal C}^2 \overline{M}^2}{2 (4\pi f)^2}
\Big[ f^{\rm (on)}_\Delta(y)
     + f^{\rm (on\, end)}_\Delta(y)
     - \frac{1}{18} f^{(\delta)}_\Delta(y)           \notag\\
& & \hspace*{1.5cm}
   +\, \frac{2 M^2 \big( \overline{M}^2 - m_\pi^2 \big)}
            {3 M_\Delta^2\, \overline{M}^2}
       \big( f^{(\delta)}_\pi(y) - \delta f^{(\delta)}_\pi(y) \big)
\Big],
\label{eq:piD-rbw}
\end{eqnarray}
where $\overline{M} = M + M_\Delta$, and ${\cal C}$ is the
meson-octet-decuplet baryon coupling, which is related to the
$\pi N \Delta$ coupling constant $g_{\pi N \Delta}$ by
	${\cal C} = \sqrt{2} f\, g_{\pi N \Delta}$
\cite{Salamu15, nonlocal-I}.
As for the $\pi N$ case, the on-shell function for the $\Delta$
intermediate state, $f^{\rm (on)}_\Delta$, is nonzero for $y > 0$,
with a shape that is qualitatively similar to $f^{\rm (on)}_N$,
but peaking at smaller $y$ because of the positive $\Delta$--nucleon
mass difference~\cite{nonlocal-I, Holtmann96, MST98}.
The on-shell end-point function, $f^{\rm (on\, end)}_\Delta$, also has
a similar shape for finite $\Lambda$, but in the $\Lambda \to \infty$
limit is associated with an end-point singularity that gives a
$\delta$-function at $y=1$.
The off-shell components of the $\Delta$ propagator induce several
terms that are proportional to $\delta$-functions at $y = 0$.
The functions $f^{(\delta)}_\pi$ and $\delta f^{(\delta)}_\pi$
are equivalent to those in Eq.~(\ref{eq:piN-rbw}), while
$f^{(\delta)}_\Delta$ is a new function that appears only for
the decuplet intermediate state~\cite{Salamu15}.

Finally, the bubble diagram contribution to the $\bar d-\bar u$
asymmetry, $f^{\rm (bub)}_\pi$, in Fig.~\ref{fig:loop_meson}(c)
is given by the same combination of basis $\delta$-function
contributions as for the rainbow diagrams,
\begin{eqnarray}
f^{\rm (bub)}_\pi(y)
&=&-\frac{2 M^2}{(4\pi f)^2}
    \Big[ f^{(\delta)}_\pi(y) - \delta f^{(\delta)}_\pi(y)
    \Big].
\label{eq:f-bub}
\end{eqnarray}
Although this term gives a nonzero PDF only at $x = 0$, since it
contributes to the integral of $\bar d-\bar u$, it will indirectly
affect the normalization for $x > 0$.
On the other hand, experimental cross sections are in practice
available only for $x > 0$, so that the $\delta$-function pieces
are generally difficult to constrain directly, especially in
regularization schemes that use different regulator parameters for
the $\delta$-function and $y > 0$ contributions~\cite{Salamu15}.
The advantage of the nonlocal approach employed here is that by
consistently introducing a vertex form factor in coordinate space
in the nonlocal Lagrangian~\cite{nonlocal-I}, the {\it same}
regulator function then appears in {\it all} splitting functions
derived from the fundamental interaction, which in our case is
parametrized through the single cutoff $\Lambda$.
Even if experimental data constrain only contributions at $x > 0$,
such as from the on-shell functions, once determined these can then
be used to compute other contributions, including those at $x = 0$.

Following Refs.~\cite{Holtmann96, XGWangPLB, XGWangPRD}, we can
constrain the parameter $\Lambda$ for the octet intermediate states
by comparing the one-pion exchange contribution with the differential
cross section for the inclusive charge-exchange process $pp \to n X$
at $y > 0$,
\begin{equation}
\sigma(pp \to n X)
= \frac{2 (D+F)^2 M^2}{(4 \pi f)^2}
  \frac{\bar{y}}{\pi}\, \hat{f}_N^{(\rm on)}(y,k_\bot^2)\,
  \sigma_{\rm tot}^{\pi^+ p}(y s),
\label{eq:nX}
\end{equation}
where $s$ is the invaraiant mass squared of the reaction.
The function $\hat{f}_N^{(\rm on)}(y,k_\bot^2)$ in Eq.~(\ref{eq:nX})
is the unintegrated on-shell nucleon splitting function, which
is related to the corresponding integrated splitting function
$f_N^{\rm (on)}(y)$ in Eq.~(\ref{eq:piN-rbw}) by
(see also Eq.~(63) in Ref.~\cite{nonlocal-I})
\begin{equation}
f_N^{(\rm on)}(y)
\equiv \int dk_\bot^2\, \hat{f}_N^{(\rm on)}(y,k_\bot^2).
\end{equation}
The cross section
  $\sigma_{\rm tot}^{\pi^+ p}(y s)$
in Eq.~(\ref{eq:nX}) is the total $\pi^+ p$ scattering cross
section evaluated at the center of mass energy $y s$.
In the numerical calculations, we use the (approximately energy
independent) empirical value
  $\sigma_{\rm tot}^{\pi^+ p} = 23.8(1)$~mb~\cite{Hufner:1992cu}.
For the SU(3) couplings we take
  $D=0.85$ and
  $F=0.41$,
which gives a triplet axial charge
  $g_A = (2 \bar\alpha^{(1)} - \bar\beta^{(1)})/3 = D+F = 1.26$
and an octet axial charge
  $g_8 = \bar\alpha^{(1)} + \bar\beta^{(1)} = 3F-D = 0.38$.

\begin{figure}[t]
\centering
\vspace*{-0.5cm}
\begin{tabular}{ccccc}
{\epsfxsize=3.62in\epsfbox{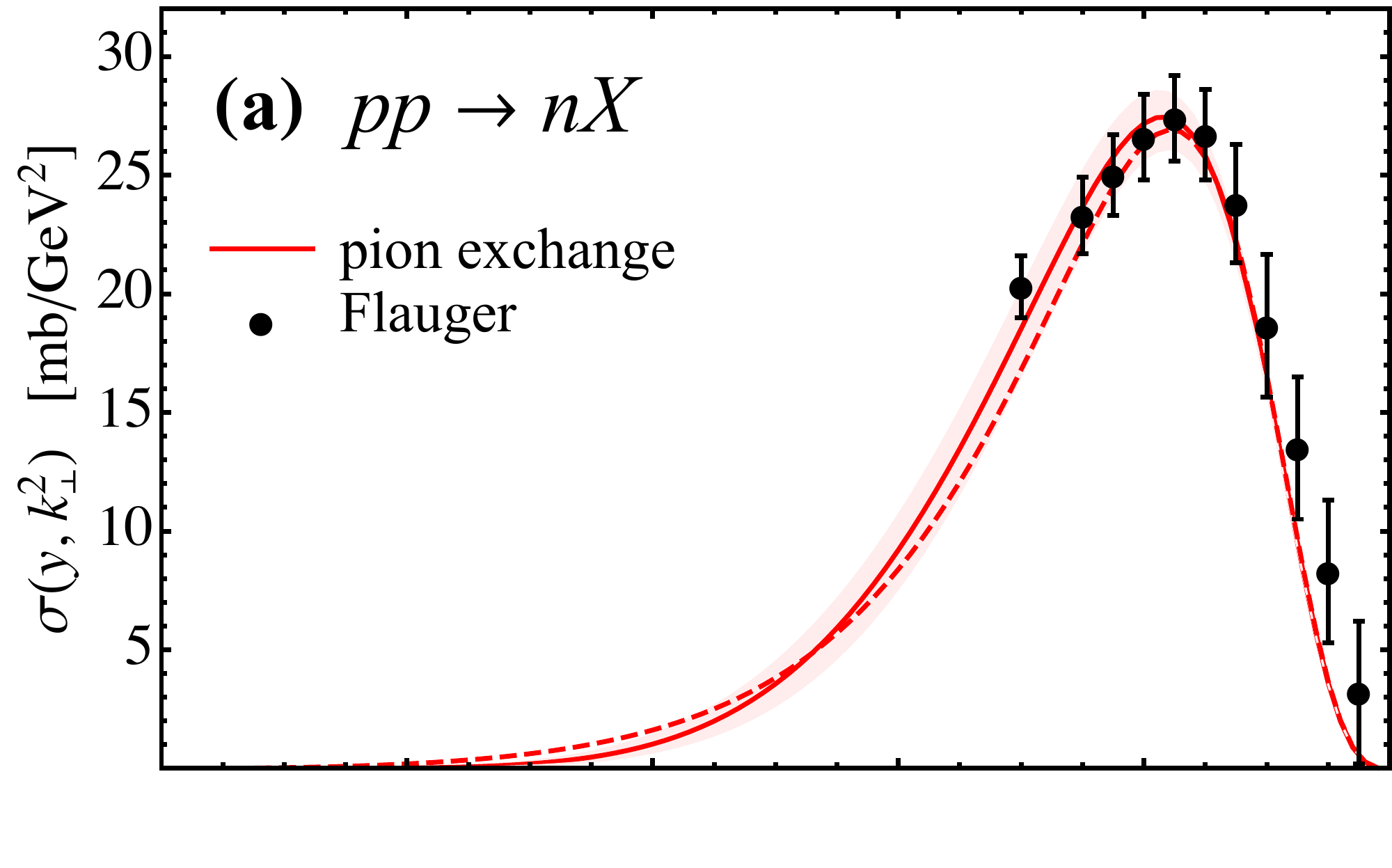}\vspace*{-0.9cm}}&\\
\hspace{-0.18cm}{\epsfxsize=3.7in\epsfbox{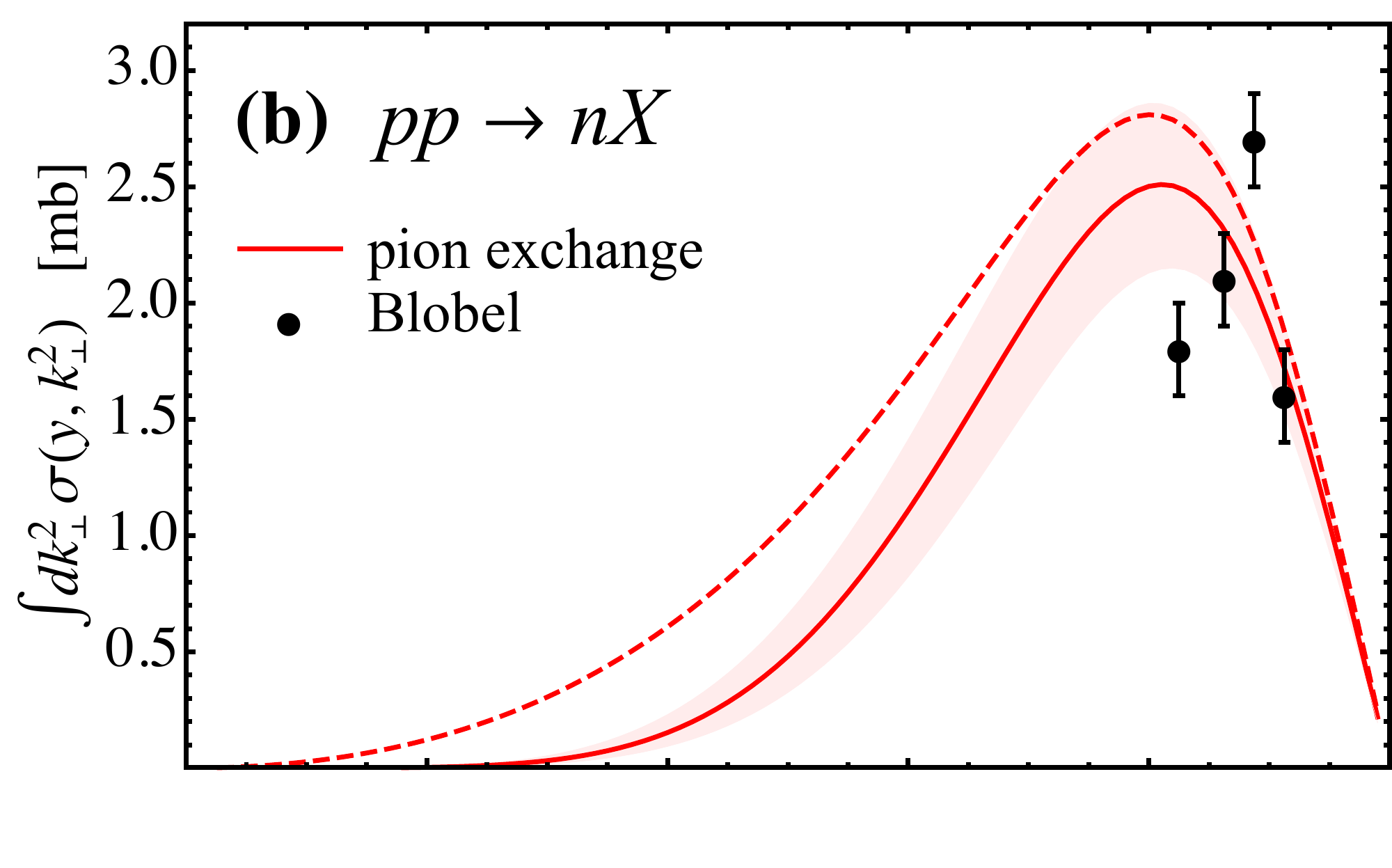}\vspace*{-0.8cm}}&\\
\hspace{0.2cm}{\epsfxsize=3.84in\epsfbox{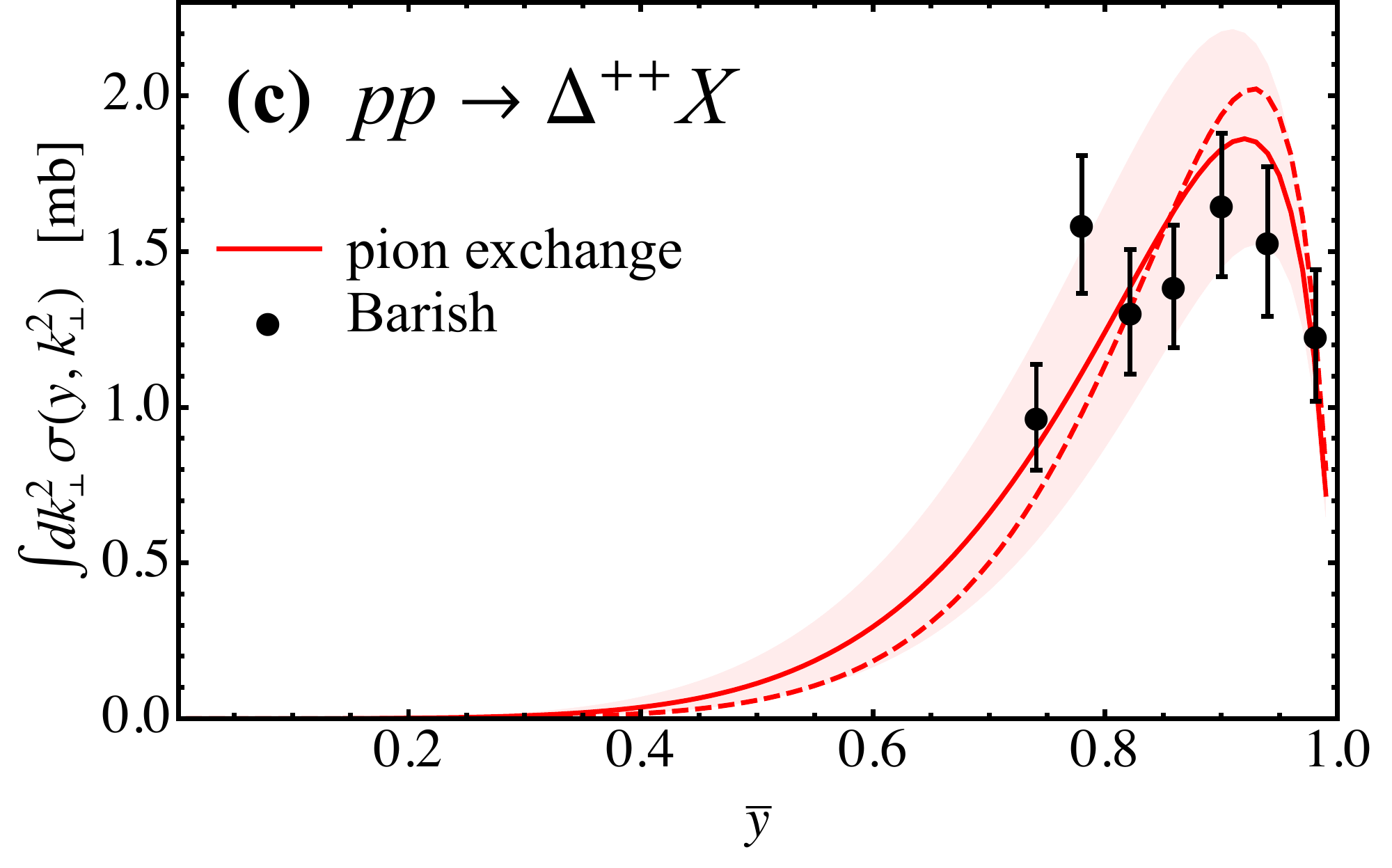}\vspace*{-0.cm}}&
\end{tabular}
\caption{Differential inclusive hadron production cross section
	$\sigma(y,k_\bot^2)$ versus $\bar{y}$ for
   {\bf (a)} $pp \to n X$ at $k_\bot^2 = 0$
	\cite{Flauger:1976ju};		
   {\bf (b)} $pp \to n X$ integrated over $k_\bot^2$
	\cite{Blobel:1978yj};		
   {\bf (c)} $pp \to \Delta^{++} X$ integrated over $k_\bot^2$
	\cite{Barish:1975uv},		
	compared with the fitted nonlocal pion exchange contributions for
	$\Lambda_{\pi N}=1.0(1)$~GeV and
	$\Lambda_{\pi\Delta}=0.9(1)$~GeV
	(solid red lines and pink $1\sigma$ uncertainty bands)
	and with Pauli-Villars regularization (dashed red lines) for
	$\Lambda_{\pi N}^{\rm (PV)}=0.3$~GeV and
	$\Lambda_{\pi\Delta}^{\rm (PV)}=0.64$~GeV.}
\label{fig:cross1}
\end{figure}

The results for the differential neutron production cross section
are shown in Fig.~\ref{fig:cross1} versus~$\bar{y}$.
The experimental data are typically presented as a function of the
ratio $2 p_L/\sqrt{s}$, where $p_L$ is the longitudinal momentum of
the produced baryon in the center of mass frame; at high energies,
however, this is equivalent to $\bar y$.
In Fig.~\ref{fig:cross1}(a) we compare our results with the
neutron production data from the ISR at CERN at energies $\sqrt{s}$
between $\approx 31$ and 63~GeV for $0^\circ$ neutron production
angles, or \mbox{$k_\bot^2=0$}~\cite{Flauger:1976ju}.
Data from the hydrogen bubble chamber experiment at the CERN proton
synchrotron at $\sqrt{s} \approx 5$ and 7~GeV~\cite{Blobel:1978yj}
are shown in Fig.~\ref{fig:cross1}(b) for the $k_\bot$-integrated
neutron cross section.
Because the pion-exchange processes is dominant only at large
$\bar{y}$~\cite{McKenney16}, with contributions from background
processes such as the exchange of heavier mesons~\cite{MT93,
Holtmann96, Kopeliovich12} becoming more important at lower
$\bar{y}$, we include data only in the region $\bar{y} > 0.7$.
Corrections from rescattering and absorption are also known to
play a role in inclusive hadron production, and are estimated
to be around 20\% at high values of $\bar{y}$~\cite{D'Alesio00,
Kopeliovich12, H1_01}.
A good description of the single and double differential neutron
data can be achieved with a cutoff parameter
$\Lambda_{\pi N} = 1.0(1)$~GeV.
A marginally larger value is found if fitting only the double
differential data, and slightly smaller value for just the
$k_\bot$-integrated cross section, but consistent within the
uncertainties.

For the inclusive production of decuplet baryons, the invariant
differential cross section for an inclusive $\Delta^{++}$ in the
final state can be written for $y > 0$ as
\begin{equation}
\sigma(pp \to \Delta^{++} X)
= \frac{{\cal C}^2 \overline{M}^2}{2 (4\pi f)^2}
  \frac{\bar{y}}{\pi}
  \Big[ \hat{f}_\Delta^{(\rm on)}(y,k_\bot^2)
      + \hat{f}_\Delta^{(\rm on\, end)}(y,k_\bot^2)
  \Big]\,
  \sigma_{\rm tot}^{\pi^- p}(y s),
\label{eq:DeltaX}
\end{equation}
where $\sigma_{\rm tot}^{p \pi^-}$ is the total $\pi^- p$
scattering cross section.
In our numerical calculations we assume this to be charge
independent, so that
  $\sigma_{\rm tot}^{\pi^- p} \approx \sigma_{\rm tot}^{\pi^+ p}$,
and for the coupling constant ${\cal C}$ we take the SU(6) symmetric
value ${\cal C} = -2D = -1.72$.
The functions $\hat{f}_\Delta^{(\rm on)}(y,k_\bot^2)$ and
$\hat{f}_\Delta^{(\rm on\, end)}(y,k_\bot^2)$ in (\ref{eq:DeltaX})
are the unintegrated decuplet on-shell and on-shell end-point
splitting functions, which are related to the corresponding
integrated splitting functions (see Eqs.~(86)--(88) in
\cite{nonlocal-I}) by the identities
\begin{subequations}
\begin{eqnarray}
f_\Delta^{(\rm on)}(y)
&\equiv& \int dk_\bot^2\, \hat{f}_\Delta^{(\rm on)}(y,k_\bot^2), \\
f_\Delta^{(\rm on\,end)}(y)
&\equiv& \int dk_\bot^2\, \hat{f}_\Delta^{(\rm on\,end)}(y,k_\bot^2),
\end{eqnarray}
\end{subequations}
respectively.
The $k_\bot^2$-integrated $\Delta^{++}$ cross section is shown in
Fig.~\ref{fig:cross1}(c) compared with hydrogen bubble chamber data
taken at Fermilab for $\sqrt{s} \approx 20$~GeV~\cite{Barish:1975uv}.
A good fit to the data is obtained with a value of the decuplet cutoff
of $\Lambda_{\pi\Delta} = 0.9(1)$~GeV, which is slightly smaller than
that for the neutron production cross sections.

To examine the model dependence of the analysis, for comparison
we also fitted the hadron production cross sections in
Figs.~\ref{fig:cross1}(a)--(c) using instead the Pauli-Villars
regularization for the local effective theory~\cite{XGWangPRD,
XGWangPLB}.
The explicit forms of the Pauli-Villars regularized octet on-shell
splitting functions can be found in \cite{XGWangPRD, XGWangPLB}.
The result for the sum of the decuplet on-shell and on-shell
end-point functions is as in Eq.~(96) of \cite{XGWangPRD},
with the integral regularized by a factor
	$(1 + 4 D_{\pi \Delta}/D_{\Lambda \Delta})$,
where $D_{\pi \Delta}$ and $D_{\Lambda \Delta}$ are momentum
dependent functions given in Eq.~(86) of \cite{XGWangPRD}.
The results for the best fit Pauli-Villars mass parameters
$\Lambda_{\pi N}^{\rm (PV)}=0.30$~GeV and
$\Lambda_{\pi \Delta}^{\rm (PV)}=0.64$~GeV
are illustrated by the dashed curves in Fig.~\ref{fig:cross1},
and are similar to those for the nonlocal calculation.
While there is some difference in the shape of the calculated
$k_\perp$-integrated neutron production cross section in
Fig.~\ref{fig:cross1}(b) at smaller values of $\bar y$, in the
region where the data provide constraints the Pauli-Villars
results lie within the uncertainty bands of the nonlocal curves.

\begin{figure}[t]
\hspace*{-0.5cm}\includegraphics[width=0.8\textwidth]{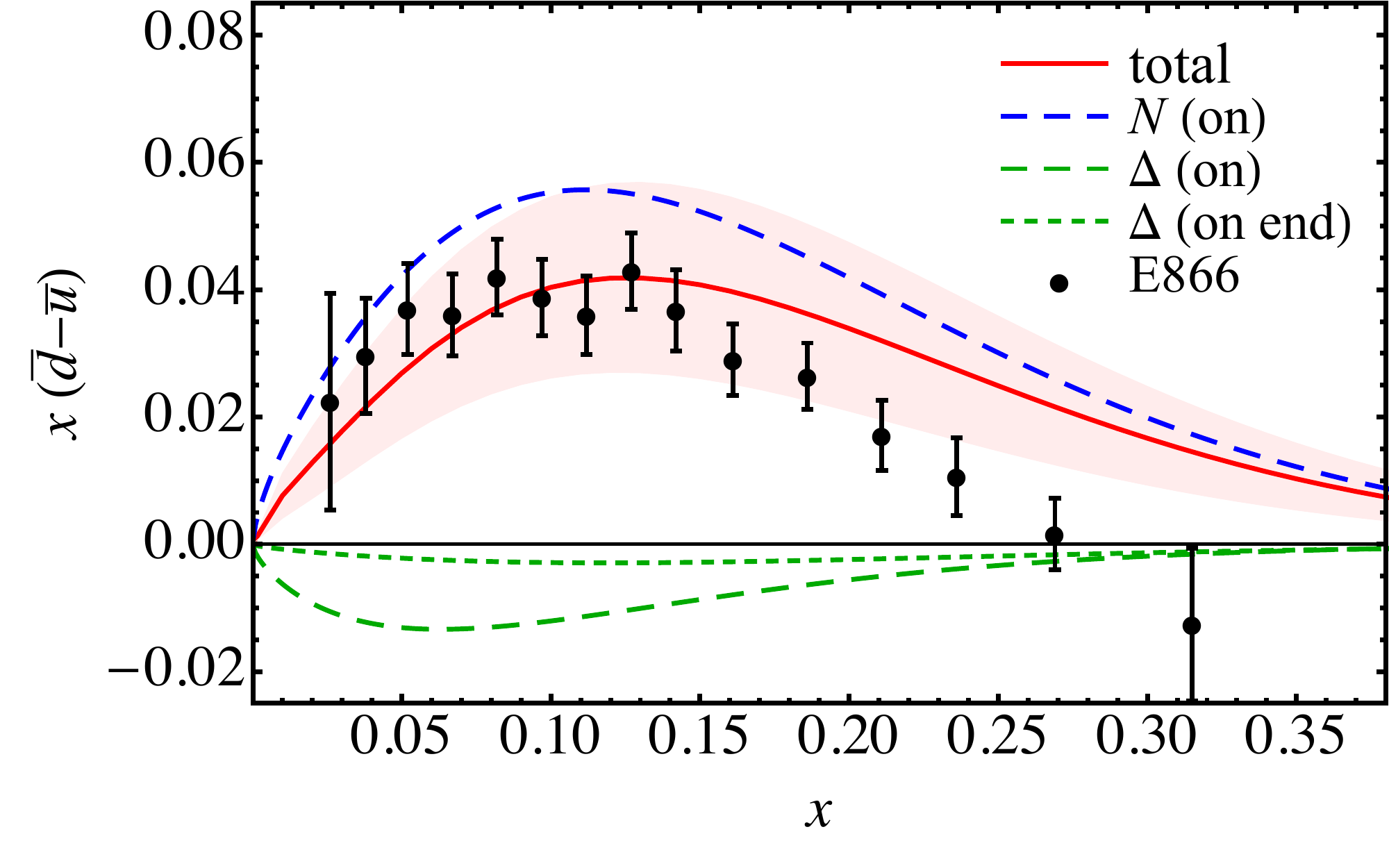}
\caption{The flavor asymmetry of the proton $x(\bar d-\bar u)$
	versus $x$ from lowest order pion exchange
	(solid red curve and pink band), with cutoff parameters
	$\Lambda_{\pi N} = 1.0(1)$~GeV and
	$\Lambda_{\pi \Delta} = 0.9(1)$~GeV, including
	nucleon on-shell (dashed blue),
	$\Delta$ on-shell (dashed green), and
	$\Delta$ end-point (dotted green) contributions,
	and compared with the asymmetry extracted from the
	Fermilab E615 Drell-Yan	experiment~\cite{Towell:2001nh}.}
\label{fig:dbub-x}
\end{figure}

Using the values of $\Lambda_{\pi N}$ and $\Lambda_{\pi \Delta}$
for our nonlocal calculation constrained by the $pp$ cross sections
in Fig.~\ref{fig:cross1}, we next evaluate the flavor asymmetry
$\bar d-\bar u$ from the convolution of the splitting functions
and the pion PDF in Eq.~(\ref{eq:dubar}).
The results for $x(\bar d-\bar u)$ are shown in Fig.~\ref{fig:dbub-x},
and compared with the asymmetry extracted from the E866 Drell-Yan
lepton-pair production data from Fermilab~\cite{Towell:2001nh}.
At nonzero $x$ values only the on-shell nucleon and $\Delta$ and
end-point $\Delta$ terms contribute to the asymmetry, each of
which is indicated in Fig.~\ref{fig:dbub-x}.
The positive nucleon on-shell term makes the largest contribution,
which is partially cancelled by the negative $\Delta$ contributions.
For the values of the cutoffs used here, the end-point term is
relatively small compared with the on-shell $\Delta$ component.

\begin{figure}[t]
\centering
\vspace*{-0.5cm}
\begin{tabular}{ccccc}
{\epsfxsize=3.52in\epsfbox{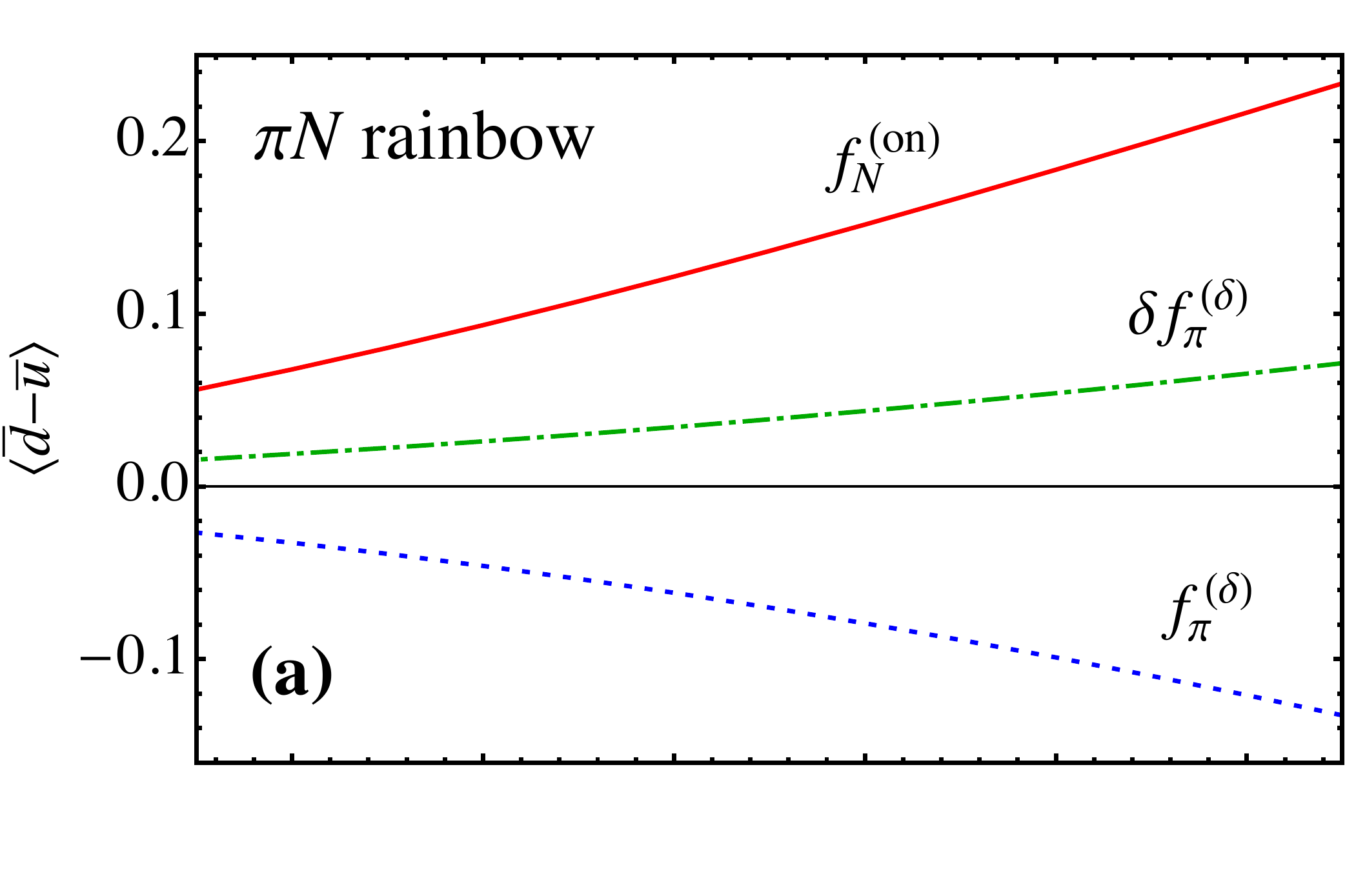} \vspace*{-1.3cm}}&\\
\hspace{-0.1cm}{\epsfxsize=3.52in\epsfbox{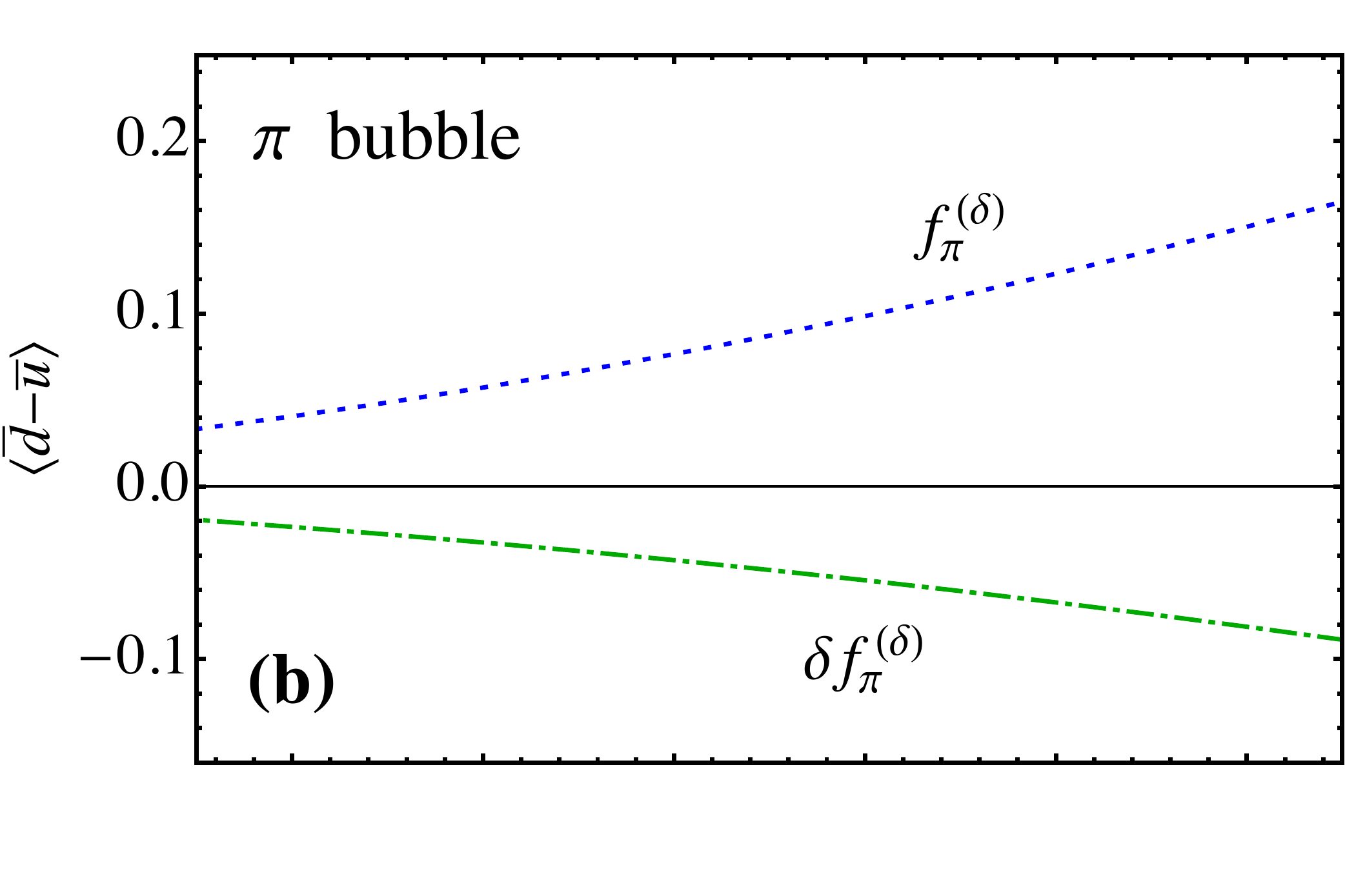}\vspace*{-1.2cm}}&\\
\hspace{-0.32cm}{\epsfxsize=3.61in\epsfbox{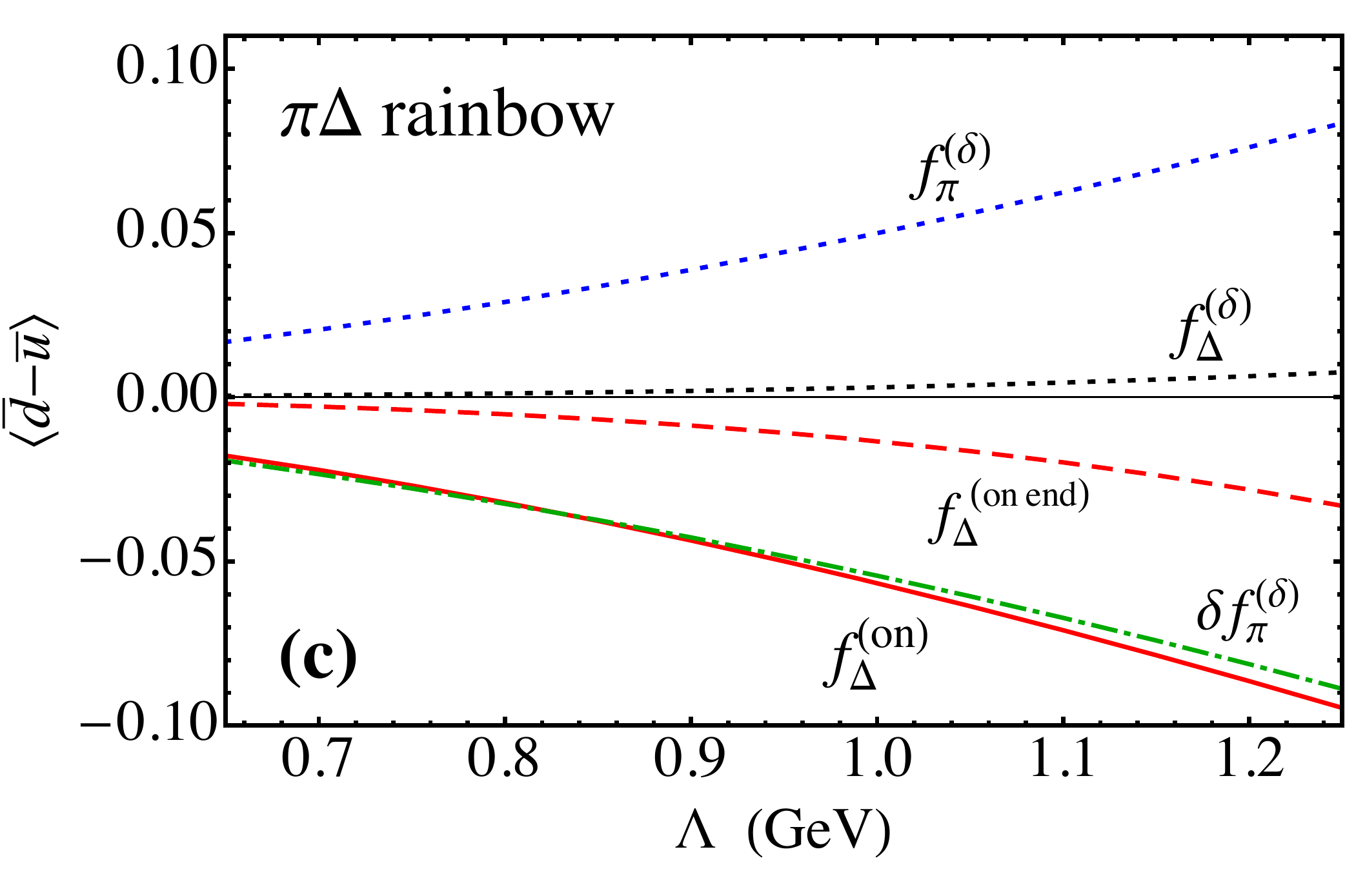}\vspace*{-0.1cm}}&
\end{tabular}
\caption{Contributions to the $\langle \bar d-\bar u \rangle$ moment
	versus the dipole cutoff parameter $\Lambda$
        ($= \Lambda_{\pi N}$ or $\Lambda_{\pi \Delta}$) from
   {\bf (a)}
	the $\pi N$ rainbow diagram [Fig.~\ref{fig:loop_meson}(a)],
	including
	on-shell (solid red curve), and
	local (dotted blue) and
	nonlocal (dot-dashed green)
	$\delta$-function terms;
   {\bf (b)}
	the pion bubble [Fig.~\ref{fig:loop_meson}(c)], including
	local (dotted blue) and
	nonlocal (dot-dashed green) $\delta$-function pieces;
   {\bf (c)}
	the $\pi \Delta$ rainbow [Fig.~\ref{fig:loop_meson}(b)],
	including on-shell (solid red),
	end-point (dashed red),
	local (dotted blue) and
	nonlocal (dot-dashed green) $\delta$-function,
	and local decuplet $\delta$-function (dotted black)
	contributions.}
\label{fig:dbub-int}
\end{figure}

Although the $\delta$-function contributions to the flavor
asymmetry are not directly visible in Fig.~\ref{fig:dbub-x},
their effect can be seen in the lowest moment of the asymmetry,
\begin{equation}
\langle \bar d - \bar u \rangle
\equiv \int_0^1 dx \big( \bar d(x)-\bar u(x) \big).
\label{eq:dbubmom}
\end{equation}
The contributions from the individual on-shell, end-point and
$\delta$-function components of the $\pi N$ and $\pi \Delta$
rainbow and the $\pi$ bubble diagrams to the moment are shown
in Fig.~\ref{fig:dbub-int} versus the dipole cutoff parameter
$\Lambda$ ($=\Lambda_{\pi N}$ or $\Lambda_{\pi \Delta}$),
for the approximate ranges of values found in the fits in
Fig.~\ref{fig:cross1}.
For the best fit values $\Lambda_{\pi N}=1.0(1)$~GeV and
$\Lambda_{\pi \Delta}=0.9(1)$~GeV, the contributions from the
individual terms in Eqs.~(\ref{eq:dubar})--(\ref{eq:f-bub}) are
listed in Table~\ref{tab:du}, along with the combined contributions
from the $x>0$ and $x=0$ terms, and the local and nonlocal
terms, to the total integrated result.
The nucleon on-shell term is the most important component,
with a contribution that is within $\approx 20\%$ of the
total integrated value
    $\langle \bar d - \bar u \rangle = 0.127^{+0.044}_{-0.042}$,
where the errors here reflect the uncertainties on the cutoff
parameters.
The on-shell and end-point $\pi \Delta$ terms yield overall negative
contributions, with magnitude $\approx 30\%$ of the on-shell $\pi N$.
The various $\delta$-function terms from all three diagrams in
Fig.~\ref{fig:loop_meson} cancel to a considerable degree, with
the $x=0$ contribution making up $\approx 20\%$ of the total.
Furthermore, the breakdown into the local and nonlocal pieces
shows that the latter is negative with magnitude $\approx 20\%$
of the local.

\begin{table}[t]
\begin{center}
\caption{Contributions to the integral
	$\langle \bar d-\bar u \rangle \equiv
	   \int_0^1 dx (\bar d-\bar u)$
	from the $\pi N$ rainbow, $\pi \Delta$ rainbow and
	$\pi$ bubble diagrams in Fig.~\ref{fig:loop_meson},
	for the best fit parameters $\Lambda_{\pi N}=1.0(1)$~GeV
	and $\Lambda_{\pi \Delta}=0.9(1)$~GeV.
	The contributions from the various terms in
	Eqs.~(\ref{eq:dubar})--(\ref{eq:f-bub}) are listed
	individually, as are the combined contributions from
	$x>0$ and $x=0$, and the local and nonlocal terms,
	to the total.
	Note that some numbers do not sum to the
	totals because of rounding.\\}
\begin{tabular}{llc}				\hline
     ~diagram~
   &
   & $\langle \bar d-\bar u \rangle$		\\ \hline
     ~$\pi N$ (rbw)~
   & ~$f_N^{\rm (on)}$~
   & ~$0.152^{+0.032}_{-0.030}$~		\\
   & ~$f_\pi^{(\delta)}$~
   & ~$-(0.079^{+0.020}_{-0.018})$~~~		\\
   & ~$\delta f_\pi^{(\delta)}$~
   & ~$0.044^{+0.010}_{-0.009}$~		\\
     ~total~$\pi N$~
   & 
   & ~$0.116^{+0.022}_{-0.022}$~		\\ \hline
     ~$\pi \Delta$ (rbw)~
   & ~$f_\Delta^{\rm (on)}$~
   & ~$-(0.044^{+0.012}_{-0.012})$~~~		\\
   & ~$f_\Delta^{\rm (on\,end)}$~
   & ~$-(0.009^{+0.004}_{-0.003})$~~~		\\
   & ~$f_\Delta^{(\delta)}$~
   & ~$0.002^{+0.001}_{-0.001}$~		\\
   & ~$f_\pi^{(\delta)}$~
   & ~$0.039^{+0.010}_{-0.010}$~		\\
   & ~$\delta f_\pi^{(\delta)}$~
   & ~$-(0.022^{+0.005}_{-0.005})$~~~		\\
     ~total~$\pi \Delta$~
   & 
   & ~$-(0.033^{+0.010}_{-0.010})$~~~		\\ \hline
     ~$\pi$~(bub)
   & ~$f_\pi^{(\delta)}$~
   & ~$0.099^{+0.025}_{-0.022}$~		\\
   & ~$\delta f_\pi^{(\delta)}$~
   & ~$-(0.054^{+0.013}_{-0.012})$~~~		\\
     ~total~$\pi$~bubble~~~
   &
   & ~$0.044^{+0.012}_{-0.010}$~		\\ \hline
     ~{\bf total}~
   & 
   & ~$\bm{0.127^{+0.044}_{-0.042}$}~		\\ \hline\hline
     ~$x\!>\!0$~
   &
   & ~$0.099^{+0.047}_{-0.046}$~		\\
     ~$x\!=\!0$~
   &
   & ~$0.028^{+0.008}_{-0.007}$~		\\ \hline
     ~local~
   &
   & ~$0.159^{+0.041}_{-0.039}$~		\\
     ~nonlocal~
   &
   & ~$-(0.032^{+0.008}_{-0.008})$~~~		\\ \hline
\end{tabular}
\label{tab:du}
\end{center}
\end{table}

Experimentally, the asymmetry at $x=0$ is of course not directly
measurable, and typically extrapolations are made to estimate
contributions from outside of the measured region.
The New Muon Collaboration, for instance, found
	$\langle \bar d - \bar u \rangle_{\rm NMC}^{\rm (exp)}
	= 0.169(32)$
from their analysis of $F_2^p-F_2^n$ in the experimentally accessible
region $0.004 \leq x \leq 0.8$, and
	$\langle \bar d - \bar u \rangle_{\rm NMC}^{\rm (tot)}
	= 0.148(39)$
when including $x \to 0$ and $x \to 1$ extrapolations~\cite{Arneodo:1994sh}.
The E866 Collaboration, on the other hand, extracted
	$\langle \bar d - \bar u \rangle_{\rm E866}^{\rm (exp)}
	= 0.080(11)$
in the experimentally measured interval $0.015 \leq x \leq 0.035$, and
	$\langle \bar d - \bar u \rangle_{\rm E866}^{\rm (tot)}
	= 0.118(12)$
for the entire $x$ range after extrapolation.
Note that the extrapolations by different analyses are often based on
different assumptions for the asymptotic $x \to 0$ and $x \to 1$
behavior, so that a direct comparison of extrapolated results is
problematic.
Nevertheless, the general magnitude of the asymmetry is comparable
with that found in our calculation, even with the uncertainties about
the $x=0$ and extrapolated contributions.

\subsection{$s-\bar{s}$ asymmetry}
\label{ssec:ssbar}

While the $\bar d-\bar u$ asymmetry is perhaps the best known
consequence of pion loops on PDFs in the nucleon, an equally
intriguing ramification of SU(3) chiral symmetry breaking is
the $s-\bar s$ asymmetry generated by kaon loops.
In analogy to the light antiquark PDFs in Eq.~(\ref{eq:dubar}),
the contribution to the antistrange PDF in the proton arising
from kaon loops in Fig.~\ref{fig:loop_meson} can be written as
\begin{equation}
\begin{split}
\bar{s}(x)
= \Big[
  \Big(   \sum_{\phi B} f^{\rm (rbw)}_{\phi B}
	+ \sum_{\phi T} f^{\rm (rbw)}_{\phi T}
	+ \sum_{\phi} f^{\rm (bub)}_\phi
  \Big) \otimes \bar{s}_K
  \Big](x),
\end{split}
\label{eq:antistrange}
\end{equation}
where here the sums are over the states
	$\phi B = \{ K^+ \Lambda, K^+ \Sigma^0, K^0 \Sigma^+ \}$
for the kaon-octet baryon rainbow diagram [Fig.~\ref{fig:loop_meson}(a)],
	$\phi T = \{ K^+ \Sigma^{*0}, K^0 \Sigma^{*+} \}$
for the kaon-decuplet baryon rainbow diagram [Fig.~\ref{fig:loop_meson}(b)],
and for the $\phi = K^+ (K^-)$ and $K^0 (\overline{K}^0)$ loop in the
bubble diagram [Fig.~\ref{fig:loop_meson}(c)].
In terms of the on-shell and $\delta$-function basis functions,
the kaon-octet baryon rainbow function can be written is given by
a form similar to that in Eq.~(\ref{eq:piN-rbw}),
\begin{subequations}
\label{eq:K-octet-rbw}
\begin{eqnarray}
f^{\rm (rbw)}_{K^+\Lambda}(y)
&=& \frac{ (D+3F)^2 (M+M_\Lambda)}{12 (4\pi f)^2}
    \Big[ f^{\rm (on)}_\Lambda(y)
        + f^{(\delta)}_K(y)
        - \delta f^{(\delta)}_K(y)
    \Big],
\label{eq:KLam-rbw}			\\
2 f^{\rm (rbw)}_{K^+\Sigma^0}(y)
= f^{\rm (rbw)}_{K^0\Sigma^+}(y)
&=& \frac{ (D\,-\,F)^2 (M+M_\Sigma)}{2 (4\pi f)^2}\,
    \Big[ f^{\rm (on)}_\Sigma(y)
        + f^{(\delta)}_K(y)
        - \delta f^{(\delta)}_K(y)
    \Big],
\label{eq:KSig-rbw}
\end{eqnarray}
\end{subequations}
for the $K \Lambda$ and $K \Sigma$ intermediate states, respectively.
For the kaon-decuplet baryon rainbow diagram, the corresponding
function is written analogously to Eq.~(\ref{eq:piD-rbw}),
\begin{eqnarray}
2 f^{\rm (rbw)}_{K^+ \Sigma^{*0}}(y)\
= f^{\rm (rbw)}_{K^0 \Sigma^{*+}}(y)
&=& \frac{{\cal C}^2 (M+M_{\Sigma^*})^2}{6 (4\pi f)^2}
\Big[ f^{\rm (on)}_{\Sigma^*}(y)
     + f^{\rm (on\, end)}_{\Sigma^*}(y)
     - \frac{1}{18} f^{(\delta)}_{\Sigma^*}(y)           \notag\\
& & \hspace*{-0.2cm}
   +\, \frac{(M+M_\Sigma)^2 \big[ (M+M_{\Sigma^*})^2 - m_K^2 \big]}
	    {6 M_{\Sigma^*}^2\, (M+M_{\Sigma^*})^2}
       \big( f^{(\delta)}_K(y) - \delta f^{(\delta)}_K(y) \big)
\Big]
\label{eq:KSigstar-rbw}
\end{eqnarray}
for the $K \Sigma^*$ states, where the coupling ${\cal C}$ is given
in the previous section.
For the bubble diagram, the splitting function for charged or neutral
kaon loops is given by a form similar to that in Eq.~(\ref{eq:f-bub}),
\begin{eqnarray}
f^{\rm (bub)}_{K^+}(y)\
 =\ 2 f^{\rm (bub)}_{K^0}(y)
&=&-\frac{ (M+M_\Sigma)^2}{(4\pi f)^2}
    \Big[ f^{(\delta)}_K(y) - \delta f^{(\delta)}_K(y) \Big].
\label{eq:f-Kbub}
\end{eqnarray}
Explicit expressions for all the basis functions are given in
Ref.~\cite{nonlocal-I}.

For the loop contributions to the strange quark PDF, the
baryon-coupling rainbow, Kroll-Ruderman and tadpole diagrams in
Fig.~\ref{fig:loop_baryon}(d)--(k) all play a role, as do the
additional gauge-link dependent diagrams that are generated by
the nonlocal Lagrangian.
Assuming that all nonperturbatively generated strangess resides in
the intermediate state hyperons, from Eq.~(\ref{eq:quark}) the loop
contributions to the strange quark PDF in the proton can be written as
\begin{equation}
\begin{split}
s(x)
&= \sum_{B\phi}
   \Big\{
   \Big[ \bar{f}^{\rm (rbw)}_{B\phi} \otimes s_B \Big](x)
 + \Big[ \bar{f}^{\rm (KR)}_B \otimes s^{\rm (KR)}_B \Big](x)
 + \Big[ \delta \bar{f}^{\rm (KR)}_B \otimes s^{(\delta)}_B \Big](x)
   \Big\}		\\
&+\, \sum_{T\phi}
   \Big\{
   \Big[ \bar{f}^{\rm (rbw)}_{T\phi} \otimes s_T \Big](x)
 + \Big[ \bar{f}^{\rm (KR)}_T \otimes s^{\rm (KR)}_T \Big](x)
 + \Big[ \delta \bar{f}^{\rm (KR)}_T \otimes s^{(\delta)}_T \Big](x)
   \Big\}		\\
&+\, \sum_\phi
   \Big\{
   \Big[ \bar{f}^{\rm (tad)}_\phi \otimes s^{\rm (tad)}_\phi \Big](x)
 + \Big[ \delta \bar{f}^{\rm (tad)}_\phi \otimes s^{(\delta)}_\phi \Big](x)
   \Big\},
\end{split}
\label{eq:strange}
\end{equation}
where the sums are over the octet bayon--meson states
	$B \phi = \{ \Lambda K^+, \Sigma^0 K^+, \Sigma^+ K^0 \}$,
decuplet baryon--meson states
	$T \phi = \{ \Sigma^{*0} K^+, \Sigma^{*+} K^0 \}$,
and mesons $\phi = K^+ (K^-)$ and $K^0 (\overline{K}^0)$ for the
tadpole contributions.
As in Eq.~(\ref{eq:quark}), the splitting functions for all the
hyperon coupling diagrams in Eq.~(\ref{eq:strange}) use the
shorthand notation ${\bar f}_j(y) \equiv f_j(1-y)$.

For the octet hyperon rainbow diagrams, Fig.~\ref{fig:loop_baryon}(d),
the individual splitting functions can be written in terms of the
on-shell, off-shell and $\delta$-function basis functions as
\begin{subequations}
\label{eq:f-rbw-Bphi}
\begin{eqnarray}
f^{\rm (rbw)}_{\Lambda K^+}(y)
&=& \frac{(D+3F)^2 (M+M_\Lambda)^2}{12(4\pi f)^2}
    \Big[ f^{\rm (on)}_\Lambda(y)
	+ f^{\rm (off)}_\Lambda(y)
	+ 4\, \delta f^{\rm (off)}_\Lambda(y)
	- f^{(\delta)}_K(y)
    \Big],				\nonumber\\
& &					\\
2 f^{\rm (rbw)}_{\Sigma^0 K^+}(y)=f^{\rm (rbw)}_{\Sigma^+ K^0}(y)
&=& \frac{(D\, -\, F)^2 (M+M_\Sigma)^2}{2(4\pi f)^2}\,
    \Big[ f^{\rm (on)}_\Sigma(y)
	+ f^{\rm (off)}_\Sigma(y)
	+ 4\, \delta f^{\rm (off)}_\Sigma(y)
	- f^{(\delta)}_K(y)
    \Big],				\nonumber\\
& &
\end{eqnarray}
\end{subequations}
where the functions $f^{\rm (on)}_{\Lambda,\Sigma}$ and $f^{(\delta)}_K$
are the same as in Eq.~(\ref{eq:K-octet-rbw}), and explicit expressions
for the off-shell functions $f^{\rm (off)}_{\Lambda,\Sigma}$ and
$\delta f^{\rm (off)}_{\Lambda,\Sigma}$ are given in Sec.~IV.B.1 of
Ref.~\cite{nonlocal-I}.
For the octet Kroll-Ruderman diagrams in Fig.~\ref{fig:loop_baryon}(e)
and \ref{fig:loop_baryon}(f), the local and nonlocal splitting functions
$f^{\rm (KR)}_{\Lambda,\Sigma}$ and $\delta f^{\rm (KR)}_{\Lambda,\Sigma}$
are given by
\begin{subequations}
\label{eq:f-KR-Bphi}
\begin{eqnarray}
f^{\rm (KR)}_\Lambda(y)
&=& \frac{(D+3F)^2 (M+M_\Lambda)^2}{12(4\pi f)^2}
    \Big[- f^{\rm (off)}_\Lambda(y)
	 + 2 f^{(\delta)}_K(y)
    \Big],				\\
2 f^{\rm (KR)}_{\Sigma^0}(y)
= f^{\rm (KR)}_{\Sigma^+}(y)
&=& \frac{(D\,-F\,)^2 (M+M_\Sigma)^2}{2(4\pi f)^2}\,
    \Big[- f^{\rm (off)}_\Sigma(y)
	 + 2 f^{(\delta)}_K(y)
    \Big].
\end{eqnarray}
\end{subequations}
and
\begin{subequations}
\label{eq:f-link-Bphi}
\begin{eqnarray}
\delta f^{\rm (KR)}_\Lambda(y)
&=& \frac{(D+3F)^2 (M+M_\Lambda)^2}{12(4\pi f)^2}
    \left[-4\, \delta f^{\rm (off)}_\Lambda(y)\,
	  -\,  \delta f^{(\delta)}_K(y)
    \right],					\\
2\, \delta f^{\rm (KR)}_{\Sigma^0}(y)
= \delta f^{\rm (KR)}_{\Sigma^+}(y)
&=& \frac{(D\,-F\,)^2 (M+M_\Sigma)^2}{2(4\pi f)^2}\,
    \left[-4\, \delta f^{\rm (off)}_\Sigma(y)\,
	  -\,  \delta f^{(\delta)}_K(y)
    \right],
\end{eqnarray}
\end{subequations}
respectively.

For the decuplet hyperon contributions, the respective splitting
functions are given by
\begin{eqnarray}
2 f^{\rm (rbw)}_{\Sigma^{*0} K^+}(y)
= f^{\rm (rbw)}_{\Sigma^{*+} K^0}(y)
&=& \frac{{\cal C}^2 (M+M_{\Sigma^*})^2}{6 (4\pi f)^2}	\notag\\
& & \hspace*{-2.0cm} \times
\Big[
  f^{\rm (on)}_{\Sigma^*}(y)
+ f^{\rm (on\, end)}_{\Sigma^*}(y)
- 2 f^{\rm (off)}_{\Sigma^*}(y)
- 2 f^{\rm (off\, end)}_{\Sigma^*}(y)
+ 4\, \delta\!f^{\rm (off)}_{\Sigma^*}(y)			\\
& & \hspace*{-1.5cm}
+ \frac{1}{18} f^{(\delta)}_{\Sigma^*}(y)
- \frac{1}{6}\, \delta\!f^{(\delta)}_{\Sigma^*}(y)
- \frac{(M+M_\Sigma)^2 [(M+M_{\Sigma^*})^2 + 3 m_K^2]}
       {6 M_{\Sigma^*}^2\, (M+M_{\Sigma^*})^2}\, f^{(\delta)}_K(y)
\Big]							\notag
\label{eq:dsigma}
\end{eqnarray}
for the decuplet rainbow diagram in Fig.~\ref{fig:loop_baryon}(g),
\begin{eqnarray}
\hspace*{-0.3cm}
2 f^{(\rm KR)}_{\Sigma^{*0}}(y)
= f^{(\rm KR)}_{\Sigma^{*+}}(y)
&=&\frac{{\cal C}^2 (M+M_{\Sigma^*})^2}{6 (4\pi f)^2}
\Big[
  2 f^{\rm (off)}_{\Sigma^*}(y)
+ 2 f^{\rm (off\, end)}_{\Sigma^*}(y)			\\
&-& \frac19
    \big( f^{(\delta)}_{\Sigma^*}(y)
        - \delta\!f^{(\delta)}_{\Sigma^*}(y)
    \big)
 +  \frac{(M+M_\Sigma)^2 [(M+M_{\Sigma^*})^2 + m_K^2]}
         {3 M_{\Sigma^*}^2\, (M+M_{\Sigma^*})^2}\, f^{(\delta)}_K(y)
\Big]							\notag
\label{eq:fTKR}
\end{eqnarray}
for the Kroll-Ruderman diagram in Fig.~\ref{fig:loop_baryon}(h), and
\begin{eqnarray}
\hspace*{-0.5cm}
2\, \delta\!f^{(\rm KR)}_{\Sigma^{*0}}(y)
 =  \delta\!f^{(\rm KR)}_{\Sigma^{*+}}(y)
&=& \frac{{\cal C}^2 (M+M_{\Sigma^*})^2}{6 (4\pi f)^2}
\Big[
  - 4\, \delta\!f^{\rm (off)}_{\Sigma^*}(y)
  + \frac{1}{18}\, \delta\!f^{(\delta)}_{\Sigma^*}(y)	\notag\\
& & \hspace*{3cm}
  -\, \frac{(M+M_{\Sigma})^2 [(M+M_{\Sigma^*})^2 - m_K^2]}
           {6 M_{\Sigma^*}^2\, (M+M_{\Sigma^*})^2}\,
    \delta\!f^{(\delta)}_K(y)
\Big]
\label{eq:fdfTKR}
\end{eqnarray}
for the nonlocal Kroll-Ruderman diagram in Fig.~\ref{fig:loop_baryon}(i).
The expressions for the decuplet basis functions
  $f^{\rm (on)}_{\Sigma^*}$,
  $f^{\rm (on\, end)}_{\Sigma^*}$,
  $f^{\rm (off)}_{\Sigma^*}$,
  $f^{\rm (off\, end)}_{\Sigma^*}$ and
  $f^{\rm (\delta)}_{\Sigma^*}$,
as well as the nonlocal functions
  $\delta f^{\rm (off)}_{\Sigma^*}$ and
  $\delta f^{\rm (\delta)}_{\Sigma^*}$,
are given in Sec.~IV.B.2 of Ref.~\cite{nonlocal-I}.

Finally, for the local and nonlocal tadpole contributions to the
strange quark PDF from Fig.~\ref{fig:loop_baryon}(j) and (k),
the splitting functions are given by
\begin{eqnarray}
\label{eq:f-tad}
f^{\rm (tad)}_{K^+}(y)
&=& 2\, f^{\rm (tad)}_{K^0}(y)\
 =\ -\frac{ (M+M_\Sigma)^2}{(4\pi f)^2}\,
    f^{(\delta)}_K(y),				\\
\delta\!f^{\rm (tad)}_{K^+}(y)
&=& 2\, \delta\!f^{\rm (tad)}_{K^0}(y)\
 =\ \frac{(M +M_\Sigma)^2}{(4\pi f)^2}\,
    \delta\!f^{(\delta)}_K(y),
\label{eq:ad45}
\end{eqnarray}
in terms of the local and nonlocal basis functions $f^{(\delta)}_K$
and $\delta\!f^{(\delta)}_K$.

To determine the regulator mass parameter for the kaon--hyperon--nucleon
vertices in Figs.~\ref{fig:loop_meson} and \ref{fig:loop_baryon},
we consider inclusive hyperon production cross sections in $pp$
collisions, in analogy with the neutron and $\Delta$ production above.
Data on inclusive $\Lambda$ production are available from the 2~m
hydrogen bubble chamber at the CERN proton synchrotron
\cite{Blobel:1978yj} and the 12-foot hydrogen bubble chamber at
ANL~\cite{Jaeger:1974pk}, and on inclusive $\Sigma^*$ production
from CERN bubble chamber experiments~\cite{Bockmann:1977sc}.
The corresponding differential cross sections for inclusive
$\Lambda$ and $\Sigma^*$ production (for $y > 0$) are given by
\begin{eqnarray}
\sigma(pp \to \Lambda X)
&=& \frac{(D+3F)^2 (M+M_\Lambda)^2}{12 (4\pi f)^2}\,
  \frac{\bar y}{\pi}\,
  {\hat f}_\Lambda^{(\rm on)}(y,k_\bot^2)\,
  \sigma_{\rm tot}^{K^+ p}(ys),
\label{eq:LambdaX}			\\
\sigma(pp \to \Sigma^{*+}X)
&=& \frac{{\cal C}^2\, (M+M_{\Sigma^*})^2}{6 (4\pi f)^2}\,
  \frac{\bar y}{\pi}\,
  \Big[   {\hat f}_{\Sigma^{*+}}^{(\rm on)}(y,k_\bot^2)
	+ {\hat f}_{\Sigma^{*+}}^{(\rm on\, end)}(y,k_\bot^2)\,
  \Big]
  \sigma_{\rm tot}^{K^0 p}(sy),
\label{eq:SigmaX}
\end{eqnarray}
where
  ${\hat f}_{\Lambda }^{(\rm on)}$,
  ${\hat f}_{\Sigma^{*+}}^{(\rm on)}$ and
  ${\hat f}_{\Sigma^{*+}}^{(\rm on\, end)}$
are the $k_\perp$-unintegrated splitting functions defined
from the on-shell and end-point basis functions
(see Eqs.~(63), (86) and (88) in Ref.~\cite{nonlocal-I})
by the relations
\begin{subequations}
\begin{eqnarray}
f_{\Lambda}^{(\rm on)}(y)
&\equiv& \int dk_\bot^2\,
	 \hat{f}_\Lambda^{(\rm on)}(y,k_\bot^2),	\\
f_{\Sigma^{*+}}^{(\rm on)}(y)
&\equiv& \int dk_\bot^2\,
	 \hat{f}_{\Sigma^{*+}}^{(\rm on)}(y,k_\bot^2),	\\
f_{\Sigma^{*+}}^{(\rm on\, end)}(y)
&\equiv& \int dk_\bot^2\,
	 \hat{f}_{\Sigma^{*+}}^{(\rm on\, end)}(y,k_\bot^2).
\end{eqnarray}
\end{subequations}
In Eqs.~(\ref{eq:LambdaX}) and (\ref{eq:SigmaX})
$\sigma_{\rm tot}^{K^+ p}$ and $\sigma_{\rm tot}^{K^0 p}$ are
the total kaon--proton scattering cross sections, evaluated at
invariant mass $sy$.
For the numerical calculations we take the empirical value for
    $\sigma_{\rm tot}^{K^+ p} = 19.9(1)$~mb
from Ref.~\cite{Hufner:1992cu}, independent of energy.
As there are no data for the $K^0 p$ total cross section, we assume
charge symmetry and relate this to the measured $K^+ n$ cross section,
    $\sigma_{\rm tot}^{K^0 p} \approx
     \sigma_{\rm tot}^{K^+ n} = 19.7(1)$~mb~\cite{Carroll:1978vq}.

\begin{figure}[t]
\centering
\vspace*{-0.5cm}
\begin{tabular}{ccccc}
{\epsfxsize=3.62in\epsfbox{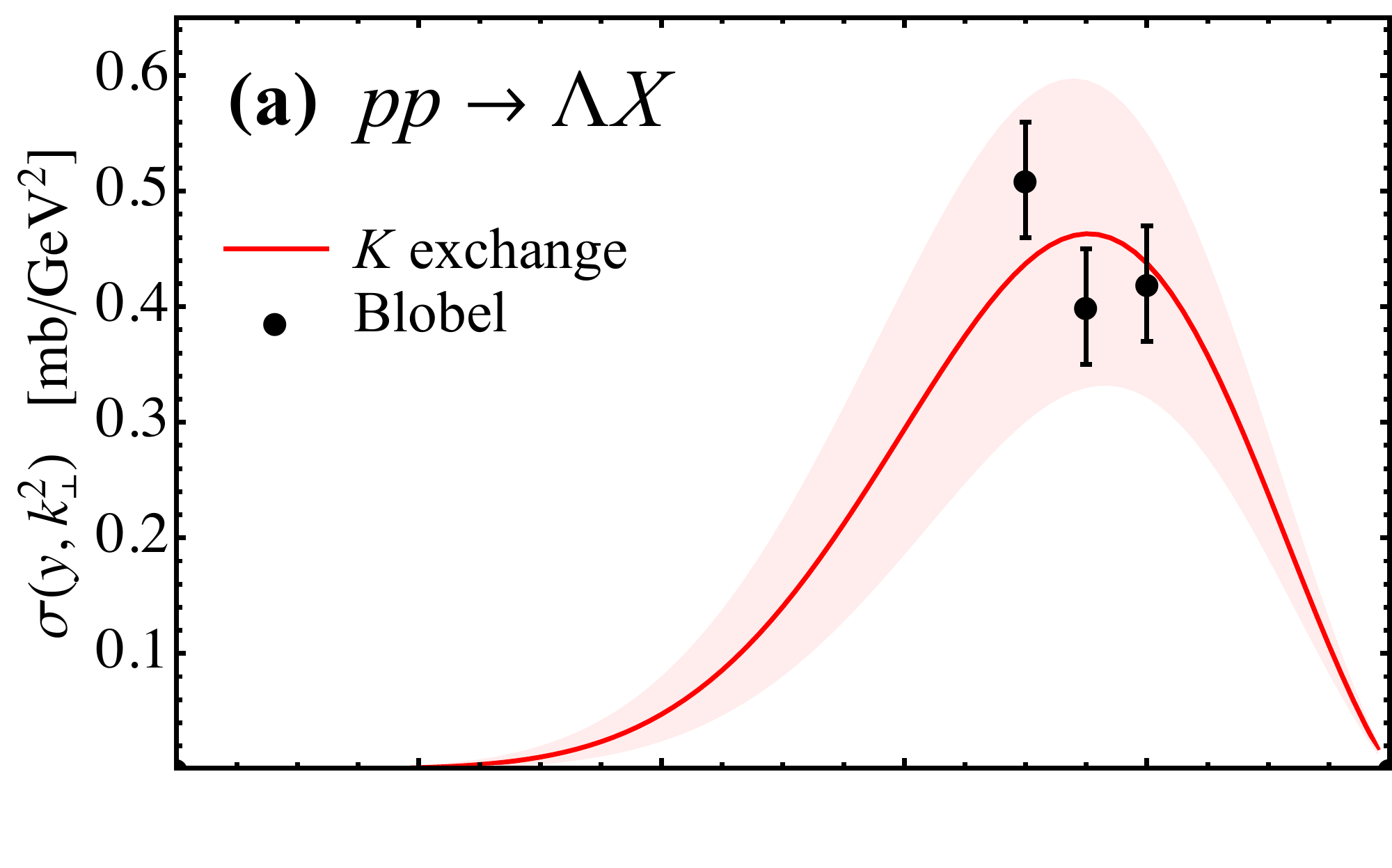}\vspace*{-1.0cm}} &\\
\hspace{-0.3cm}{\epsfxsize=3.732in\epsfbox{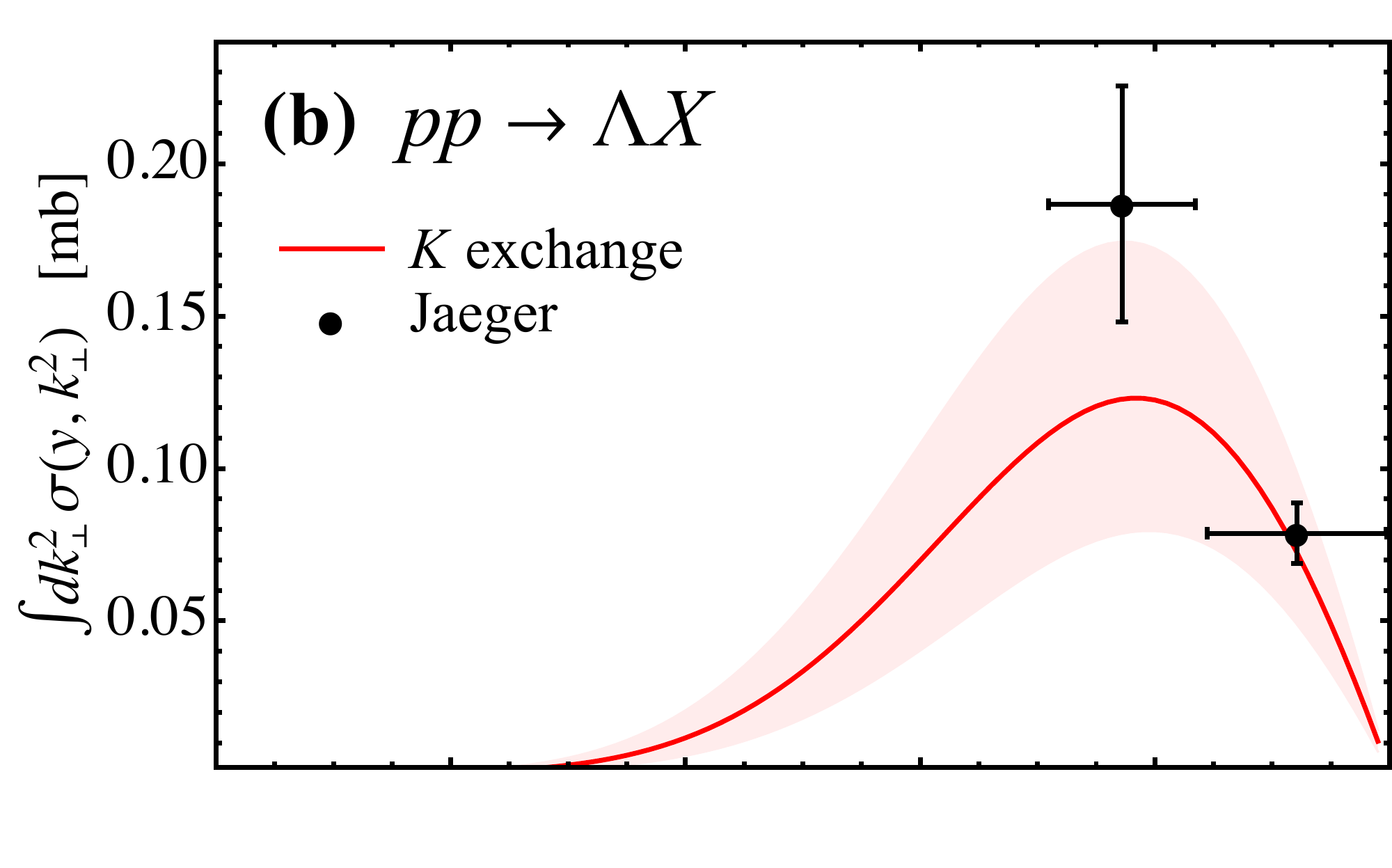}\vspace*{-0.795cm}}&\\
\hspace{-0.06cm}{\epsfxsize=3.81in\epsfbox{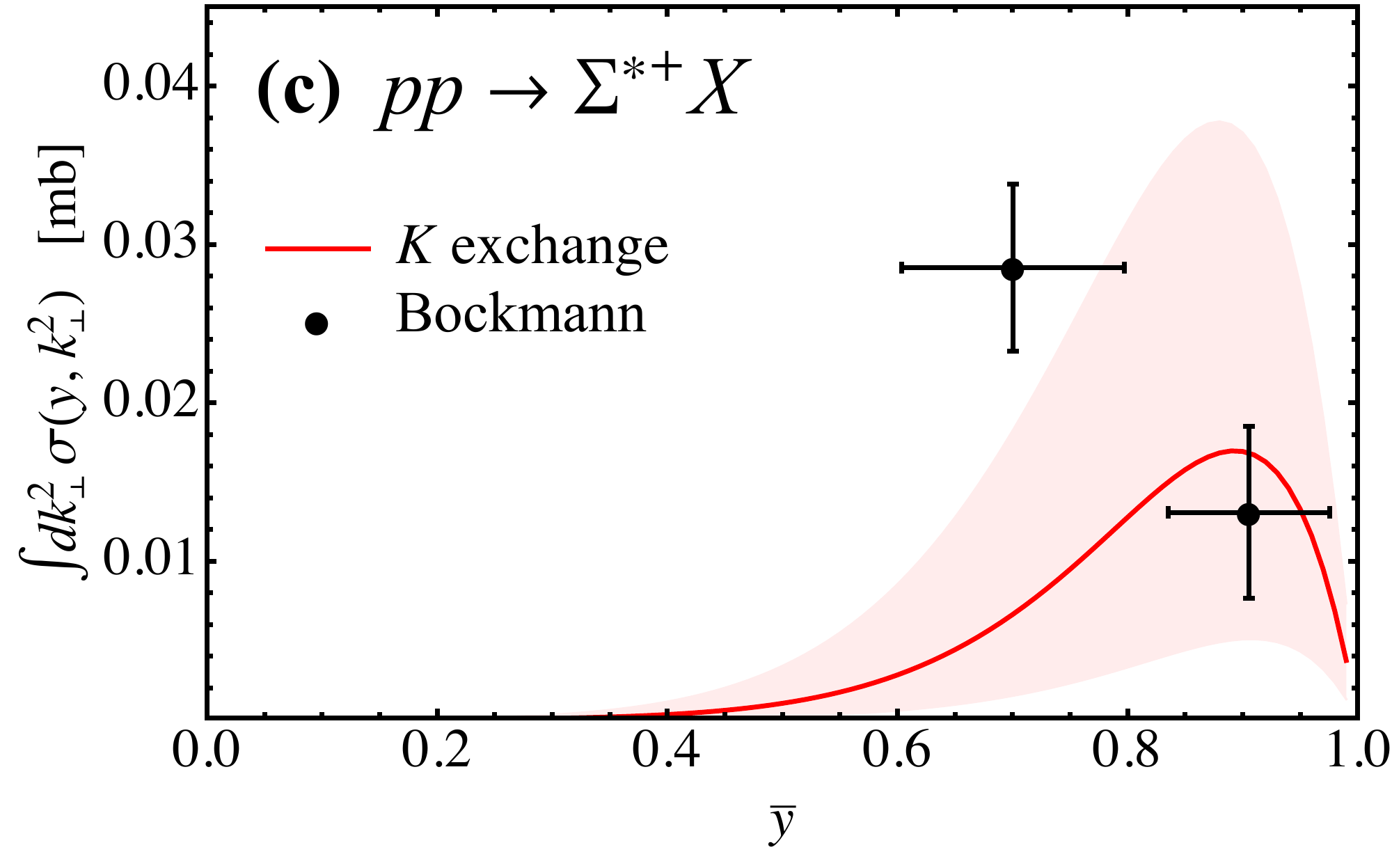}}&
\end{tabular}
\caption{Differential inclusive hadron production cross section
	$\sigma(y,k_\bot^2)$ versus $\bar{y}$ for
   {\bf (a)} $pp \to \Lambda X$ at $k_\bot = 0.075$~GeV
	\cite{Blobel:1978yj};		
   {\bf (b)} $pp \to \Lambda X$ integrated over $k_\bot^2$
	\cite{Jaeger:1974pk};		
   {\bf (c)} $pp \to \Sigma^{*+} X$ integrated over $k_\bot^2$
	\cite{Bockmann:1977sc},		
	compared with the fitted nonlocal kaon exchange
	contributions for dipole regulator parameters
	$\Lambda_{K\Lambda}=1.1(1)$~GeV and
	$\Lambda_{K\Sigma^*}=0.8(1)$~GeV
	(solid red lines and pink $1\sigma$ uncertainty bands).}
\label{fig:cross2}
\end{figure}

In a similar vein to the pion exchange analysis of neutron and
$\Delta$ production discussed above, in Fig.~\ref{fig:cross2}
we compare the inclusive $pp \to \Lambda X$ and $\Sigma^{*+} X$
cross sections for $\bar{y} > 0.7$ with the kaon exchange contributions
calculated from Eqs.~(\ref{eq:LambdaX}) and (\ref{eq:SigmaX}).
The best fit to the CERN bubble chamber $\Lambda$ production
data from Ref.~\cite{Blobel:1978yj} at $k_\bot = 0.075$~GeV
[Fig.~\ref{fig:cross2}(a)] and the $k_\perp$-integrated data
from Ref.~\cite{Jaeger:1974pk} [Fig.~\ref{fig:cross2}(b)]
yields a dipole regulator mass $\Lambda_{K\Lambda} = 1.1(1)$~GeV,
similar to the value found for the $\pi N$ cutoff parameter from
the inclusive neutron production data in Fig.~\ref{fig:cross1}.
Comparison of the singly differential decuplet $\Sigma^{*+}$
production data at large~$\bar{y}$ [Fig.~\ref{fig:cross2}(c)]
with the kaon exchange cross section in Eq.~(\ref{eq:SigmaX})
gives a best fit for the decuplet regulator mass of
$\Lambda_{K\Sigma^*} = 0.8(1)$~GeV.	
The cutoff parameter for the decuplet baryon is again slightly
smaller than that for the octet baryon, as was found for the pion
exchange contributions to the neutron and $\Delta$ cross sections
in Fig.~\ref{fig:cross1}.

\begin{figure}[t]
\centering
\begin{tabular}{ccc}
\hspace{-0.7cm}{\epsfxsize=3.33in\epsfbox{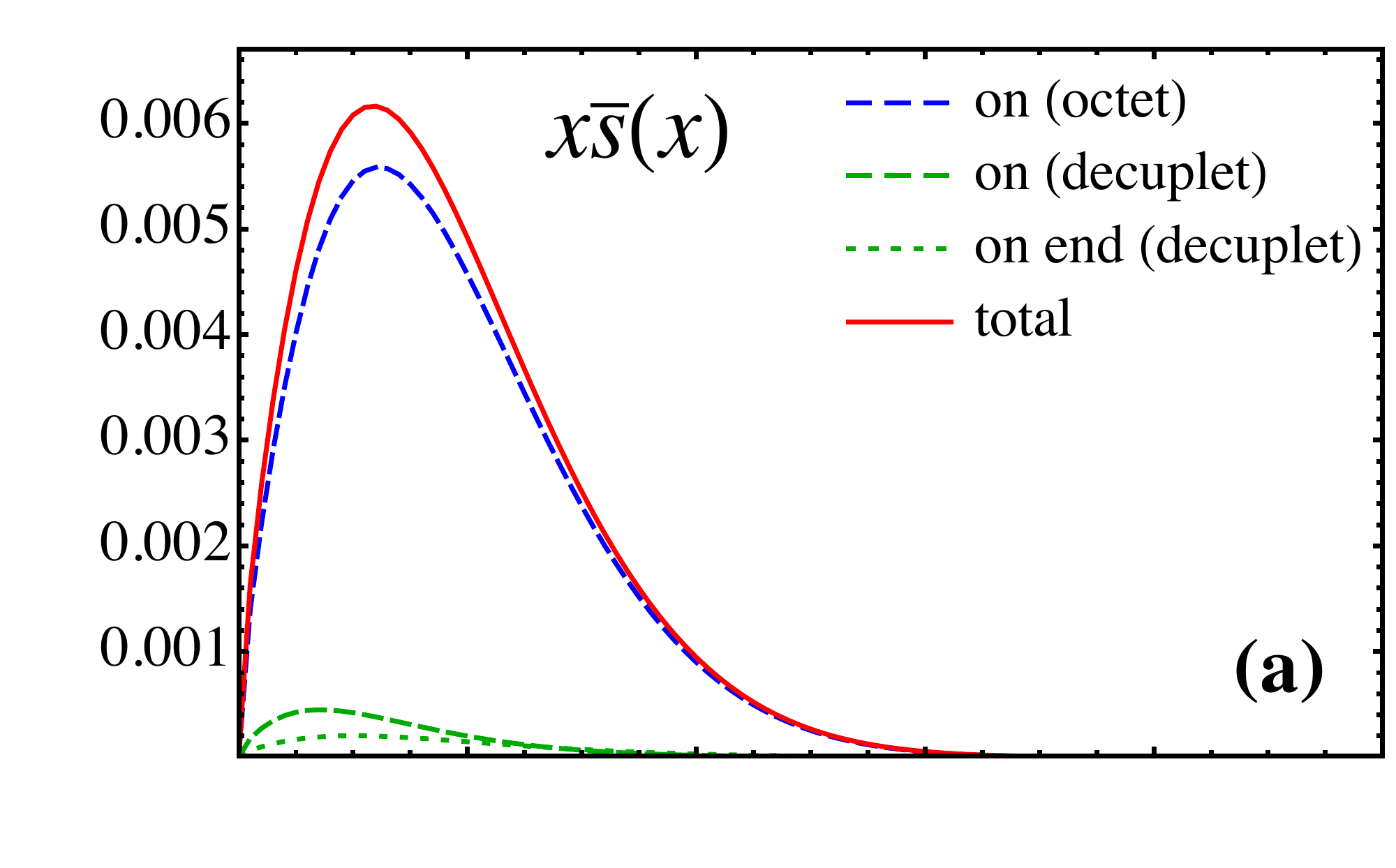}\vspace*{-0.9cm}}&
\hspace{-1.0cm}{\epsfxsize=3.49in\epsfbox{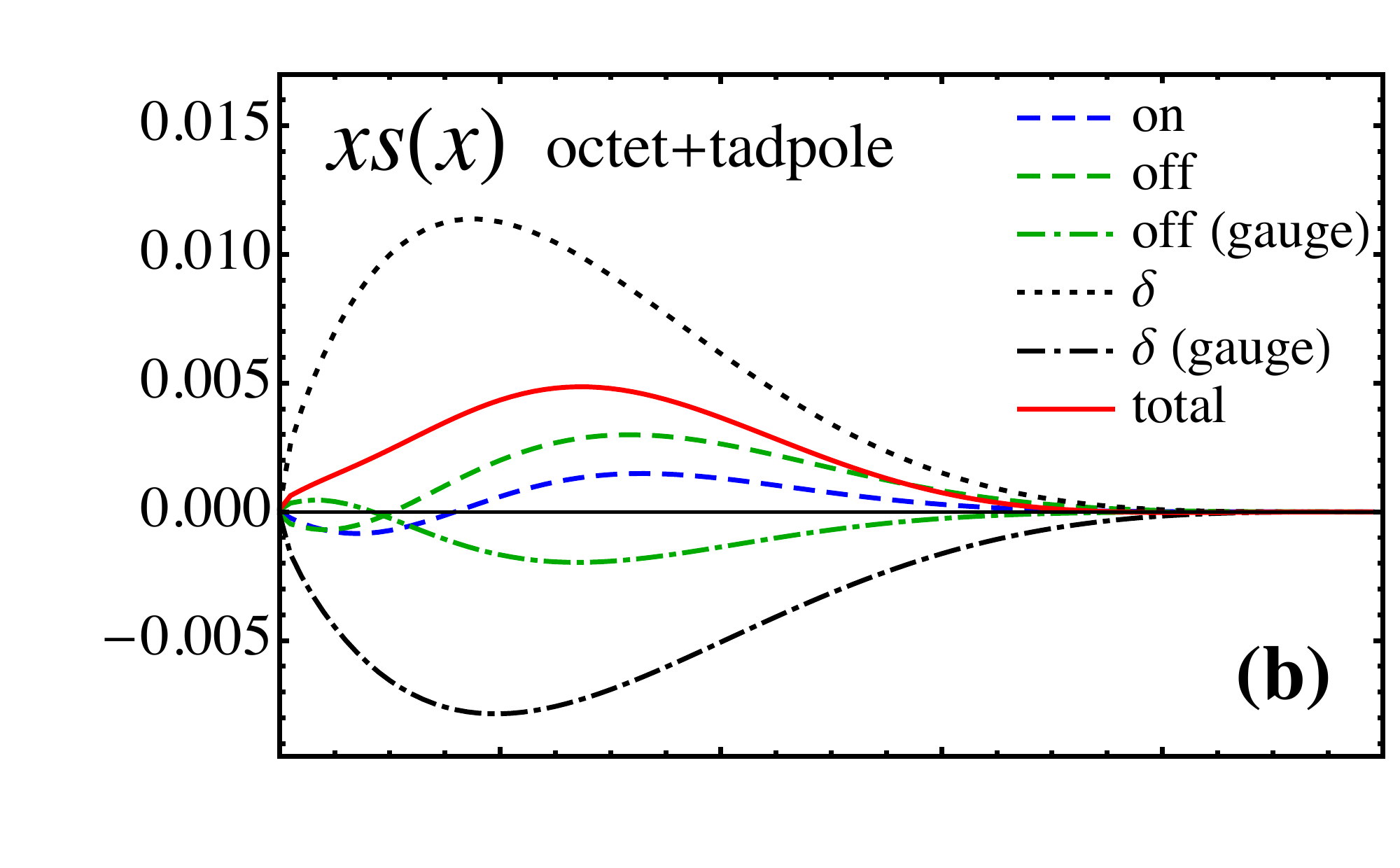}}&\\
\hspace{-0.84cm}{\epsfxsize=3.51in\epsfbox{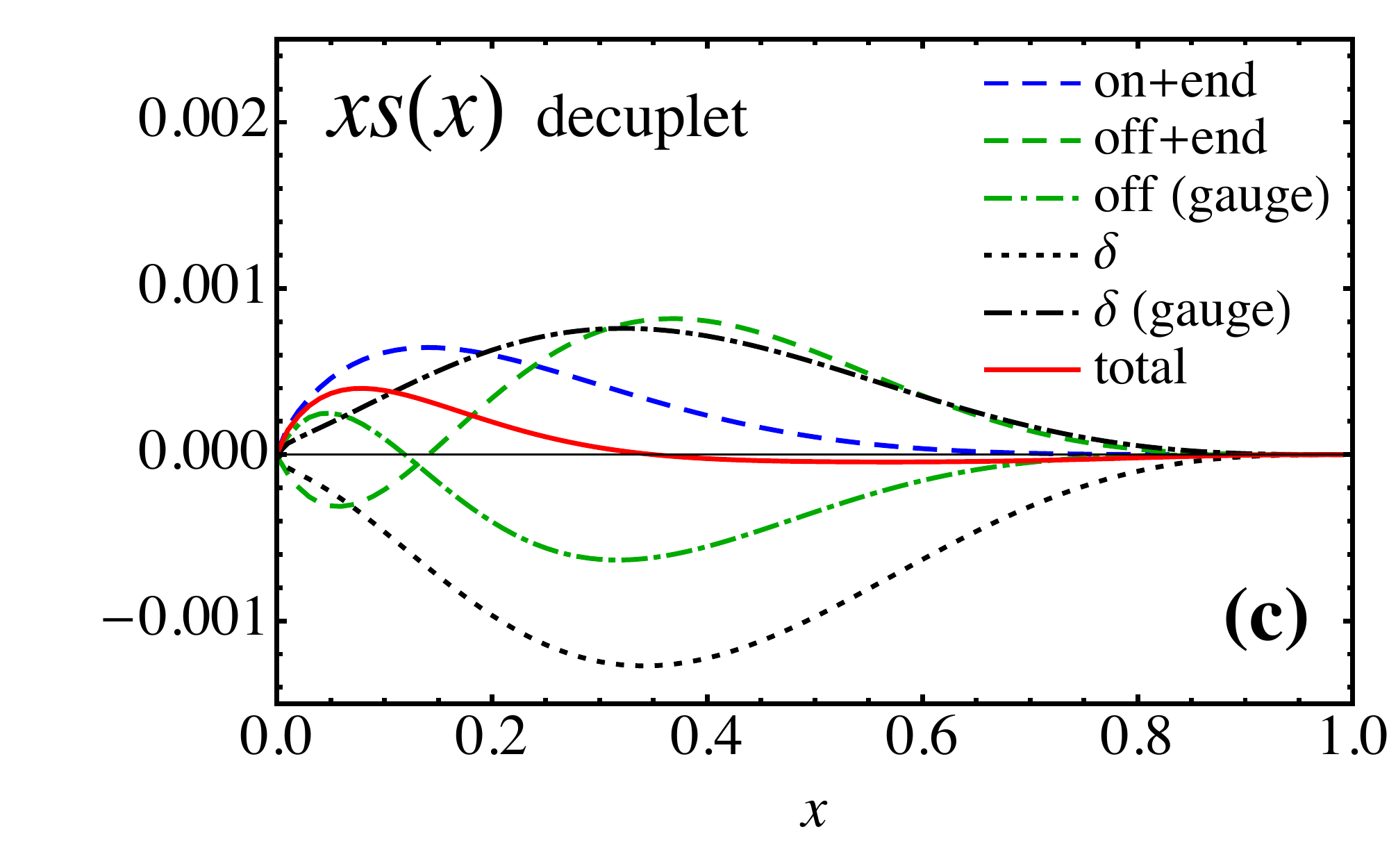}}&
\hspace{-0.65cm}{\epsfxsize=3.46in\epsfbox{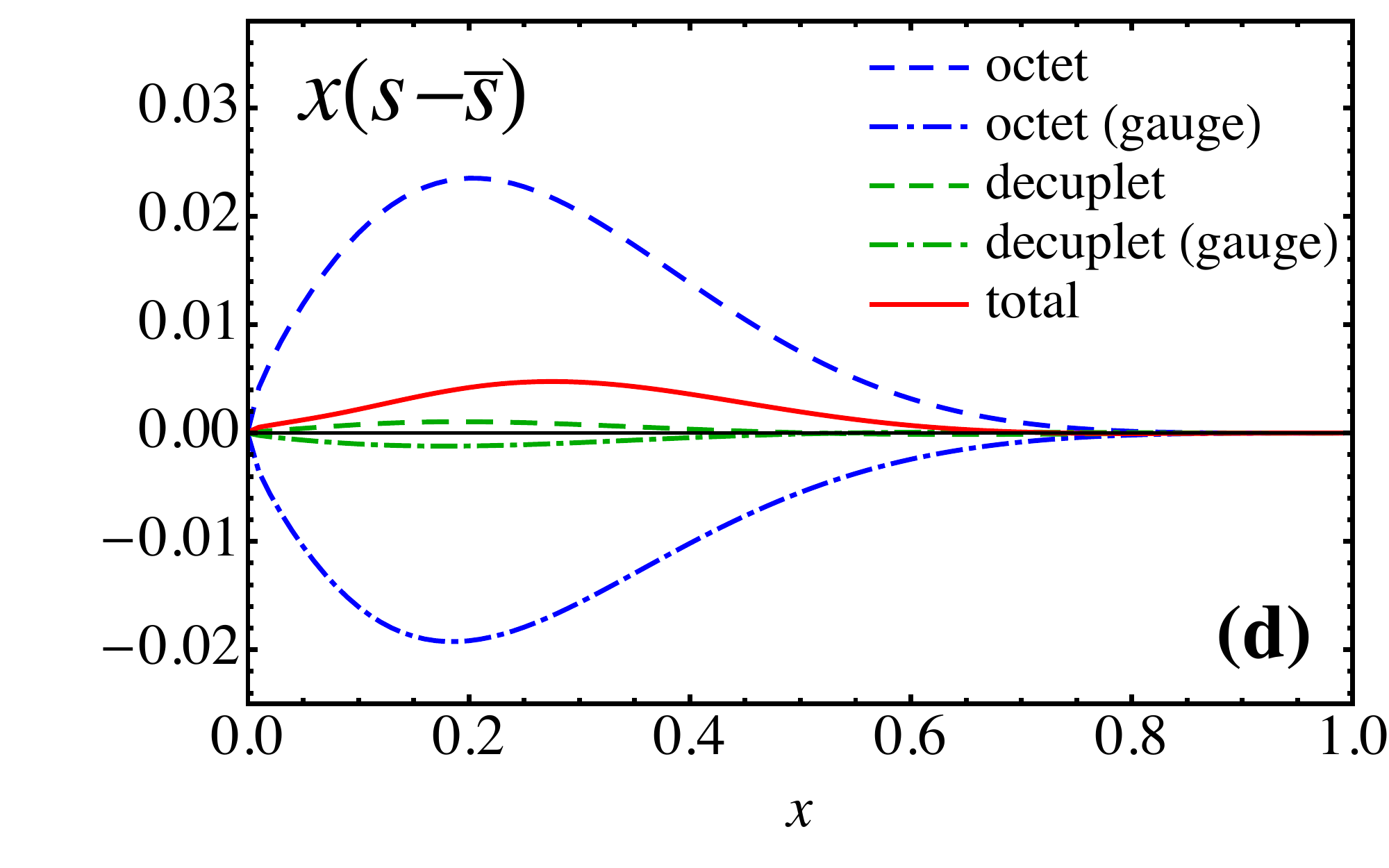}}&
\end{tabular}
\caption{Kaon loop contributions to
   {\bf (a)} antistrange PDF $x \bar s$ from the octet and decuplet
	rainbow diagrams [Fig.~\ref{fig:loop_meson}(a)-(b)];
   {\bf (b)} strange quark PDF $xs$ from the
	octet rainbow [Fig.~\ref{fig:loop_baryon}(d)],
	Kroll-Ruderman [Fig.~\ref{fig:loop_baryon}(e)-(f)],
	and tadpole [Fig.~\ref{fig:loop_baryon}(j)-(k)]	diagrams;
   {\bf (c)} strange PDF $xs$ from the
	decuplet rainbow [Fig.~\ref{fig:loop_baryon}(g)]
	and Kroll-Ruderman [Fig.~\ref{fig:loop_baryon}(h)-(i)] diagrams;
   {\bf (d)} strange asymmetry $x(s-\bar{s})$, showing the local
	and nonlocal (gauge) octet and decuplet contributions,
	along with the total asymmetry.
	The PDFs are computed with the best fit regulator parameters
	$\Lambda_{K\Lambda} = \Lambda_{K\Sigma} = 1.1$~GeV and
        $\Lambda_{K\Sigma^*} = 0.8$~GeV.}
\label{fig:xstrange}
\end{figure}

With these values of the cutoffs, we can compute the kaon loop
contributions to the strange and antistrange distributions in the
proton, and estimate the shape and magnitude of the strange asymmetry
$s-\bar s$.
In Fig.~\ref{fig:xstrange} the various octet and decuplet contributions
to $xs$ and $x\bar s$ are shown for the best fit parameters
$\Lambda_{K\Lambda} = \Lambda_{K\Sigma} = 1.1$~GeV and
$\Lambda_{K\Sigma^*} = 0.8$~GeV.
For the $x \bar s$ PDF in Fig.~\ref{fig:xstrange}(a), the octet on-shell
contribution from the rainbow diagram [Fig.~\ref{fig:loop_meson}(a)]
dominates over the decuplet on-shell and end-point terms from the
decuplet rainbow [Fig.~\ref{fig:loop_meson}(b)].
The resulting $x\bar s$ distribution peaks at $x \approx 0.1$
and essentially vanishes beyond $x \approx 0.6$.
The $\delta$-function terms from the rainbow diagrams as well as
from the kaon bubble diagram [Fig.~\ref{fig:loop_meson}(c)]
contribute to $\bar s$ only at $x=0$ and so do not appear in
Fig.~\ref{fig:xstrange}(a).

In contrast, for the strange quark distribution, from the
convolution in Eq.~(\ref{eq:strange}) one finds that all terms
from each of the rainbow, Kroll-Ruderman and tadpole diagrams
in Fig.~\ref{fig:loop_baryon}(d)--(k) have nonzero contributions
at $x > 0$.
Since there are many individual terms, we display ones
involving octet+tadpole and decuplet baryons separately
in Fig.~\ref{fig:xstrange}(b) and \ref{fig:xstrange}(c),
respectively.
Unlike the on-shell term dominance of the antistrange PDF,
for the strange distribution there are sizeable contributions
from many of the terms, with nontrivial cancellations between
them.
For the octet baryons, the on-shell and off-shell terms change
sign at around $x \approx 0.1$, with significant cancellation
occurring between the local and nonlocal (gauge link dependent)
off-shell contributions.
The (positive) local and (negative) nonlocal $\delta$-function
terms come with the largest magnitudes, but mostly cancel
among themselves, leaving a total octet contribution that
is positive and peaks around $x=0.2-0.3$, with a similar
order of magnitude as the $x\bar s$ distribution.

A qualitatively similar scenario is evident in 
Fig.~\ref{fig:xstrange}(c) for the decuplet intermediate state 
contributions to $xs$, where the individual on-shell, off-shell,
$\delta$-function and gauge link terms are shown.
(Note that the on-shell and off-shell terms include also
the respective end-point pieces.)
The predominantly positive on-shell, off-shell and nonlocal
$\delta$-function contributions at $x \gtrsim 0.2$ largely
cancel with the predominantly negative local $\delta$-function
and nonlocal off-shell terms, resulting in a very small overall
decuplet contribution to $xs$, peaking at $x \sim 0.1$, that is
an order of magnitude smaller than the octet.

Finally, the resulting asymmetry $x(s-\bar s)$ in
Fig.~\ref{fig:xstrange}(d) reflects the interplay between
the $\bar s$ PDF, which is dominant at low $x$, and the
$s$-quark PDF, which extends to larger values of $x$.
A key feature of this result is the strong cancellations
between positive local and negative nonlocal, gauge-link
dependent contributions, in both the octet and decuplet channels.
The net effect is then a small positive $x(s-\bar s)$ asymmetry,
peaking at $x \approx 0.2-0.3$, and about an order of magnitude
smaller than the asymmetry between the $\bar d$ and $\bar u$ PDFs
resulting from pion loops.

\begin{table}[p]
\begin{center}
\caption{Contributions from octet $Y=\Lambda, \Sigma^0, \Sigma^+$
	and decuplet $Y^*=\Sigma^{*0}, \Sigma^{*+}$ hyperons to the
	average number
	(in units of $10^{-2}$)
	and momentum carried
	(in units of $10^{-3}$)
	by $s$ and $\bar s$ quarks in the nucleon
	from diagrams in Figs.~\ref{fig:loop_meson}
	and \ref{fig:loop_baryon},
	for dipole regulator mass parameters
	$\Lambda_{KY} = 1.1(1)$~GeV and
        $\Lambda_{KY^*} = 0.8(1)$~GeV.
	Note that some of the numbers do not sum to the
	totals because of rounding.\\}
{\scriptsize
\begin{tabular}{llcc|llcc}	\hline
   &
   & $\langle \bar s \rangle$
   & $\langle x \bar s \rangle$
   &
   &
   & $\langle s \rangle$
   & $\langle x s \rangle$	\\
   &
   & ~$(\times 10^{-2})$~
   & ~$(\times 10^{-3})$~
   &
   &
   & ~$(\times 10^{-2})$~
   & ~$(\times 10^{-3})$~	\\ \hline
    ~$KY{\rm (rbw)}$~
  & ~$f_Y^{\rm (on)}$~
  & $1.39^{+0.69}_{-0.54}$
  & $1.33^{+0.74}_{-0.56}$
  & ~$YK{\rm (rbw)}$~
  & ~$f_Y^{\rm (on)}$~
  & $1.39^{+0.69}_{-0.54}$
  & $1.67^{+0.78}_{-0.63}$			\\
  & ~$f_K^{(\delta)}$~
  & $-(1.66^{+0.79}_{-0.63})$~~
  & 0
  &
  & ~$f_Y^{\rm (off)}$~
  & $-(4.01^{+1.68}_{-1.42})$~~
  & $-(5.35^{+2.12}_{-1.83})$~~			\\
  & ~$\delta f_K^{(\delta)}$~
  & $1.12^{+0.50}_{-0.41}$
  & 0
  &
  & ~$\delta f_Y^{\rm (off)}$~
  & $2.70^{+1.07}_{-0.92}$
  & $3.12^{+1.13}_{-1.02}$			\\
  &
  &
  &
  &
  & ~$ f_K^{(\delta)}$~
  & $1.66^{+0.79}_{-0.63}$
  & $2.82^{+1.35}_{-1.07}$			\\
  &
  &
  &
  & ~$YK{\rm (KR)}$~
  & ~$f_Y^{\rm (off)}$~
  & $4.01^{+1.68}_{-1.42}$
  & $6.29^{+2.50}_{-2.15}$			\\
  &
  &
  &
  &
  & ~$f_K^{(\delta)}$~
  & $-(3.31^{+1.58}_{-1.26})$~~
  & $-(6.66^{+3.18}_{-2.53})$~~			\\
  &
  &
  &
  & ~$YK{\rm (\delta KR)}$~
  & ~$\delta f_Y^{\rm (off)}$~
  & $-(2.70^{+1.07}_{-0.92})$~~
  & $-(3.68^{+1.33}_{-1.20})$~~			\\
  &
  &
  &
  &
  & ~$\delta f_K^{(\delta)}$~
  & $1.12^{+0.50}_{-0.41}$
  & $2.24^{+1.01}_{-0.82}$			\\ 
   ~total~octet
  &
  & $0.85^{+0.40}_{-0.32}$
  & $1.33^{+0.74}_{-0.56}$
  & ~total~octet
  &
  & $0.85^{+0.40}_{-0.32}$
  & $0.46^{+0.14}_{-0.14}$			\\ \hline
    ~$K {\rm (bub)}$~
  & ~$f_K^{(\delta)}$~
  & $4.85^{+2.32}_{-1.84}$
  & 0
  & ~$K {\rm (tad)}$~
  & ~$f_K^{(\delta)}$~
  & $4.85^{+2.32}_{-1.84}$
  & $7.87^{+3.76}_{-2.98}$               	\\
  & ~$\delta f_K^{(\delta)}$~
  & $-(3.27^{+1.47}_{-1.20})$~~
  & 0
  & ~$K {\rm (\delta tad)}$~
  & ~$\delta f_K^{(\delta)}$~
  & $-(3.27^{+1.47}_{-1.20})$~~
  & $-(5.30^{+2.38}_{-1.94})$~~			\\ 
   ~total~bubble
  &
  & $1.59^{+0.85}_{-0.64}$
  & 0
  & ~total~tadpole
  &
  & $1.59^{+0.85}_{-0.64}$
  & $2.57^{+1.38}_{-1.04}$			\\ \hline
    ~$KY^*{\rm (rbw)}$~
  & ~$f_{Y^*}^{\rm (on)}$~
  & $0.09^{+0.13}_{-0.07}$
  & $0.06^{+0.09}_{-0.04}$
  & ~$Y^*K{\rm (rbw)}$~
  & ~$f_{Y^*}^{\rm (on)}$~
  & $0.09^{+0.13}_{-0.07}$
  & $0.10^{+0.14}_{-0.08}$			\\
  & ~$f_{Y^*}^{\rm (on\,end)}$~
  & $0.04^{+0.07}_{-0.03}$
  & $0.03^{+0.06}_{-0.03}$
  &
  & ~$f_{Y^*}^{\rm (on\,end)}$~
  & $0.04^{+0.07}_{-0.03}$
  & $0.04^{+0.07}_{-0.03}$			\\
  & ~$f_{Y^*}^{(\delta)}$~
  & $-(0.01^{+0.01}_{-0.01})$~~
  & 0
  &
  & ~$f_{Y^*}^{\rm (off)}$~
  & $-(0.59^{+0.72}_{-0.42})$~~
  & $-(0.75^{+0.89}_{-0.52})$~~			\\
  & ~$f_K^{(\delta)}$~
  & $-(0.15^{+0.20}_{-0.11})$~~
  & 0
  &
  & ~$f_{Y^*}^{\rm (off\,end)}$~
  & $0.17^{+0.23}_{-0.12}$
  & $0.21^{+0.29}_{-0.15}$			\\
  & ~$\delta f_K^{(\delta)}$~
  & $0.11^{+0.14}_{-0.08}$
  & 0
  &
  & ~$\delta f_{Y^*}^{\rm (off)}$~
  & $0.34^{+0.45}_{-0.24}$
  & $0.38^{+0.47}_{-0.27}$			\\
  &
  &
  &
  &
  & ~$f_K^{(\delta)}$~
  & $0.18^{+0.24}_{-0.13}$
  & $0.26^{+0.34}_{-0.19}$			\\
  &
  &
  &
  &
  & ~$f_{Y^*}^{(\delta)}$~
  & $0.01^{+0.01}_{-0.01}$
  & $0.01^{+0.02}_{-0.01}$			\\
  &
  &
  &
  &
  & ~$\delta f_{Y^*}^{(\delta)}$~
  & $-(0.07^{+0.11}_{-0.05})$~~
  & $-(0.10^{+0.16}_{-0.07})$~~			\\
  &
  &
  &
  & ~$Y^* K{\rm (KR)}$~
  & ~$f_{Y^*}^{\rm (off)}$~
  & $0.59^{+0.72}_{-0.42}$
  & $1.02^{+1.21}_{-0.71}$			\\
  &
  &
  &
  &
  & ~$f_{Y^*}^{\rm (off\,end)}$~
  & $-(0.17^{+0.23}_{-0.12})$~~
  & $-(0.29^{+0.39}_{-0.21})$~~			\\
  &
  &
  &
  &
  & ~$f_K^{(\delta)}$~
  & $-(0.34^{+0.44}_{-0.24})$~~
  & $-(0.65^{+0.85}_{-0.47})$~~			\\
  &
  &
  &
  &
  & ~$f_{Y^*}^{(\delta)}$~
  & $-(0.02^{+0.03}_{-0.01})$~~
  & $-(0.03^{+0.05}_{-0.02})$~~			\\
  &
  &
  &
  &
  & ~$\delta f_{Y^*}^{(\delta)}$~
  & $0.05^{+0.08}_{-0.03}$
  & $0.09^{+0.15}_{-0.07}$			\\
  &
  &
  &
  & ~$Y^* K{\rm (\delta KR)}$~
  & ~$\delta f_{Y^*}^{\rm (off)}$~
  & $-(0.34^{+0.45}_{-0.24})$~~
  & $-(0.51^{+0.63}_{-0.36})$~~			\\
  &
  &
  &
  &
  & ~$\delta f_K^{(\delta)}$~
  & $0.11^{+0.14}_{-0.08}$
  & $0.22^{+0.27}_{-0.15}$			\\
  &
  &
  &
  &
  & ~$\delta f_{Y^*}^{(\delta)}$~
  & $0.02^{+0.04}_{-0.02}$
  & $0.05^{+0.07}_{-0.03}$			\\ 
   ~total~decuplet
  &
  & $0.08^{+0.12}_{-0.06}$
  & $0.09^{+0.15}_{-0.07}$
  & ~total~decuplet
  &
  & $0.08^{+0.12}_{-0.06}$
  & $0.04^{+0.04}_{-0.03}$			\\ \hline
   ~{\bf total}
  &
  & \bm{$2.51^{+1.36}_{-1.02}$}
  & \bm{$1.42^{+0.89}_{-0.62}$}
  &~{\bf total}
  &
  & \bm{$2.51^{+1.36}_{-1.02}$}
  & \bm{$3.08^{+1.55}_{-1.20}$}			\\ \hline\hline
   ~non $\delta$-function~
  &
  & $1.52^{+0.88}_{-0.64}$
  & $1.42^{+0.89}_{-0.62}$
  &~non $\delta$-function~
  &
  & $1.52^{+0.88}_{-0.64}$
  & $2.25^{+1.21}_{-0.92}$			\\
%
   ~$\delta$-function~
  &
  & $0.99^{+0.60}_{-0.50}$
  & 0
  &~$\delta$-function~
  &
  & $0.99^{+0.60}_{-0.50}$
  & $0.82^{+0.67}_{-0.63}$			\\ \hline
   ~local
  &
  & $4.55^{+2.23}_{-1.77}$
  & $1.42^{+0.89}_{-0.62}$
  &~local
  &
  & $4.55^{+2.23}_{-1.77}$
  & $6.58^{+3.14}_{-2.60}$			\\
   ~nonlocal
  &
  & $-(2.04^{+1.04}_{-0.92})$~~
  & 0
  &~nonlocal
  &
  & $-(2.04^{+1.04}_{-0.92})$~~
  & $-(3.50^{+1.66}_{-1.46})$~~			\\ \hline 
\end{tabular}
}
\label{tab:ssbar}
\end{center}
\end{table}

In addition to the shape, it is instructive also to examine the
contributions of the various terms to the lowest moments of the
$s$ and $\bar s$ PDFs, in particular, the average number of
strange and antistrange quarks,
\begin{equation}
\langle s \rangle = \int_0^1 dx\, s(x), \ \ \ \
\langle \bar s \rangle = \int_0^1 dx\, \bar s(x),
\label{eq:S0}
\end{equation}
and the average momentum carried by them,
\begin{equation}
\langle x s \rangle = \int_0^1 dx\, x s(x), \ \ \ \
\langle x \bar s \rangle = \int_0^1 dx\, x \bar s(x).
\label{eq:S1}
\end{equation}
The conservation of strangeness of course requires equal numbers
of $s$ and $\bar s$ quarks in the nucleon,
	$\langle s \rangle = \langle \bar s \rangle$,
as a direct consequence of local gauge invariance, although the
shapes of the $s$ and $\bar s$ distributions themselves are
obviously rather different.
The zero net strangeness can be verified by explicitly summing the
contributions to $\langle s \rangle$ and $\langle \bar s \rangle$
from the various diagrams in Figs.~\ref{fig:loop_meson} and
\ref{fig:loop_baryon}, as Table~\ref{tab:ssbar} indicates.
Note also that the conservation of strangeness holds for the
octet and decuplet contributions individually, as well as for
the tadpole and bubble diagrams,
\begin{equation}
\langle s \rangle_{\rm oct} = \langle \bar s \rangle_{\rm oct},\ \ \
\langle s \rangle_{\rm dec} = \langle \bar s \rangle_{\rm dec},\ \ \
\langle s \rangle_{\rm tad} = \langle \bar s \rangle_{\rm bub}.
\label{eq:S0indiv}
\end{equation}

Although the contribution to the total strange and antiquark quark
number from the decuplet intermediate states is about an order of
magnitude smaller than that from octet intermediate states,
the role of the kaon bubble and tadpole terms is more significant,
making up $\approx 60\%$ of the total.
For the antistrange moment, $\langle \bar s \rangle$, including
the $\delta$-functions contributions from the rainbow diagrams,
some 40\% of the total moment comes from $x=0$.
For the strange $\langle s \rangle$ moment, on the other hand,
the structure of the convolution in Eq.~(\ref{eq:strange}) means
that all of the contributions to $s(x)$ are at $x > 0$, including
ones involving $\delta$-function splitting functions.
Interestingly, significant cancellation occurs between the
local terms and the gauge link-dependent nonlocal contributions,
which turn out to be negative and about half as large in magnitude
as the local.

While the lowest moments of the $s$ and $\bar s$ are constrained
to be equal, there is no such requirement for higher moments,
including the $x$-weighted moment corresponding to the momentum
carried by $s$ and $\bar s$ quarks.
Since the total $s-\bar s$ asymmetry is found to be mostly positive
over the range of $x$ relevant in this analysis, not surprisingly
the total $\langle x (s-\bar s) \rangle$ moment is also positive.
Including the uncertainties on the kaon-nucleon-hyperon vertex
regulator parameters from Fig.~\ref{fig:cross2}, the combined
asymmetry in our analysis is
\begin{equation}
\langle x (s - \bar s) \rangle
= (1.66^{\, +\, 0.81}_{\, -\, 0.74})\, \times 10^{-3}.
\label{eq:xs-sbar}
\end{equation}

\begin{figure}[t]
\centering
\vspace*{-0.5cm}
\begin{tabular}{ccccc}
{\epsfxsize=3.5in\epsfbox{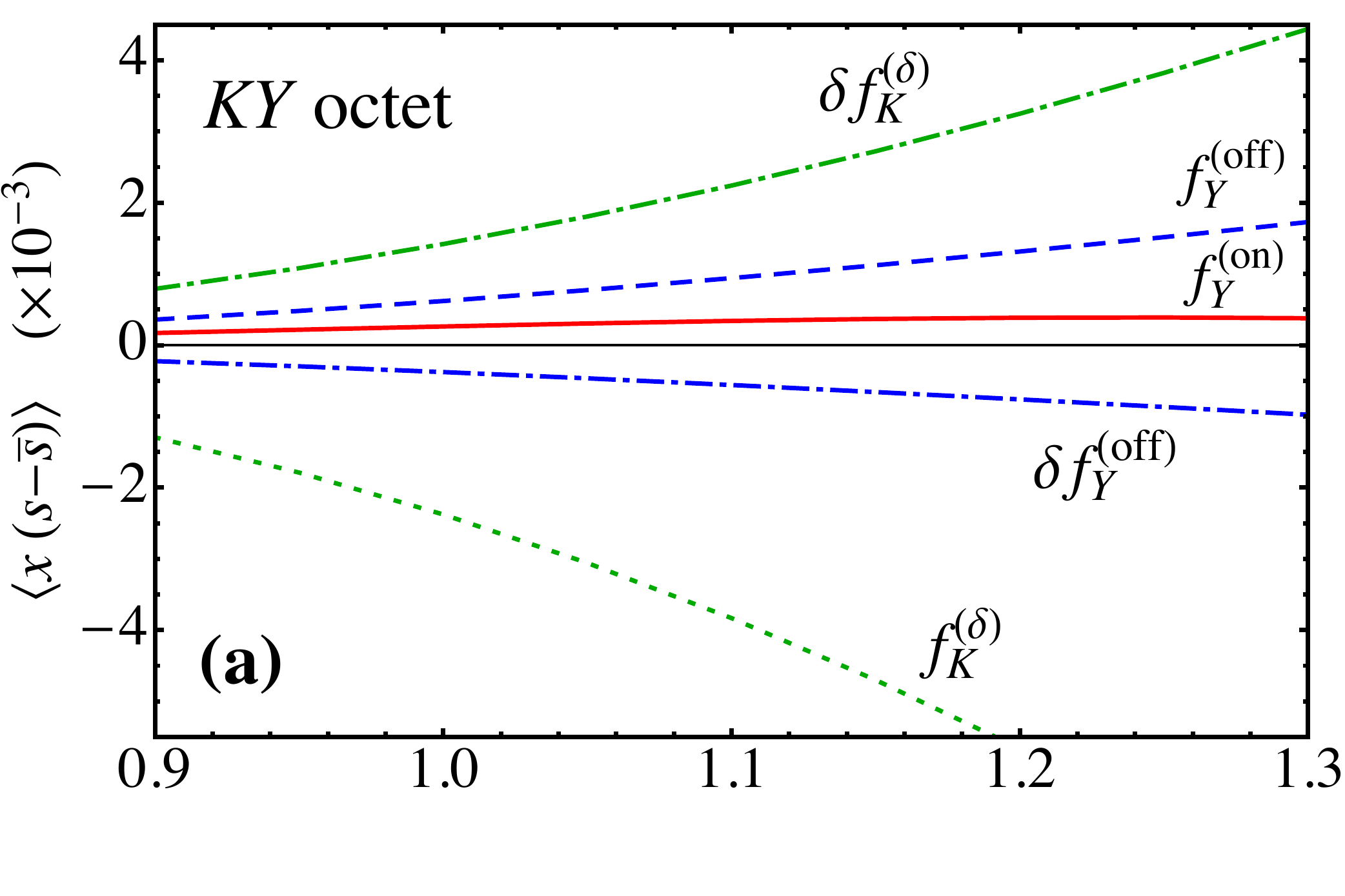} \vspace*{-1cm}}&\\
\hspace{-0.3cm}{\epsfxsize=3.54in\epsfbox{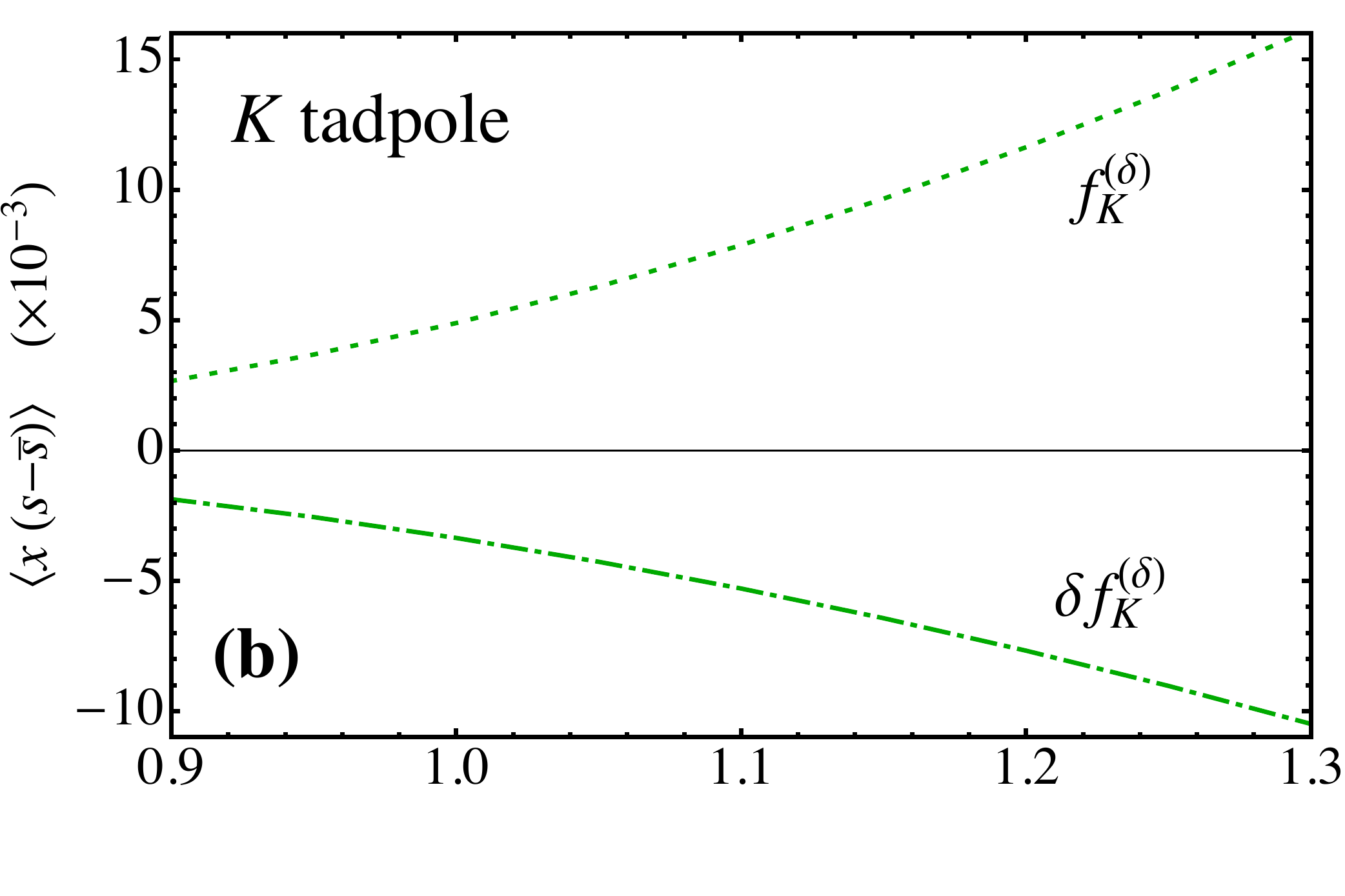}\vspace*{-1cm}}&\\
\hspace{-0.55cm}{\epsfxsize=3.495in\epsfbox{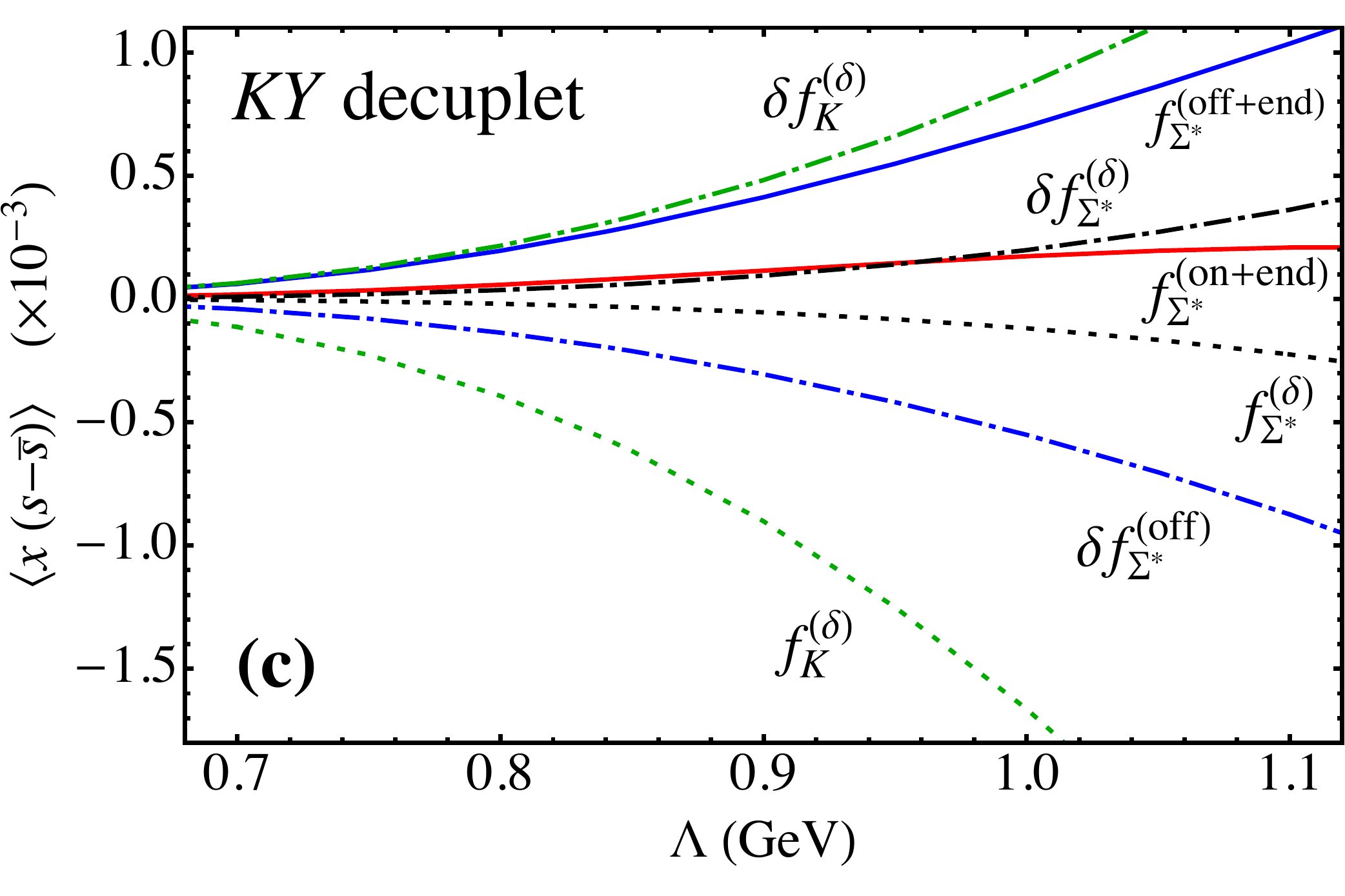}\vspace*{0.2cm}}&
\end{tabular}
\caption{Contributions to the $\langle x (s-\bar s) \rangle$
	moment versus the dipole cutoff parameter $\Lambda$
	($=\Lambda_{KY}$, for $Y=\Lambda$ or $\Sigma$ hyperons,
        or $\Lambda_{K\Sigma^*}$) from
   {\bf (a)}
	the $KY$ octet rainbow
	[Figs.~\ref{fig:loop_meson}(a) and \ref{fig:loop_baryon}(d)]
	and Kroll-Ruderman [Fig.~\ref{fig:loop_baryon}(e)-(f)] diagrams;
   {\bf (b)}
	the $K$ bubble [Fig.~\ref{fig:loop_meson}(c)] and tadpole
	[Fig.~\ref{fig:loop_baryon}(j) and (k)] diagrams;
   {\bf (c)}
	the $K\Sigma^*$ decuplet rainbow
	[Figs.~\ref{fig:loop_meson}(b) and \ref{fig:loop_baryon}(g)]
	and Kroll-Ruderman [Fig.~\ref{fig:loop_baryon}(h)-(i)]
	diagrams.}
\label{fig:xs-sbar}
\end{figure}

It is instructive, however, to observe the origin of the asymmetry
in our chiral effective theory formulation.
As mentioned above, there are no contributions to the momentum
carried by $\bar s$ quarks from any of the $\delta$-function terms
from the rainbow or kaon bubble diagrams, so that only the on-shell
and on-shell end-point terms are nonzero.
In contrast, all terms, including $\delta$-function, contribute to
the momentum carried by $s$ quarks.
The result is a relatively small asymmetry that survives the
cancellation of the (positive) on-shell $s$ and $\bar s$ terms,
with large contributions from individual off-shell and
$\delta$-function terms.
As illustrated in Fig.~(\ref{fig:xs-sbar}) for the various
contributions to the strange momentum asymmetry versus the
regulator cutoff mass, the largest of these in magnitude is the
(negative) $f_K^{(\delta)}$ term from the Kroll-Ruderman diagram
[Fig.~\ref{fig:loop_baryon}(e)], with comparably large (positive)
gauge link contributions $\delta f_K^{(\delta)}$ from the rainbow
[Fig.~\ref{fig:loop_baryon}(d)] and nonlocal Kroll-Ruderman
[Fig.~\ref{fig:loop_baryon}(f)] diagrams.
After the cancellations of various terms, the octet baryon
contribution to the strange momentum asymmetry is actually
negative,
  $\langle x(s-\bar s) \rangle_{\rm oct} \approx -0.87 \times 10^{-3}$.
Terms involving decuplet hyperon states give relatively small
absolute contributions, with significant cancellations that lead
to a negligible overall strange decuplet asymmetry,
  $\langle x(s-\bar s) \rangle_{\rm dec} \approx -0.05 \times 10^{-3}$.

Interestingly, the most significant role played here is by the
kaon tadpole terms [Fig.~\ref{fig:loop_baryon}(j)-(k)].
With strong cancellations between the positive local $f_K^{(\delta)}$
and negative nonlocal $\delta f_K^{(\delta)}$ terms, the total
asymmetry from the tadpole,
  $\langle x(s-\bar s) \rangle_{\rm tad} \approx 2.57 \times 10^{-3}$,
is still about 3 times larger in magnitude than that from the rainbow
and Kroll-Ruderman diagrams.  The result is an overall asymmetry in
Eq.~(\ref{eq:xs-sbar}) that is positive.

Experimentally, identifying an asymmetry of this size will be
challenging, but not impossible.
Traditionally, inclusive charm meson production in charged current
neutrino and antineutrino DIS from nuclei has been used to provide
information about the $s$ and $\bar s$ PDFs in the nucleon, and
analyses of data from neutrino experiments at BEBC~\cite{BEBC94},
CDHS~\cite{CDHS84}, CDHSW~\cite{CDHSW91}, CCFR~\cite{CCFR95} and
NuTeV~\cite{NuTeV03, NuTeV07} have yielded values in the range
  $\langle x (s - \bar s) \rangle \sim (0-3)\, \times 10^{-3}$
\cite{Barone00, Davidson02, Bentz10}.
Unfortunately, the neutrino--nucleus data are known to be affected by
uncertainties in nuclear medium effects when relating nuclear structure
functions to those of free nucleons~\cite{Kalantarians:2017mkj},
and in the nuclear dependence of charm quark energy loss and
$D$-meson interactions during hadronization in the nuclear medium
\cite{Accardi09, Majumder11}.

Alternatively, the $s$ and $\bar s$ distributions can be constrained
by $K^\pm$ meson production data from semi-inclusive deep-inelastic
scattering (SIDIS) off protons and deuterons, such as from the
HERMES~\cite{HERMES08, HERMES14} or COMPASS~\cite{COMPASS17} experiments.
In a first of its kind global analysis, the JAM Collaboration recently
fitted both the SIDIS and inclusive DIS data, along with other high
energy scattering data, within a Bayesian likelihood analysis using
Monte Carlo techniques to simultaneously determine both the
spin-averaged PDFs and parton-to-hadron fragmentation functions
\cite{JAM19}.
The analysis found a suppressed strange content in the nucleon at
large $x$, and found no clear evidence for a nonzero $s-\bar s$
asymmetry within relatively large uncertainties.
In future, high-precision SIDIS data from the Jefferson Lab 12~GeV
program or from the planned Electron-Ion Collider should provide
better constraints on the $s$ and $\bar s$ PDFs, as may $W\, +$ charm
production data from $pp$ collisions at the LHC~\cite{ATLAS_Wc14,
CMS_Wc14, CSKK18}.

An important consequence of a better determination of the $s-\bar s$
asymmetry in the nucleon is more robust constraints on the weak mixing
angle $\sin^2\theta_W$ extracted from the NuTeV data on $\nu$ and
$\bar \nu$ nuclear cross sections~\cite{NuTeV02}.
For the total strange asymmetries range found in this analysis,
	$0.9 \times 10^{-3}
	\lesssim \langle x(s-\bar s) \rangle
	\lesssim 2.5 \times 10^{-3}$,
the resulting correction to the weak angle lies in the range
	$-0.9 \times10^{-3}
	\lesssim \Delta\sin\theta^2_W
	\lesssim 2.4 \times10^{-3}$,
or between 18\% and 49\% of the total quoted discrepancy
\cite{NuTeV02, NuTeV03} (see Refs.~\cite{Londergan05}
for a review and further discussion).

\newpage
\section{Conclusion}
\label{sec.conclusion}

In this paper we have calculated the contributions to the sea quark
distributions in the proton which are generated within a nonlocal
chiral effective field theory. Both octet and decuplet intermediate
states were included in the one-loop calculation using a 4-dimensional
dipole regulator to deal with the ultraviolet divergences.
This regulator was introduced explicitly in the nonlocal Lagrangian
density, with gauge invariance ensured through the presence of
gauge links.  A consequence of the introduction of the regulator
are additional diagrams [Figs.~\ref{fig:loop_baryon}(f),
\ref{fig:loop_baryon}(i) and~\ref{fig:loop_baryon}(k)] that arise
from the expansion of the gauge links to lowest order in the
electromagnetic coupling.

The free parameters entering the calculation, namely the dipole
regulator masses, have been determined by fitting the available
inclusive differential $pp \to n X$, $pp \to \Delta X$,
$pp \to \Lambda X$ and $pp \to \Sigma^* X$ cross section data.
Using the fitted values of the dipole masses,
\mbox{$\Lambda = \{ 1.0(1), 0.9(1), 1.1(1), 0.8(1) \}$~GeV}
for the
\mbox{$\{ \pi N, \pi \Delta, K \Lambda, K \Sigma^* \}$}
states, respectively, we computed the $x$ dependence of the sea quark
asymmetry $\bar{d}-\bar{u}$, which is dominated at $x>0$ by the
on-shell contribution involving a nucleon intermediate state.
The general shape and magnitude of the asymmetry extracted from the
E866 Drell-Yan data~\cite{Towell:2001nh} are described quite well,
with the exception of the apparent change sign at higher $x$ values,
which is practically impossible to accommodate within the current
theoretical framework.  On the other hand, preliminary data from
the SeaQuest experiment at Fermilab~\cite{SeaQuest} suggest that
the extracted $\bar d/\bar u$ ratio may flatten out at large $x$
values and remain above unity.
The integrated $\bar{d}-\bar{u}$ asymmetry was found to lie in the
range between $\langle \bar{d}-\bar{u} \rangle \approx 0.09$ and 0.17,
which encompasses the values extracted by the
New Muon~\cite{Arneodo:1994sh} and E866~\cite{Towell:2001nh}
Collaborations of $\approx 0.15$ and 0.12, respectively.
Remarkably, some 30\% of our calculated value is associated with
a $\delta$-function contribution at $x=0$, which is not accessible
experimentally at finite energy.

For the strange distributions in the proton, both the $s$ and
$\bar{s}$ PDFs were found to be positive at all values of $x>0$.
Interestingly in this case, while the $\bar s$ distribution receives
$\delta$-function contributions also at $x=0$ (around 2/3 of the total),
the $s$ PDF vanishes at $x=0$; both integrate to the same value,
however, to ensure zero total strangeness,
	$\langle s \rangle = \langle \bar s \rangle$.
Again, the contributions from the octet baryon intermediate states
are dominant, with decuplet baryon contributions about an order
of magnitude smaller.
The gauge link dependent terms play a significant role in the
nonlocal formulation of the chiral theory, contributing about half
of the total $\langle s \rangle$ and $\langle \bar s \rangle$,
but of opposite sign.

Large cancellations also appear in the $x$-weighted asymmetry
$x(s-\bar{s})$, which remains small but positive across all $x$,
with the integrated value lying in the range 
        $0.9 \times 10^{-3}
	 \lesssim \langle x(s-\bar{s}) \rangle \lesssim
	 2.5 \times 10^{-3}$.
This is broadly consistent with previous determinations from neutrino
scattering experiments~\cite{CCFR95, NuTeV07}, although the uncertainties
on the empirical bounds are rather large.
A nonzero moment $\langle x(s-\bar{s}) \rangle$ leads to a
correction~\cite{Bentz10} to the NuTeV extraction of
$\sin^2\theta_W$~\cite{NuTeV02}.
Our result supports the idea that the strange--antistrange quark
asymmetry may indeed reduce the NuTeV anomaly by up to one standard
deviation, which, along with other corrections such as charge symmetry
breaking in the nucleon sea~\cite{Rodionov:1994cg, Londergan:2003ij,
Londergan:2009kj} and the isovector EMC effect~\cite{Cloet:2009qs},
may account for the apparent anomaly entirely in terms of Standard
Model physics.

Future progress on constraining the $s-\bar s$ asymmetry experimentally
is expected to come on several fronts.  Parity-violating inclusive DIS
and semi-inclusive kaon electroproduction from hydrogen at Jefferson Lab
and at a future Electron-Ion Collider will provide independent
combinations of flavor PDFs, with the $s$ and $\bar s$ distributions
weighted by different electroweak charges and fragmentation functions,
respectively.  At higher energies, data on inclusive $W\, +$ charm
production in $pp$ collisions at the LHC~\cite{ATLAS_Wc14, CMS_Wc14}
can also provide sensitivity to differences between the $s$ and $\bar s$
PDFs at small values of $x$, complementing the constraints at higher $x$
values from fixed target experiments.

\section*{Acknowledgments }

This work is supported by NSFC under Grant No.~11475186,
the Sino-German CRC 110 ``Symmetries and the Emergence of
Structure in QCD'' project by NSFC under the grant No.~11621131001,
the DOE Contract No.~DE-AC05-06OR23177, under which Jefferson
Science Associates, LLC operates Jefferson Lab, DOE Contract
No.~DE-FG02-03ER41260, and by the Australian Research Council
through the ARC Centre of Excellence for Particle Physics at
the Terascale (CE110001104) and Discovery Projects DP151103101
and DP180100497.

\newpage


\begin{thebibliography}{99}

\bibitem{Arneodo:1994sh}
  M.~Arneodo {\it et al.}, 
  Phys. Rev. D {\bf 50}, R1 (1994).

\bibitem{Ackerstaff:1998sr}
  K.~Ackerstaff {\it et al.}, 
  Phys. Rev. Lett. {\bf 81}, 5519 (1998).

\bibitem{Baldit}
  A.~Baldit {\it et al.}, 
  Phys. Lett. B {\bf 332}, 244 (1994).

\bibitem{Towell:2001nh}
  R.~S.~Towell {\it et al.}, 
  Phys. Rev. D {\bf 64}, 052002 (2001).

\bibitem{Thomas:1983fh}
  A.~W.~Thomas,
  Phys. Lett. {\bf 126} B, 97 (1983).

\bibitem{Thomas:1982kv}
  A.~W.~Thomas,
  Adv. Nucl. Phys. {\bf 13}, 1 (1984).

\bibitem{Thomas:1981vc}
  A.~W.~Thomas, S.~Theberge and G.~A.~Miller,
  Phys. Rev. D {\bf 24}, 216 (1981).

\bibitem{Geesaman:2018ixo}
  D.~F.~Geesaman and P.~E.~Reimer,
  arXiv:1812.10372 [nucl-ex].

\bibitem{Ashman}
  J.~Ashman {\it et al.},
  Nucl. Phys. {\bf B328}, 1 (1989).

\bibitem{Ellis}
  J.~R.~Ellis and R.~L.~Jaffe,
  Phys. Rev. D {\bf 9}, 1444 (1974).

\bibitem{Catani04}
  S.~Catani, D.~de Florian, G.~Rodrigo and W.~Vogelsang,
  Phys. Rev. Lett. {\bf 93}, 152003 (2004).

\bibitem{Signal87}
  A.~I.~Signal and A.~W.~Thomas,
  Phys. Lett. B {\bf 191}, 205 (1987).

\bibitem{Malheiro97}
  W.~Melnitchouk and M.~Malheiro,
  Phys. Rev. C {\bf 55}, 431 (1997).

\bibitem{Sufian18}
  R.~S.~Sufian {\it et al.},
  Phys. Rev. D {\bf 98}, 114004 (2018).

\bibitem{nonlocal-I}
  Y.~Salamu, C.-R.~Ji, W.~Melnitchouk, A.~W.~Thomas and P.~Wang,
  Phys. Rev. D {\bf 99}, 014041 (2019).

\bibitem{He1}
  F.~He and P.~Wang,
  Phys. Rev. D {\bf 97}, 036007 (2018).

\bibitem{He2}
  F.~He and P.~Wang,
  Phys. Rev. D {\bf 98}, 036007 (2018).

\bibitem{Chen02}
  J.-W.~Chen and X.~Ji,
  Phys. Rev. Lett. {\bf 87}, 152002 (2001);
  {\bf 88}, 249901(E) (2002).

\bibitem{XGWangPRD}
  X.~G.~Wang, C.-R.~Ji, W.~Melnitchouk, Y.~Salamu, A.~W.~Thomas and P.~Wang,
  Phys. Rev. D {\bf 94}, 094035 (2016).

\bibitem{XGWangPLB}
  X.~G.~Wang, C.~R.~Ji, W.~Melnitchouk, Y.~Salamu, A.~W.~Thomas and P.~Wang,
  Phys. Lett. B {\bf 762}, 52 (2016).

\bibitem{Ji:2013bca}
  C.~R.~Ji, W.~Melnitchouk and A.~W.~Thomas,
  Phys. Rev. D {\bf 88}, 076005 (2013).

\bibitem{Shanahan:2013xw}
  P.~E.~Shanahan, A.~W.~Thomas and R.~D.~Young,
  Phys. Rev. D {\bf 87}, 114515 (2013).

\bibitem{Hacker:2005fh}
  C.~Hacker, N.~Wies, J.~Gegelia and S.~Scherer,
  Phys. Rev. C {\bf 72}, 055203 (2005).

\bibitem{Nath:1971wp}
  L.~M.~Nath, B.~Etemadi and J.~D.~Kimel,
  Phys. Rev. D {\bf 3}, 2153 (1971).

\bibitem{Moiseeva13}
  A.~M.~Moiseeva and A.~A.~Vladimirov,
  Eur. Phys. J. A {\bf 49}, 23 (2013).

\bibitem{Labrenz:1996jy}
  J.~N.~Labrenz and S.~R.~Sharpe,
  Phys. Rev. D {\bf 54}, 4595 (1996).

\bibitem{XDWang_spin}
  X.~G.~Wang {\it et al.}
  (to be published).

\bibitem{Jenkins:1991ts}
  E.~E.~Jenkins,
  Nucl. Phys. {\bf B368}, 190 (1992).

\bibitem{Burkardt13}
  M.~Burkardt, K.~S.~Hendricks, C.-R.~Ji, W.~Melnitchouk and A.~W.~Thomas,
  Phys. Rev. D {\bf 87}, 056009 (2013).

\bibitem{Salamu15}
  Y.~Salamu, C.-R.~Ji, W.~Melnitchouk and P.~Wang,
  Phys. Rev. Lett. {\bf 114}, 122001 (2015).

\bibitem{Holtmann96}
  H.~Holtmann, A.~Szczurek and J.~Speth,
  Nucl. Phys. {\bf A596}, 631 (1996).

\bibitem{MST98}
  W.~Melnitchouk, J.~Speth and A.~W.~Thomas,
  Phys. Rev. D {\bf 59}, 014033 (1998).

\bibitem{Flauger:1976ju}
  W.~Flauger and F.~Monnig,
  Nucl. Phys. {\bf B109}, 347 (1976).

\bibitem{Blobel:1978yj}
  V.~Blobel {\it et al.},
  Nucl. Phys. {\bf B135}, 379 (1978).

\bibitem{Barish:1975uv}
  S.~J.~Barish {\it et al.},
  Phys. Rev. D {\bf 12}, 1260 (1975).

\bibitem{Jaeger:1974pk}
  K.~Jaeger {\it et al.},
  Phys. Rev. D {\bf 11}, 1756 (1975).

\bibitem{Bockmann:1977sc}
  K.~Bockmann {\it et al.},
  Nucl. Phys. {\bf B143}, 395 (1978).

\bibitem{Aicher:2010cb}
  M.~Aicher, A.~Sch\"afer and W.~Vogelsang,
  Phys. Rev. Lett. {\bf 105}, 252003 (2010).

\bibitem{Martin:1998sq}
  A.~D.~Martin, R.~G.~Roberts, W.~J.~Stirling and R.~S.~Thorne,
  Eur. Phys. J. C {\bf 4}, 463 (1998).

\bibitem{Leader:2010rb}
  E.~Leader, A.~V.~Sidorov and D.~B.~Stamenov,
  Phys. Rev. D {\bf 82}, 114018 (2010).

\bibitem{Barry:2018ort}
  P.~C.~Barry, N.~Sato, W.~Melnitchouk and C.~R.~Ji,
  Phys. Rev. Lett. {\bf 121}, 152001 (2018).

\bibitem{Owens84}
  J.~F.~Owens,
  Phys. Rev. D {\bf 30}, 943 (1984).

\bibitem{SMRS92}
  P.~J.~Sutton, A.~D.~Martin, R.~G.~Roberts and W.~J.~Stirling,
  Phys. Rev. D {\bf 45}, 2349 (1992).

\bibitem{GRV92}
  M.~Gl\"uck, E.~Reya and A.~Vogt,
  Z. Phys. C {\bf 53}, 651 (1992).

\bibitem{Wijesooriya05}
  K.~Wijesooriya, P.~E.~Reimer and R.~J.~Holt,
  Phys. Rev. C {\bf 72}, 065203 (2005).

\bibitem{JMO13}
  P.~Jimenez-Delgado, W.~Melnitchouk and J.~F.~Owens,
  J. Phys. G: Nucl. Part. Phys. {\bf 40}, 093102 (2013).

\bibitem{ForteWatt13}
  S.~Forte and G.~Watt,
  Ann. Rev. of Nucl. Part. Sci. {\bf 63}, (2013).

\bibitem{Hufner:1992cu}
  J.~Hufner and B.~Povh,
  Phys. Rev. D {\bf 46}, 990 (1992).

\bibitem{McKenney16}
  J.~R.~McKenney, N.~Sato, W.~Melnitchouk and C.R.~Ji,
  Phys. Rev. D {\bf 93}, 054011 (2016).

\bibitem{MT93}
  W.~Melnitchouk and A.~W.~Thomas,
  Phys. Rev. D {\bf 47}, 3794 (1993).

\bibitem{Kopeliovich12}
  B.~Z.~Kopeliovich, I.~K.~Potashnikova, B.~Povh and I.~Schmidt,
  Phys. Rev. D {\bf 85}, 114025 (2012).

\bibitem{D'Alesio00}
  U.~D'Alesio and H.~J.~Pirner,
  Eur. Phys. J. A {\bf 7}, 109 (2000).

\bibitem{H1_01}
  C.~Adloff {\it et al.},
  Nucl. Phys. {\bf B619}, 3 (2001).

\bibitem{Carroll:1978vq}
  A.~S.~Carroll {\it et al.},
  Phys. Lett. {\bf 80} B, 423 (1979).

\bibitem{Barone:2006xj}
  V.~Barone {\it et~al.},
  JHEP {\bf 01} (2006) 006.

\bibitem{BEBC94}
  G.~T.~Jones {\it et al.},
  Z. Phys. C {\bf 62}, 575 (1994).

\bibitem{CDHS84}
  H.~Abramowicz {\it et al.},
  Z. Phys. C {\bf 25}, 29 (1984).

\bibitem{CDHSW91}
  P.~Berge {\it et al.},
  Z. Phys. C {\bf 49}, 187 (1991).

\bibitem{CCFR95}
  A.~O.~Bazarko {\it et al.},
  Z. Phys. C {\bf 65}, 189 (1995).

\bibitem{NuTeV03}
  G.~P.~Zeller {\it et al.},
  Phys. Rev. D {\bf 65}, 111103(R) (2002);
  Erratum: {\it ibid} {\bf 67}, 119902 (2003).

\bibitem{NuTeV07}
  D.~Mason {\it et al.},
  Phys. Rev. Lett. {\bf 99}, 192001 (2007).

\bibitem{Barone00}
  V.~Barone, C.~Pascaud and F.~Zomer,
  Eur. Phys. J. C {\bf 12}, 243 (2000).

\bibitem{Davidson02}
  S.~Davidson, S.~Forte, P.~Gambino, N.~Rius and A.~Strumia,
  JHEP {\bf 0202}, 037 (2002).

\bibitem{Bentz10}
  W.~Bentz, I.~C.~Clo\"{e}t, J.~T.~Londergan and A.~W.~Thomas,
  Phys. Lett. B {\bf 693}, 462 (2010).

\bibitem{Kalantarians:2017mkj}
  N.~Kalantarians, C.~Keppel and M.~E.~Christy,
  Phys. Rev. C {\bf 96}, 032201 (2017).

\bibitem{Accardi09}
  A.~Accardi, F.~Arleo, W.~K.~Brooks, D.~D'Enterria and V.~Muccifora,
  Riv. Nuovo Cim. {\bf 32}, 439 (2010).

\bibitem{Majumder11}
  A.~Majumder and M.~Van~Leeuwen,
  Prog. Part. Nucl. Phys. A {\bf 66}, 41 (2011).

\bibitem{HERMES08}
  A.~Airapetian {\it et al.},
  Phys. Lett. B {\bf 666}, 446 (2008).

\bibitem{HERMES14}
  A.~Airapetian {\it et al.},
  Phys. Rev. D {\bf 89}, 097101 (2014).

\bibitem{COMPASS17}
  C.~Adolph {\it et al.}, 
  Phys. Lett. B {\bf 767}, 133 (2017).

\bibitem{JAM19}
  N.~Sato, C.~Andres, J.J.~Ethier and W.~Melnitchouk,
  arXiv:1905.03788 [hep-ph].

\bibitem{ATLAS_Wc14}
  G.~Aad {\it et al.},
  JHEP {\bf 05} (2014) 068.

\bibitem{CMS_Wc14}
  S.~Chatrchyan {\it et al.},
  JHEP {\bf 02} (2014) 013.

\bibitem{CSKK18}
  A.~M.~Cooper-Sarkar and K.~Wichmann,
  Phys. Rev. D {\bf 98}, 014027 (2018).

\bibitem{NuTeV02}
  G.~P.~Zeller {\it et al.},
  Phys. Rev. Lett. {\bf 88}, 091802 (2002);
  Erratum: {\it ibid} {\bf 90}, 239902 (2003).

\bibitem{Londergan05}
  J.~T.~Londergan,
  Nucl. Phys. Proc. Suppl. {\bf 141}, 68 (2005).

\bibitem{SeaQuest}
  A.~Tadepalli, {\it ``Recent progress on studies of light quark flavor
  asymmetry at SeaQuest experiment''}, talk presented at APS April Meeting,
  Denver, Colorado (April 16, 2019),
  {\tt http://meetings.aps.org/Meeting/APR19/Session/X09.7}.

\bibitem{Rodionov:1994cg}
  E.~N.~Rodionov, A.~W.~Thomas and J.~T.~Londergan,
  Mod. Phys. Lett. A {\bf 9}, 1799 (1994).

\bibitem{Londergan:2003ij} 
  J.~T.~Londergan and A.~W.~Thomas,
  Phys. Rev. D {\bf 67}, 111901 (2003).

\bibitem{Londergan:2009kj}
  J.~T.~Londergan, J.-C.~Peng and A.~W.~Thomas,
  Rev. Mod. Phys. {\bf 82}, 2009 (2010).

\bibitem{Cloet:2009qs} 
  I.~C.~Clo\"et, W.~Bentz and A.~W.~Thomas,
  Phys. Rev. Lett. {\bf 102}, 252301 (2009).

\end{thebibliography}
\end{document}